\documentclass[12pt]{article}
\usepackage{amsfonts, graphicx, verbatim, mathtools,amssymb, amsthm, mathrsfs}
\usepackage{tikz}
\usetikzlibrary{arrows.meta,positioning}
\usepackage{epstopdf}
\usepackage{graphicx}

\usepackage{epsfig,amsmath,amssymb,mathrsfs}
\usepackage[utf8]{inputenc}
\usepackage[colorlinks=true,citecolor=blue,,linktocpage=true,linkcolor=blue,urlcolor=black]{hyperref}
\usepackage{mathtools,slashed}
\usepackage{array}
\usepackage{xcolor}
\usepackage[force]{feynmp-auto}
\usepackage{caption}
\usepackage{subcaption}
\usepackage{overpic,youngtab}
\usepackage[latin,english]{babel}
\usepackage{amsmath}
\usepackage{amssymb}
\usepackage{epstopdf}
\usepackage{graphics,psfrag,rotating}
\usepackage{mathabx}
\usepackage{graphicx}
\usepackage{dcolumn}
\usepackage{float}
\usepackage{pdflscape}
\usepackage{array}
\usepackage{booktabs}
\usepackage{amscd}
\usepackage{mathtools}
\usepackage{fancybox}
\usepackage{fix-cm}
\usepackage[colorlinks=true,citecolor=blue,,linktocpage=true,linkcolor=blue,urlcolor=black]{hyperref}
\usepackage{tikz}
\usetikzlibrary{matrix}
\usetikzlibrary{positioning}
\usetikzlibrary{arrows}

\renewenvironment{itemize}
  {\begin{list}%
     {}%
     {\setlength{\topsep}{0pt}%
      \setlength{\partopsep}{0pt}%
      \setlength{\itemsep}{-4pt}%
      \setlength{\labelsep}{5pt}%
      \setlength{\itemindent}{0pt}%
     }%
  }%
  {\end{list}}%

\makeatletter
\renewcommand{\section}{\setcounter{equation}{0}\@startsection
 {section}%
 {1}%
 {0pt}%
 {-1\baselineskip}%
 {0.4\baselineskip}%
 {\bfseries\large}}%
\renewcommand{\subsection}{\@startsection
 {subsection}%
 {2}%
 {0pt}%
 {-0.75\baselineskip}%
 {0.2\baselineskip}%
 {\bfseries}}%
\renewcommand{\subsubsection}{\@startsection
 {subsubsection}%
 {3}%
 {0pt}%
 {-0.5\baselineskip}%
 {0.1\baselineskip}%
 {\sc}}%
\makeatother

\DeclareMathAlphabet{\mathpzc}{OT1}{pzc}{m}{it}

\usepackage{tikz}
\usetikzlibrary{patterns,snakes}
\usetikzlibrary{shapes.misc}
\usetikzlibrary{decorations.markings,decorations.pathmorphing}
\usetikzlibrary{calc}
\tikzstyle{spring}=[line width=0.8,black,snake=coil,segment amplitude=4.25,segment length=4.75,line cap=round]

\def\be{\begin{equation}}
\def\ee{\end{equation}}

\def\g5{\gamma_{5}}

\def\id3k{\int\!\! \dfrac{d^3\!\vec{k}}{(2\pi)^3 }}

\newcommand{\bea}{\begin{eqnarray}}
\newcommand{\eea}{\end{eqnarray}}
\newcommand{\beann}{\begin{eqnarray*}}
\newcommand{\eeann}{\end{eqnarray*}}
\newcommand{\ba}{\begin{array}}
\newcommand{\ea}{\end{array}}

\newcommand{\tr}{\mathbf{tr}}

 \def\g {\gamma}

\newcommand{\email}[1]{\href{mailto:#1}{\tt #1}}

\begin{document}

\rightline{\scriptsize{FT/UCM 112/2025}}
\vglue 50pt

\begin{center}

{\LARGE  \bf Quantum noncommutative ABJM field theory:\\ four- \& six-point functions}\\
\vskip 1.0true cm
{\Large \bf Carmelo P. Martin$^1$, Josip Trampeti\' c$^{2,3}$\\ and Jiangyang You$^{4}$
}
\\
\vskip .7cm
{
    {$^1$Universidad Complutense de Madrid (UCM), Departamento de Física Teórica and IPARCOS, Facultad de Ciencias Físicas, 28040 Madrid, Spain\\
$^2$Rudjer Bo\v skovi\' c Institute, Divisions of Experimental Physics, HR-10002 Zagreb, Croatia\\
$^3$Max-Planck-Institut f\"ur Physik, (Werner-Heisenberg-Institut), Boltzmannstraße 8, 85748 Garching bei München,  Germany\\
$^4$Rudjer Bo\v skovi\' c Institute, Divisions of Physical Chemistry, HR-10002 Zagreb, Croatia	}
	
	\vskip .5cm
	\begin{minipage}[l]{.9\textwidth}
		\begin{center}
			\textit{E-mail:}
			\email{carmelop@fis.ucm.es, josip.trampetic@irb.hr, jiangyang.you@irb.hr}.
			
		\end{center}
	\end{minipage}
}
\end{center}
\thispagestyle{empty}

\begin{abstract}
Following our previous paper “Quantum noncommutative ABJM theory: first steps” [JHEP {\bf 1804} (2018) 070)] \cite{Martin:2017nhg}, in this article we investigate 1-loop 1PI 4- \& 6-point functions by using the component formalism in the Landau gauge and show that they are UV finite and have well defined $\theta^{\mu\nu}\rightarrow 0$ limit. Those results also hold  for all 1-loop functions which are UV finite by power-counting. In summary, taking into account results from previous \cite{Martin:2017nhg} and this paper, we conclude that, at least at the 1-loop  order, the NCABJM theory is free from the noncommutative UV and IR instabilities, and that in the limit $\theta^{\mu\nu}\rightarrow 0$  it flows to the ordinary ABJM theory.
\end{abstract}

\vfill
\clearpage

\section{Introduction}

Quantum field theory (QFT) constructed by Aharony, Bergman, Jafferis and Maldacena (known under authors names ABJM theory \cite{Aharony:2008ug}), spanned on the Moyal noncommutative (NC) space was introduced in \cite{Martin:2017nhg} as NCABJM QFT, and shown to be ${\cal N}=6$ supersymmetric, like the undeformed ordinary ABJM theory. In \cite{Martin:2017nhg} we have computed all 1-loop 1PI 2- and 3-point functions, and shown that they are UV finite and have well-defined commutative limits $\theta^{\mu\nu}\rightarrow 0$, corresponding exactly to the 1PI functions of the ordinary ABJM theory. This article represents further analysis of one-loop 1PI four- \& six-point functions of the NCABJM field theory. In the above the $\theta^{\mu\nu}=c^{\mu\nu}/\Lambda^2_{\rm NC}$ represents the NC deformation parameter, antisymmetric $4\times4$ matrix with ${\rm dimension[\theta^{\mu\nu}]}=[mass^{-2}]=[length^2]$. The $c^{\mu\nu}$ are dimensionless coefficients of order one,  while $\Lambda_{\rm NC}$ is the scale of noncommutativity. Note that proposed ordinary ABJM quantum field theory \cite{Aharony:2008ug},
is a relative\footnote{However, for ${\cal N}$ = 4 NCSYM the spacetime dimension is $D=4$, whereas it is $D=3$ for  NCABJM; hence, it cannot be stated that both theories must have the same properties regarding their UV and IR behaviour.} of the ${\cal N}$ = 4 Supersymmetric Yang-Mills (SYM) theory, and it represents holographic dual of the M theory on the $AdS_4\times S^7/Z_k$ space, contributing to a deeper understanding of celebrated gauge/gravity duality conjecture \cite{Maldacena:1997re}. This NCQFT needs to be analyzed on the quantum level because it gives possibility to search for quantum gravity on four dimensional Minkowski spacetime \cite{Drukker:2010nc,Dabholkar:2014wpa}. In addition, the NCABJM QFT may be useful as the effective theory describing condensed matter systems where physics is determined by the Chern-Simons action \cite{Asano:2004hy,Ferrari:2005kx,Ammon:2015wua,Bea:2014yda}.

The ABJM field theory, at the level $\kappa$ is the super-conformal ${\cal N} =6$ symmetric \cite{Bandres:2008ry},  and this symmetry is enhanced to ${\cal N} =8$, when  $\kappa=1$ or $\kappa=2$ \cite{Kwon:2009ar}.  Also, the  ABJM theory was formulated on the ${\cal N} =3$ harmonic superspace \cite{Buchbinder:2008vi}, and shown that such superfield perturbation theory is UV finite \cite{Buchbinder:2009dc}.

Important to note that the on-shell scattering amplitude techniques \cite{Arkani-Hamed:2017mur} have been used to work out some tree- and one-loop level scattering amplitudes in the ABJM field theory too \cite{Bargheer:2010hn,Bargheer:2012cp,Bianchi:2012cq}. These computations have uncover algebraic ---the Yangian of the corresponding super-conformal algebra \cite{Lee:2010du}--- and geometric ---the orthogonal Grassmannian  \cite{Elvang:2014fja}--- structures that  play a big role in the analysis of such theory integrability \cite{Beisert:2017pnr}.

Noncommutative quantum field theories (NCQFT)  \cite{Seiberg:1999vs,Gomis:2000zz,Aharony:2000gz,Trampetic:2003fu,Szabo:2001kg,Trampetic:2005ib} spanned on the Moyal space, is well established research field in the High Energy Theoretical Physics, containing recent applications of the scattering amplitudes technique to it \cite{Raju:2009yx,Huang:2010fc,Martin:2016hji,Martin:2016saw,Mizera:2019blq,Latas:2020nji,Trampetic:2021awu,Trampetic:2022tij,Pisarski:1985yj}. In the early days, by investigating the 1-loop quantum properties of the Moyal NC Yang-Mills (NCYM) theories \cite{Martin:1999aq}, it was discovered the appearance of the IR divergences leading to the celebrated effect of the UV/IR mixing  \cite{Bigatti:1999iz,Matusis:2000jf,Seiberg:2000gc,Seiberg:2000ms,Minwalla:1999px,Hayakawa:1999yt,VanRaamsdonk:2000rr,VanRaamsdonk:2001jd,Ferrari:2004ex,Schupp:2008fs,Horvat:2011bs,Horvat:2013rga,Horvat:2015aca,Martin:2016zon,Martin:2020ddo}.  On the $\kappa$-Minkowski and the Snyder noncommutative spaces for the $\phi^4$ theory the same mixing effect was discovered too \cite{Grosse:2005iz,Magnen:2008pd,Blaschke:2009aw,Meljanac:2011cs,Meljanac:2017grw,Meljanac:2017jyk}. A bit later it was explicitly demonstrated that the quadratic (tachyonic) IR divergences get eliminated in the Supersymmetric version of the Moyal NCYM theory \cite{Hanada:2014ima}. 
Finally, very recently the NCQFT was constructed and studied as a more general twist-noncommutative deformations of gauge theory relevant to the AdS/CFT \cite{Meier:2023kzt,Meier:2023lku}. We learned that the $\phi^4$ theory on a quadratically-noncommutative spacetime, obtained from a Drinfeld twist-noncommutative deformations of gauge theory, does show the UV/IR mixing effect \cite{ST}, which together with the same effect on Moyal, $\kappa$-Minkowski and Snyder quantum-plane types, increases possibility that the effect of UV/IR mixing is an universal quantum property of any NCQFT’s. However, the relevant answer to that fundamental question is still at large.

In searching for the signal of spacetime noncommutativity at high energies it was applied the most simple model of the NCQFT to the particle and astro particle physics phenomenology in various physical environments (from Earth’s  laboratories to Cosmology), for various physical processes (forbidden, invisible and rare particle decays, scatterings, productions, annihilations), and generally for various areas of physical manifestations in: the Early Universe, Big Bang Nucleosynthesis, Reheating Phase After Inflation, Vacuum Birefringence, Ultra High Energy Cosmic Rays, Cosmogenic Left and Right (Sterile) Neutrinos, Holography, Weak Gravity Conjecture, Hierarchy Problem, and the Entanglement, \cite{Hewett:2000zp,Mathews:2000we,Behr:2002wx,Schupp:2002up,Minkowski:2003jg,Abel:2006wj,Horvat:2010sr,Horvat:2010km,Horvat:2011iv,Horvat:2011wh,Horvat:2011qn,Horvat:2011qg,Horvat:2012vn,Trampetic:2015zma,Horvat:2017gfm,Horvat:2017aqf,Selvaganapathy:2019jkm,Horvat:2020ycy,Trampetic:2023qfv,Lust:2017wrl,Craig:2019zbn,Cribiori:2025oek,Martin:2025hzm}, respectively. From those prospectives note that quantum ABJM field theory on the Moyal noncommutative spacetime could be helpful --through the gauge/gravity correspondence-- in studying the noncommutative gravity in four dimensions and in the study of the Fractional Quantum Hall effect \cite{Fradkin:2002qw}.

The main purpose of this paper is to rigorously show at the one-loop level, that the four-point and six-point functions of the noncommutative ABJM theory have a well-defined  $\theta^{\mu\nu}\to 0$ limit and that this limit is given by the ordinary/commutative ABJM theory \cite{Aharony:2008ug}. We do this by applying the Lebesgue’s dominated convergence theorem of Fourier transformed theory when the Feynman integrals we meet are both UV and IR divergent for non-exceptional momenta, on the one side \cite{Velo,Rudin}. On the other side we explicitly compute dangerous contributions --which, of course, are not finite by power-counting (see ref. \cite{Velo})-- that may jeopardise the existence of  vanishing $\theta^{\mu\nu}$ limit. The occurrence of these dangerous contributions makes {\it the existence of a well-defined $\theta^{\mu\nu}\rightarrow 0$ limit a highly nontrivial issue}. We shall carry out our analysis in the component formalism in the Landau gauge to avoid the spurious IR divergent behaviour that  occurs when the superfield formalism is used with a local gauge-fixing term\footnote{In reference \cite{Leoni:2010az}, it is discussed the way to cure this IR divergent behaviour by using a non-local gauge-fixing term which depends on the  dimensional regularization regulator $\epsilon=(3-D)/2$.} and to be able to apply the Lebesgue's theorem (see ref. \cite{Rudin}) in a straightforward way. We do this by computing and analysing all 1-loop 1PI functions in the noncommutative variant of the $\rm U(1)_{\kappa}\times U(1)_{-\kappa}$ theory, following our first steps paper \cite{Martin:2017nhg}.

It is advisable to make some comments regarding the relation between the existence of the limit $\theta^{\mu\nu}\to 0$ of the 1PI Green functions of the NCABJM theory and the UV finiteness of the ordinary ABJM theory established in \cite{Buchbinder:2009dc}. Since, the Moyal phases introduce a partial regularization of the 1PI Green functions of the theory, one may be tempted to conclude that if the corresponding ordinary theory is UV finite, then, the limit in which this partial regularization is removed should exist. A priori, there is no guarantee that this will be the case, for the ABJM theory is not UV finite by power-counting; hence, the individual Feynman diagrams in general are not UV finite, and UV finiteness is achieved as a result of the cancellation among several UV divergent contributions produced by summing appropriate sets of Feynman diagrams (see subsection 5.3). As consequence of the partial regularization not being Lorentz invariant, the cancellation we have just mentioned may leave behind a bounded, in the limit $\theta^{\mu\nu}\to 0$, contribution which has not well-defined limit. An example of an integral which is finite --in the sense that remains bounded-- in the limit $\theta^{\mu\nu}\to 0$ can be found in (\ref{6.2}); of course, the limit $\theta^{\mu\nu}\to 0$ of that integral does not exist. Hence, we stress again that, even at one-loop, {\it the existence of the limit $\theta^{\mu\nu}\to 0$ is a highly nontrivial issue}, which has to be analyzed carefully. 

We should also mention that it has not been shown that for noncommutative Chern-Simons-Matter theories --which are formulated at $D=3$-- the noncommutative IR divergences which occur as a result of the UV/IR mixing cancel if there is enough supersymmetry; in particular, it has not been proved that the logarithmic noncommutative IR divergences vanish if the beta functions of the corresponding ordinary theory also vanish. This is unlike the situation for the NCYM theories in $D=4$ --see \cite{Matusis:2000jf}.

For convenience, from \cite{Martin:2017nhg} we repeat the field content and the free field propagators of the noncommutative $\rm U(1)_{\kappa}\times U(1)_{-\kappa}$ ABJM quantum field theory actions in sections 2, and 3, respectively.  Additional Feynman rules relevant to our computations are given in the appedix A. Let us point out that we quantize the theory in the Landau gauge for two reasons: First, the Chern-Simons propagator is simpler and second, it does not contain contributions with a dangerous IR behaviour --see section III of ref. \cite{Pisarski:1985yj}. In sections 4 and further we compute and discuss all 1PI 1-loop 4-point functions of the $\rm U(1)_{\kappa}\times U(1)_{-\kappa}$ NCABJM quantum field theory. Finally in section 8 we have also discussed the scalar 1-loop 6-point functions, and their limiting properties too. \footnote{The 1-loop n-point functions are sometimes shortly named as the n-correlators, respectively.}

\section{The $\rm U(N)_{\kappa}\times U(N)_{-\kappa}$ quantum field theory action}

Field content of the $\rm U(N)_{\kappa}\times U(N)_{-\kappa}$ ABJM theory consists of four $\rm N \times N$ matrices of complex scalars $(X_A)^a{}_{\dot{a}}$ and their adjoints $(X^A)^{\dot{a}}{}_a$. These transform as $({\bf
\bar N}, {\bf N})$ and $({\bf N}, {\bf \bar N})$ representations of
the gauge group, respectively.  Similarly, the spinor fields are
matrices $(\Psi^A)^a{}_{\dot{a}}$ and their adjoints $(\Psi_A)^{\dot{a}}{}_a$. The $\rm U(N)$ gauge fields are hermitian matrices $A^a{}_b$ and $\hat A^{\dot{a}}{}_{\dot{b}}$. In matrix notation, the covariant derivatives are
\begin{equation}
D_\mu X_A = \partial_\mu X_A  +i (A_\mu X_A - X_A\hat A_\mu),\;\;
D_\mu X^A = \partial_\mu X^A  + i( \hat A_\mu X^A - X^A A_\mu),
\label{5}
\end{equation}
with similar formulas for spinors. Infinitesimal gauge transformations with respect to the gauge ($A_\mu$),
hgauge ($\hat A_\mu$), and the scalar ($X_A$) fields are given by
\begin{equation}
\delta A_\mu = D_\mu \Lambda = \partial_\mu \Lambda+i [A_\mu,\Lambda],\;
\delta \hat A_\mu = D_\mu \hat\Lambda = \partial_\mu\hat\Lambda +i [\hat A_\mu,\hat\Lambda],\;
\delta X_A =-i\Lambda X_A +i X_A \hat\Lambda,
\label{8}
\end{equation}
and so forth. The action consists of terms that are straightforward generalizations of those of  ordinary $\rm {U(1)_{\kappa} \times U(1)_{-\kappa}}$ ABJM field theory, as well as the new interaction terms that vanish for N=1. The noncommutative Chern--Simons plus kinetic term with additional four and six fields terms actions in three dimensions are:
\begin{eqnarray}
S&=&S_{\rm CS}+S_{\rm kin}+S_{4}+S_{6},
\label{Action}\\
S_{\rm CS} &=& \frac{\kappa}{2\pi} \int d^3 x  \, \epsilon^{\mu\nu\lambda} \tr\bigg(
\frac{1}{2} A_\mu\star \partial_\nu A_\lambda  + \frac{i}{3} A_\mu\star A_\nu\star A_\lambda-\frac{1}{2} \hat A_\mu\star \partial_\nu \hat A_\lambda - \frac{i}{3} \hat A_\mu\star \hat A_\nu\star \hat A_\lambda\bigg),
\nonumber\\
\label{ACS}\\
S_{\rm kin} &=& \frac{\kappa}{2\pi} \int d^3 x \,\tr\left( - D^\mu X^A\star D_\mu
X_A +i\bar\Psi_A \star\slashed{D} \Psi^A\right),
\label{Akin}\\
S_{4}&=&S_{4a}+S_{4b}+S_{4c},
\label{AS4}\\
S_{4a}&=&\frac{i\kappa}{2\pi} \int d^3 x\,\tr\Big[
\epsilon^{ABCD} (\bar\Psi_A\star  X_B\star \Psi_C\star  X_D)
-\epsilon_{ABCD}( \bar\Psi^A \star X^B \star\Psi^C \star X^D )\Big],
\label{AS4a}\\
S_{4b}&=&\frac{i\kappa}{2\pi} \int d^3 x\,\tr\Big[
 \bar\Psi^A \star\Psi_A \star X_B \star X^B -  \bar\Psi_A
\star\Psi^A \star X^B \star X_B\Big],
\label{AS4b}\\
S_{4c}&=&\frac{i\kappa}{2\pi} \int d^3 x\,\tr\Big[
2 ( \bar\Psi_A\star \Psi^B\star X^A\star X_B)
- 2( \bar\Psi^A\star\Psi_B \star X_A \star X^B)\Big],
\label{AS4c}
\end{eqnarray}
\begin{eqnarray}
S_6&=&  -\frac{1}{6}\frac{\kappa}{2\pi} \int d^3 x\,\tr(N^{IA}\star N^I_A)
\nonumber\\
&=&  \frac{1}{3}\frac{\kappa}{2\pi} \int d^3 x\,\tr \Big[
X^A\star X_A\star X^B\star X_B\star X^C\star X_C
+ X_A\star X^A\star X_B\star X^B\star X_C\star X^C
 \nonumber\\
&&\phantom{xx}
+4 X_A\star X^B\star X_C\star X^A\star X_B\star X^C
-6 X^A\star X_B\star X^B\star X_A\star X^C\star X_C \Big],
 \label{AS6}
\end{eqnarray}
where
\begin{eqnarray}
N^{IA}&=&\tilde\Gamma^{IAB}\Big(X_C\star X^C\star X_B-X_B\star X^C\star X_C\Big)-2\tilde\Gamma^{IBC}X_B\star X^A\star X_C,
\nonumber\\
 N^I_A&=&\Gamma^I_{AB}\Big(X^C\star X_C\star X^B-X^B\star X_C\star X^C\Big)-2\Gamma^{IBC}X^B\star X_A\star X_C,
\label{AS6N}
\end{eqnarray}
with $\Gamma^I_{AB}$ being $4\times 4$ matrices, generators of the $\rm SO(6)$ group, satisfying:
\begin{eqnarray}
\Gamma^I_{AB}&=&-\Gamma^I_{BA},\,\forall I=1,...,6, \;
\big\{\Gamma^I \Gamma^J+\Gamma^J \Gamma^I\big\}=2\delta^{IJ},\;\;N^I_A=\big(N^{IA}\big)^\dagger.
\nonumber\\
\tilde\Gamma^I&=&(\Gamma^I)^\dagger\;
\Longleftrightarrow\;
\tilde\Gamma^{IAB}=(\Gamma^I_{BA})^{*}=-(\Gamma^I_{AB})^{*}=
\frac{1}{2}\epsilon^{ABCD}\Gamma^I_{CD},
\label{AS6N}
\end{eqnarray}
The coefficients in three possible structures for the $\Psi^2 X^2$ terms are chosen so that it gives correct result required by supersymmetry. Certain properties are discussed and demonstrated in details in the main text and the  Appendix of ref. \cite{Bandres:2008ry}.

Next we give the noncommutative gauge-fixing plus ghost terms explicitly:
\begin{eqnarray}
S_{\rm gf+ghost}=-\frac{\kappa}{2\pi} \int d^3 x \Big[\frac{1}{2\xi}\partial_\mu A^\mu\star\partial_\nu A^\nu-\bar\Lambda\star\partial_\mu D^\mu\Lambda-\frac{1}{2\xi}\partial_\mu \hat A^\mu\star\partial_\nu \hat A^\nu+\bar{\hat\Lambda}\star\partial_\mu D^\mu\hat\Lambda\Big],
\nonumber\\
\label{GfGh}
\end{eqnarray}
where the above covariant derivative is defined as usual: $D^\mu \Phi=\partial^\mu\Phi+i[A^\mu,\Phi]$, for arbitrary field $\Phi$.

\section{Free field propagators from the action}

In this paper we shall use the Landau gauge which amounts to the following setting of the gauge parameter: $\xi=0$, after having worked out free gauge propagators.

\begin{figure}[t]
\begin{center}
\includegraphics[width=15cm]{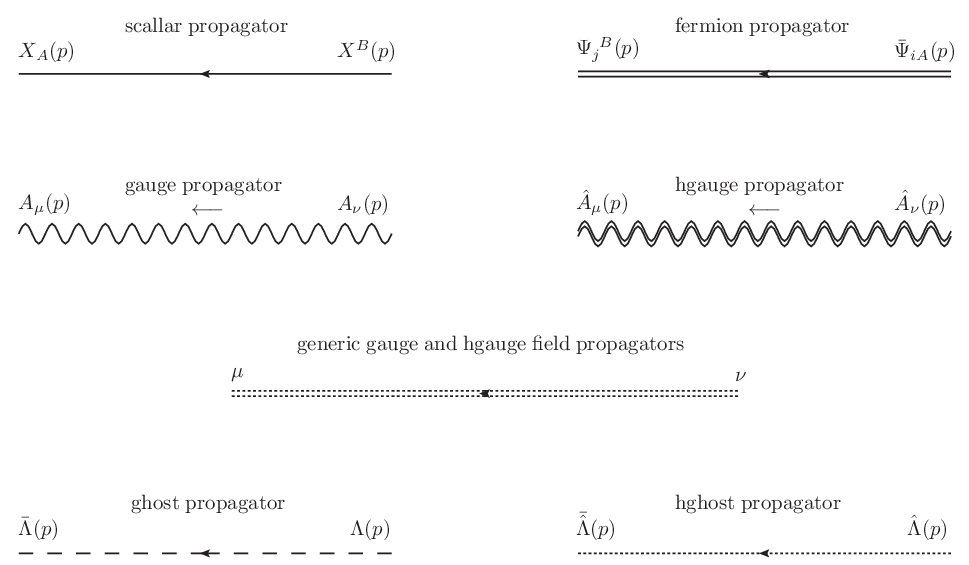}
\end{center}
\caption{Notations and the propagators of the relevant fields. Separate arrows in (h)gauge fields indicate the momentum flow.}
\label{fig:notationprop1}
\end{figure}

Diagramatic notations of the relevant fields in our theory in accord with figure \ref{fig:notationprop1},
like free gauge field $A^\mu(\xi=0)$, hgauge field $\hat A^\mu(\xi=0)$, ghost $\Lambda$ and hghost $\hat\Lambda$, scalar $X^A$, and finally fermion $\psi^A_i$ fields, together with their propagators in momentum space are given next, respectively:
\begin{eqnarray}
A_\nu\to A_\mu:\;\;&\Longrightarrow& \frac{2\pi}{\kappa}\Big(\frac{-\epsilon_{\mu\nu\rho}p^\rho}{p^2}\Big),\;\;\,\,
\phantom{xxx}
\hat A_\nu\to\hat A_\mu:\;\;\Longrightarrow \frac{2\pi}{\kappa}\Big(\frac{\epsilon_{\mu\nu\rho}p^\rho}{p^2}\Big),
\label{prophA}\\
\Lambda\to\bar\Lambda:\;\;&\Longrightarrow& \frac{2\pi}{\kappa}\Big(\frac{-i}{p^2}\Big),\;\;\;\,
\phantom{xxxxxxxxx}
\hat\Lambda\to\bar{\hat\Lambda}:\;\;\Longrightarrow \frac{2\pi}{\kappa}\Big(\frac{-i}{p^2}\Big),
\label{prophL}\\
X^B\to X_A:\;\;&\Longrightarrow& \frac{2\pi}{\kappa}\Big(\frac{-i}{p^2}\Big)\delta_A{}^B,\;\;\,\,
\phantom{xx.}
\bar\Psi_{Ai}\to\Psi_j{}^B:\;\;\Longrightarrow \frac{2\pi}{\kappa}\Big(\frac{-i{\slashed p}_{ij}}{p^2}\Big)\delta_A{}^B.
\label{prophS,F}
\end{eqnarray}
The interaction vertices Feynman rules are derived following the conventional procedure. Results are listed in the appendix D of \cite{Martin:2017nhg}, and in the appendix A of this paper. 
We use clockwise circles for the star-product ordering.

\section{Scalar field four-point functions}

Total one-loop scalar field 4-correlators $S_{4X}\equiv\big<X^{A_1}X_{A_2}X_{A_3}X^{A_4}\big>$
is the sum of the contributions from the sum of diagrams given in figures \ref{fig:Fig21}-\ref{fig:Fig4-6tad}
\begin{equation}
S_{4X}=S^{tad}+S^{bub}+S^{tri}+S^{box},
\label{S4X}
\end{equation}
where $S^{box}$, $S^{tri}$, and $S^{bub}$ denotes contributions from eight box, eight triangle,
and nine bubble diagrams, represented partly in figures \ref{fig:Fig21}-\ref{fig:Fig41}, and denoted as:
\begin{equation}
S^{box}=\sum_{r=1}^8 S^{box}_r,\;\;S^{tri}=\sum_{r=1}^8 S^{tri}_r,\;\;S^{bub}=\sum_{r=1}^9 S^{bub}_r.
\label{btb}
\end{equation}
There is also six scalar fields tadpole contribution $S^{tad}$ from a single loop diagram, [up to the $\pi$-,$\sigma$-permutations of all fields in (\ref{AS6})], given in figure \ref{fig:Fig4-6tad}, respectively. Note that number of contributions within (\ref{S4X}) actually do not exists due to the absence of relevant contributing terms in the complete action $S$ (\ref{Action}).

\subsection{Box diagram contributions to the 4 scalar field 4-point functions}

\begin{figure}[t]
\begin{center}
\includegraphics[width=15cm,angle=0]{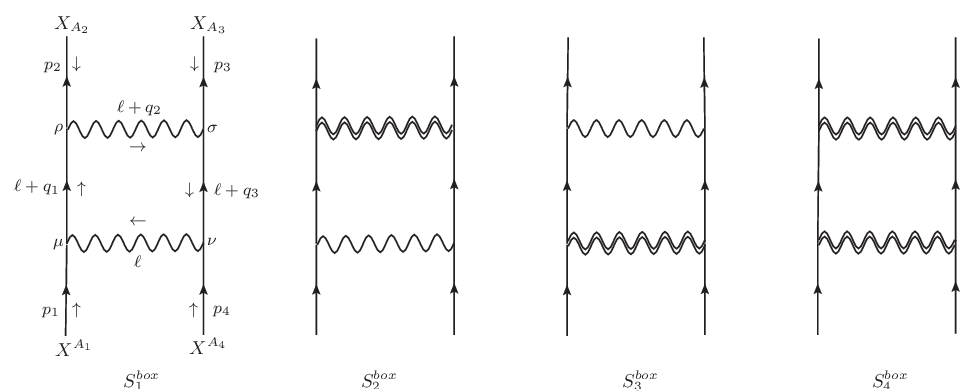}
\end{center}
\caption{Scalar field 4-correlator with (h)gauge-scalar box-loop diagrams $S^{box}_{1,2,3,4}$ defined in (\ref{box1})-(\ref{N4}). Arrows on diagram lines show the flow of charge, while separate arrows indicate the flow of incoming momenta on all four diagrams.}
\label{fig:Fig21}
\end{figure}

From figure \ref{fig:Fig21} we obtain the following integrals 
\begin{eqnarray}
&&S^{box}_r(q_1,q_2,q_3;\theta)=\int \frac{d^D \ell}{(2\pi)^D}
\frac{N^{box}_r(\ell,q_1,q_2,q_3;\theta)}{\ell^2(\ell+q_1)^2(\ell+q_2)^2(\ell+q_3)^2},\;\;r=1,...,4,
\label{box1}\\
&&N^{box}_1(\ell,q_1,q_2,q_3;\theta)=+16\cdot\delta^{A_1}_{A_2}\delta^{A_4}_{A_3}
\,e^{-\frac{i}{2}(q_1+q_3)\theta q_2}e^{-i\ell\theta q_2} f_q^{box}(\ell,q_1,q_2,q_3),
\label{N1}\\
&&N^{box}_2(\ell,q_1,q_2,q_3;\theta)=-16\cdot\delta^{A_1}_{A_2}\delta^{A_4}_{A_3}
\,e^{\frac{i}{2}(q_1+q_3)\theta q_2}e^{-i\ell\theta (q_1-q_2+q_3)} f_q^{box}(\ell,q_1,q_2,q_3),
\label{N2}\\
&&N^{box}_3(\ell,q_1,q_2,q_3;\theta)=-16\cdot\delta^{A_1}_{A_2}\delta^{A_4}_{A_3}
\,e^{-\frac{i}{2}(q_1+q_3)\theta q_2}e^{i\ell\theta (q_1-q_2+q_3)} f_q^{box}(\ell,q_1,q_2,q_3),
\label{N3}\\
&&N^{box}_4(\ell,q_1,q_2,q_3;\theta)=+16\cdot\delta^{A_1}_{A_2}\delta^{A_4}_{A_3}
\,e^{\frac{i}{2}(q_1+q_3)\theta q_2}e^{i\ell\theta q_2} f_q^{box}(\ell,q_1,q_2,q_3),
\label{N4}
\\
&&f_q^{box}(\ell,q_1,q_2,q_3)=\varepsilon(\ell,q_1,q_3)\Big[\varepsilon(\ell,q_1,q_2)-
\varepsilon(\ell,q_1,q_3)+\varepsilon(\ell,q_2,q_3)+\varepsilon(q_1,q_2,q_3)\Big],
\label{boxfq}
\end{eqnarray}
by employing FORM. There is the following shorthand notation $\varepsilon(a,b,c)=\epsilon_{\mu\nu\rho}a^\mu b^\nu c^\rho$ in equation (\ref{boxfq}). The $f_q^{box}(\ell,q_1,q_2,q_3)$ is polynomial in momenta, with maximal power 2 of loop momentum $\ell$, which ensures that $S_r^{box}, \forall r=1,...,4$, is finite by power-counting. Summing up four above numerators (\ref{N1})-(\ref{N4}) we have
\begin{eqnarray}
&&\sum_{r=1}^4 S^{box}_r(q_1,q_2,q_3;\theta)=(+16)\delta^{A_1}_{A_2}\delta^{A_4}_{A_3}
\int \frac{d^D \ell}{(2\pi)^D}
\frac{f_q^{box}(\ell,q_1,q_2,q_3;\theta)}{\ell^2(\ell+q_1)^2(\ell+q_2)^2(\ell+q_3)^2}
\label{BOX1}\\
&&\cdot\bigg\{e^{\frac{i}{2}(q_1+q_3)\theta q_2}\Big[e^{i\ell\theta q_2}-e^{-i\ell\theta(q_1-q_2+q_3)}\Big]
+e^{-\frac{i}{2}(q_1+q_3)\theta q_2}\Big[e^{-i\ell\theta q_2}-e^{i\ell\theta(q_1-q_2+q_3)}\Big]
\bigg\}.
\nonumber
\end{eqnarray}

Next, according to figure \ref{fig:Fig21} we express needed three internal momenta $(q_1,q_2,q_3)$ used in the first diagram in terms of external momenta $(p_1,p_2,p_3,p_4)$ and vice-versa, necessary to perform the
$f^{box}_q(\ell,q_1,q_2,q_3)$ (\ref{boxfq}) computation in FORM:
\begin{eqnarray}
q_1&=&p_1,\;\;q_2=p_1+p_2,\;\;q_3=p_1+p_2+p_3, \;\;\rm and
\nonumber\\
p_2&=&q_2-q_1,\;\;p_3=q_3-q_2,\;\;p_4=-q_3, \;\;\;p_1+p_2+p_3+p_4=0.
\label{momq}
\end{eqnarray}
Now 
we prepare the following numerators
\begin{eqnarray}
N^{box}_1(\ell,p_1,p_2,p_3;\theta)&=&+16\cdot\delta^{A_1}_{A_2}\delta^{A_4}_{A_3}
\,e^{-\frac{i}{2}(p_1\theta p_2-p_3\theta p_4)}e^{-i\ell\theta(p_1+p_2)} f_q^{box}(\ell,p_1,p_2,p_3),
\nonumber\\
N^{box}_2(\ell,p_1,p_2,p_3;\theta)&=&-16\cdot\delta^{A_1}_{A_2}\delta^{A_4}_{A_3}
\,e^{\frac{i}{2}(p_1\theta p_2-p_3\theta p_4)}e^{-i\ell\theta (p_1+p_3)}f_q^{box}(\ell,p_1,p_2,p_3),
\nonumber\\
N^{box}_3(\ell,p_1,p_2,p_3;\theta)&=&-16\cdot\delta^{A_1}_{A_2}\delta^{A_4}_{A_3}
\,e^{-\frac{i}{2}(p_1\theta p_2-p_3\theta p_4)}e^{i\ell\theta (p_1+p_2)}f_q^{box}(\ell,p_1,p_2,p_3),
\nonumber\\
N^{box}_4(\ell,p_1,p_2,p_3;\theta)&=&+16\cdot\delta^{A_1}_{A_2}\delta^{A_4}_{A_3}
\,e^{\frac{i}{2}(p_1\theta p_2-p_3\theta p_4)}e^{i\ell\theta (p_1+p_3)}f_q^{box}(\ell,p_1,p_2,p_3),
\label{Nb}
\end{eqnarray}
where $f_q^{box}(\ell;p_1,p_2,p_3)\equiv f_q^{box}$ is complicated and long expression needed to be handled by the computer.\\
\begin{figure}[t]
\begin{center}
\includegraphics[width=15cm,angle=0]{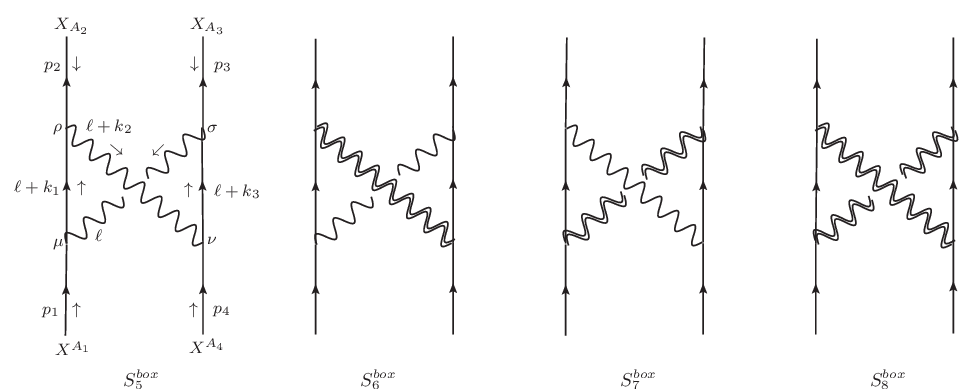}
\end{center}
\caption{Scalar field 4-correlator with (h)guage-scalar box-loop diagrams $S^{box}_{5,6,7,8}$ defined in (\ref{box5})-(\ref{N8}). Arrows on diagram lines indicate the flow of charge, while separate arrows show the flow of all incoming momenta.}
\label{fig:Fig22}
\end{figure}

Second, from figure \ref{fig:Fig22} we compute the following cross integrals using again the above new method
\begin{eqnarray}
S^{box}_r(k_1,k_2,k_3;\theta)&=&\int \frac{d^D \ell}{(2\pi)^D}
\frac{ N_r(\ell,k_1,k_2,k_3;\theta)}{\ell^2(\ell+k_1)^2(\ell+k_2)^2(\ell+k_3)^2},\,\;r=5,...,8,
\label{box5}\\
N^{box}_5(\ell,k_1,k_2,k_3;\theta)&=&+16\cdot\delta^{A_1}_{A_2}\delta^{A_4}_{A_3}
\,e^{-\frac{i}{2}(k_1-k_3)\theta k_2} f_k^{box}(\ell,k_1,k_2,k_3),
\label{N5}\\
N^{box}_6(\ell,k_1,k_2,k_3;\theta)&=&-16\cdot\delta^{A_1}_{A_2}\delta^{A_4}_{A_3}
\,e^{\frac{i}{2}(k_1-k_3)\theta k_2}e^{-i\ell\theta (k_1-k_3)} f_k^{box}(\ell,k_1,k_2,k_3),
\label{N6}\\
N^{box}_7(\ell,k_1,k_2,k_3;\theta)&=&-16\cdot\delta^{A_1}_{A_2}\delta^{A_4}_{A_3}
\,e^{-\frac{i}{2}(k_1-k_3)\theta k_2}e^{i\ell\theta (k_1-k_3)} f_k^{box}(\ell,k_1,k_2,k_3),
\label{N7}\\
N^{box}_8(\ell,k_1,k_2,k_3;\theta)&=&+16\cdot\delta^{A_1}_{A_2}\delta^{A_4}_{A_3}
\,e^{\frac{i}{2}(k_1-k_3)\theta k_2} f_k^{box}(\ell,k_1,k_2,k_3),
\label{N8}
\end{eqnarray}
where $f_k^{box}(\ell,k_1,k_2,k_3)$ is the same as $f_q^{box}(\ell,q_1,q_2,q_3)$ after replacement $q_i\to k_i$.
Summing up four numerators (\ref{N5})-(\ref{N8})
we have
\begin{eqnarray}
&&\sum_{r=5}^8 S^{box}_r(k_1,k_2,k_3;\theta)=(+16)\delta^{A_1}_{A_2}\delta^{A_4}_{A_3}
\int\frac{d^D \ell}{(2\pi)^D} \frac{f_k^{box}(\ell,k_1,k_2,k_3;\theta)}{\ell^2(\ell+k_1)^2(\ell+k_2)^2(\ell+k_3)^2}
\nonumber\\
&&\cdot\bigg\{2\cos{\frac{(k_1-k_3)\theta k_2}{2}}
-e^{\frac{i}{2}(k_1-k_3)\theta k_2}e^{-i\ell\theta (k_1-k_3)}-e^{-\frac{i}{2}(k_1-k_3)\theta k_2}e^{i\ell\theta (k_1-k_3)}\bigg\}
\label{BOX2}
\end{eqnarray}
where cos term gives planar integral, while other two with exponential $\ell\theta (k_1-k_3)$ dependence are non-planar integrals.

Here according to figure \ref{fig:Fig22} we give definitions of three internal momenta $k_i$ in terms of external momenta $p_i$ and vice-versa, needed during the $f^{box}_k(\ell,k_1,k_2,k_3)$ computation in FORM:
\begin{eqnarray}
p_1&=&k_1,\;\;p_2=k_2-k_1,\;\;p_3=-k_3,\;\;p_4=k_3-k_2,\;\;{\rm and}
\nonumber\\
k_1&=&p_1\equiv q_1,\;\;k_2=p_1+p_2\equiv q_2,\;\;k_3=p_1+p_2+p_4\equiv q_3.
\label{momk}
\end{eqnarray}
Using the above definitions of internal momenta $k_i$ in terms of external momenta we obtain
\begin{eqnarray}
N^{box}_5(\ell, p_1,p_2,p_3;\theta)&=&+16\cdot\delta^{A_1}_{A_2}\delta^{A_4}_{A_3}
\,e^{-\frac{i}{2}(p_1\theta p_2-p_3\theta p_4)} f_k^{box}(\ell,p_1,p_2,p_4),
\nonumber\\
N^{box}_6(\ell, p_1,p_2,p_3;\theta)&=&-16\cdot\delta^{A_1}_{A_2}\delta^{A_4}_{A_3}
\,e^{\frac{i}{2}(p_1\theta p_2-p_3\theta p_4)}e^{-i\ell\theta (p_1+p_3)}f_k^{box}(\ell,p_1,p_2,p_4),
\nonumber\\
N^{box}_7(\ell, p_1,p_2,p_3;\theta)&=&-16\cdot\delta^{A_1}_{A_2}\delta^{A_4}_{A_3}
\,e^{-\frac{i}{2}(p_1\theta p_2-p_3\theta p_4)}e^{i\ell\theta (p_1+p_3)}f_k^{box}(\ell,p_1,p_2,p_4),
\nonumber\\
N^{box}_8(\ell, p_1,p_2,p_3;\theta)&=&+16\cdot\delta^{A_1}_{A_2}\delta^{A_4}_{A_3}
\,e^{\frac{i}{2}(p_1\theta p_2-p_3\theta p_4)}f_k^{box}(\ell,p_1,p_2,p_4),
\label{Nbox8}
\end{eqnarray}
where we had to switch $p_3\to p_4$ to obtain needed  $f_k^{box}(\ell;p_1,p_2,p_4)\equiv f_k^{box}$.

Some important comments are in order. Constructing $\star$-orientated scalar-gauge boson box-loop diagrams in figures \ref{fig:Fig21} and \ref{fig:Fig22}, contributing to the 1-loop 4-point functions, we notice that box diagrams are self-orientated so that diagrams with identical gauge boson lines are either planar or non-planar. All integrals from diagrams with mixed fields $S^{box}_{2,3,6,7}$ --one wavy and one double wavy lines-- belong to the non-planar case.

Finally for any further computation/analysis of the box diagrams sum (figures \ref{fig:Fig21} and \ref{fig:Fig22}) at $\theta\not=0$, from eqs.(\ref{box1})-(\ref{Nbox8}) we are writing total $S^{box}$ formulae in terms of external momenta:
\begin{eqnarray}
S^{box}&=&\sum_{r=1}^8 S^{box}_r=\int \frac{d^3 \ell}{(2\pi)^3}\sum_{r=1}^4
\frac{N^{box}_r(\ell,p_1,p_2,p_3;\theta)}{\ell^2(\ell+p_1)^2(\ell+p_1+p_2)^2
(\ell+p_1+p_2+p_3)^2}
\nonumber\\
&+&\int \frac{d^3 \ell}{(2\pi)^3}\sum_{r=5}^8\frac{N^{box}_r(\ell,p_1,p_2,p_3;\theta)}{\ell^2(\ell+p_1)^2(\ell+p_1+p_2)^2(\ell+p_1+p_2+p_4)^2},
\label{BOX3}\\
\sum_{r=1}^4 N^{box}_r&=&-i32\delta^{A_1}_{A_2}\delta^{A_4}_{A_3}
\nonumber\\
&&\cdot\Big[e^{-\frac{i}{2}(p_1\theta p_2-p_3\theta p_4)}\sin\ell\theta (p_1+p_2)
+e^{\frac{i}{2}(p_1\theta p_2-p_3\theta p_4)}\sin\ell\theta (p_1+p_3)\Big]f_q^{box},
\nonumber\\
\sum_{r=5}^8 N^{box}_r&=&16\delta^{A_1}_{A_2}\delta^{A_4}_{A_3}
\nonumber\\
&&\cdot\Big[e^{-\frac{i}{2}(p_1\theta p_2-p_3\theta p_4)}
\Big(1-e^{i\ell\theta (p_1+p_3)}\Big)+e^{\frac{i}{2}(p_1\theta p_2-p_3\theta p_4)}\Big(1-e^{-i\ell\theta (p_1+p_3)}\Big)
\Big]f_k^{box},
\nonumber
\end{eqnarray}
Here one can see trivially from the above sums that at  $\theta^{\mu\nu}=0$ each term is zero by itself, yielding
\begin{equation}
\lim_{\theta\to 0}\;\sum_{r=1}^4 N^{box}_r (p_1,p_2,p_3;\theta)=0,\quad
\lim_{\theta\to 0}\;\sum_{r=5}^8 N^{box}_r (p_1,p_2,p_3;\theta)=0.
\label{limNbox}
\end{equation}
These results, upon using Lebgesgue's dominated convergence theorem, lead to
\begin{equation}
\lim_{\theta\to 0}\; S^{box}=0.
\label{limSbox}
\end{equation}

Indeed, first, we have the following inequalities
\begin{equation}
\Big|\sum_{r=1}^4 N^{box}_r\Big|\leq 64\,\delta^{A_1}_{A_2}\delta^{A_4}_{A_3}\,\big|f_q^{box}\big|,\quad
\Big|\sum_{r=5}^8 N^{box}_r \Big|\leq 64\,\delta^{A_1}_{A_2}\delta^{A_4}_{A_3}\,\big|f_k^{box}\big|;
\label{inequalities}
\end{equation}
where $|c|$ stands for the modulus of the complex number expression denoted by $c$. Secondly, the momentum dependent expressions
\begin{equation*}
\frac{f_q^{box}}{\ell^2(\ell+p_1)^2(\ell+p_1+p_2)^2(\ell+p_1+p_2+p_3)^2},\quad
\frac{f_k^{box}}{\ell^2(\ell+p_1)^2(\ell+p_1+p_2)^2(\ell+p_1+p_2+p_4)^2}
\end{equation*}
(with polinomial $f_q^{box}$ and $f_k^{box}$ being defined earlier in the text line after Eqs. (\ref{Nb}) and (\ref{Nbox8}), respectively) are absolutely integrable functions, at $D=3$ and for the non-exceptional momenta, as a consequence of the power-counting theorem \cite{Velo} for Feynman integrals in Euclidean space. This result and the inequalities in (\ref{inequalities}) lead to the conclusion that the integrals in (\ref{BOX3}) satisfy the hypothesis of Lebesgue's dominated convergence theorem \cite{Rudin} and, hence, using (\ref{limNbox}) we have
\begin{eqnarray}
&{\;\;}&\lim_{\theta\rightarrow 0}\,\int \frac{d^3 \ell}{(2\pi)^3}\sum_{r=1}^4
\cfrac{N^{box}_r(\ell,p_1,p_2,p_3;\theta)}{\ell^2(\ell+p_1)^2(\ell+p_1+p_2)^2
(\ell+p_1+p_2+p_3)^2}
\nonumber\\
&=&\int \frac{d^3 \ell}{(2\pi)^3}\,\lim_{\theta\rightarrow 0}\,\sum_{r=1}^4
\cfrac{N^{box}_r(\ell,p_1,p_2,p_3;\theta)}{\ell^2(\ell+p_1)^2(\ell+p_1+p_2)^2
(\ell+p_1+p_2+p_3)^2}\,=\,0,
\label{LDTbox1}
\end{eqnarray}
\begin{eqnarray}
&{\;\;}&\lim_{\theta\rightarrow 0}\,\int \frac{d^3 \ell}{(2\pi)^3}\sum_{r=5}^8\cfrac{N^{box}_r(\ell,p_1,p_2,p_3;\theta)}{\ell^2(\ell+p_1)^2(\ell+p_1+p_2)^2(\ell+p_1+p_2+p_4)^2}
\nonumber\\
&=&\int \frac{d^3 \ell}{(2\pi)^3}\,\lim_{\theta\rightarrow 0}\,\sum_{r=5}^8\cfrac{N^{box}_r(\ell,p_1,p_2,p_3;\theta)}{\ell^2(\ell+p_1)^2(\ell+p_1+p_2)^2(\ell+p_1+p_2+p_4)^2}=0,
\label{LDTbox2}
\end{eqnarray}
which imply (\ref{limSbox}).

Since the commutative limits exist for the each sum in (\ref{BOX3}) of box diagrams from figures \ref{fig:Fig21} and \ref{fig:Fig22} the eq. (\ref{limSbox}) is proven, i.e. the commutative limit $\theta^{\mu\nu}\to 0$ exist for the sum of all eight diagrams contributing to the $S^{box}$ from (\ref{btb}).\\
$\phantom{XXXXXXXXXXXXXXXXXXXXXXXXXXXXXXXXXXXXXXXxi}$Q.E.D.\\

Finally, to compute $\theta$-nonvanishing and/or $\theta$-final 4-point function $S^{box}$ (\ref{BOX3}) we need integrals $I_2^0,I_3^0,I_4^0$, given in appendices B, C, D, respectively. However, inspecting completed formulae for $\theta\not=0$ in $S^{box}$ we realize that it is too lengthy, very much cumbersome and non-transparent, thus the complete computation of $\theta$-nonvanishing part has to be performed by the computer.

\subsection{Triangle diagram contributions to the 4 scalar field 4-point functions}

\begin{figure}[t]
\begin{center}
\includegraphics[width=15cm,angle=0]{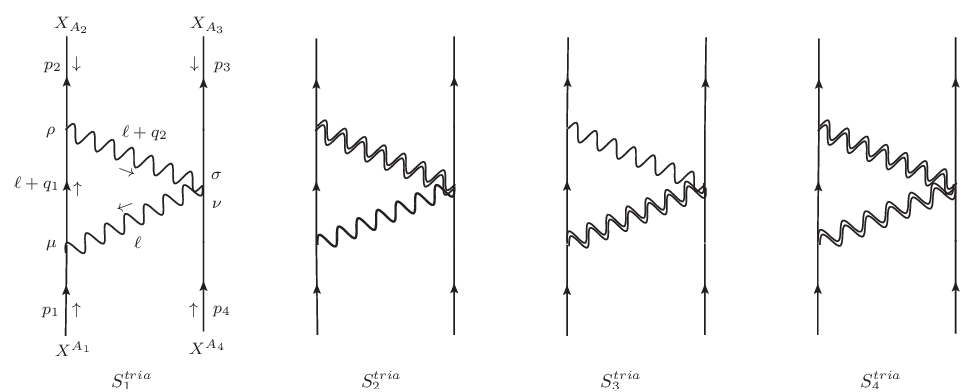}
\end{center}
\caption{Scalar field 4-correlator with triangle-(h)gauge loop diagrams $S^{tri}_{1,2,3,4}$. Arrows on diagram lines show the flow of charge, while separate arrows indicate the flow of incoming momenta.}
\label{fig:Fig31}
\end{figure}

From figure \ref{fig:Fig31} we compute the following contributions also using new method
\begin{eqnarray}
&&S^{tri}_r(q_1,q_2,q_3;\theta)=\int \frac{d^D \ell}{(2\pi)^D}
\frac{N^{tri}_r(\ell,q_1,q_2,q_3;\theta)}{\ell^2(\ell+q_1)^2(\ell+q_2)^2},\;\;r=1,...,4,
\label{tria1}\\
&&N^{tri}_1(\ell,q_1,q_2,q_3;\theta)=+4\cdot\delta^{A_1}_{A_2}\delta^{A_4}_{A_3}
\,e^{-\frac{i}{2}(q_1+q_3)\theta q_2}\Big[1+e^{-i\ell\theta q_2} \Big] f_q^{tri}(\ell,q_1,q_2,q_3),
\label{Nt1}\\
&&N^{tri}_2(\ell,q_1,q_2,q_3;\theta)=-8\cdot\delta^{A_1}_{A_2}\delta^{A_4}_{A_3}
\,e^{\frac{i}{2}(q_1+q_3)\theta q_2}e^{-i\ell\theta(q_1-q_2+q_3)} f_q^{tri}(\ell,q_1,q_2,q_3),
\label{Nt2}\\
&&N^{tri}_3(\ell,q_1,q_2,q_3;\theta)=-8\cdot\delta^{A_1}_{A_2}\delta^{A_4}_{A_3}
\,e^{-\frac{i}{2}(q_1+q_3)\theta q_2}e^{i\ell\theta(q_1-q_2+q_3)} f_q^{tri}(\ell,q_1,q_2,q_3),
\label{Nt3}\\
&&N^{tri}_4(\ell,q_1,q_2,q_3;\theta)=+4\cdot\delta^{A_1}_{A_2}\delta^{A_4}_{A_3}
\,e^{\frac{i}{2}(q_1+q_3)\theta q_2}\Big[1+e^{i\ell\theta q_2} \Big] f_q^{tri}(\ell,q_1,q_2,q_3),
\label{Nt4}\\
&&f_q^{tri}(\ell,q_1,q_2,q_3)=\varepsilon(\ell,q_1,q_3)\Big[\varepsilon(\ell,q_1,q_3)-
\varepsilon(\ell,q_2,q_3)-\varepsilon(q_1,q_2,q_3)\Big].
\label{trifq}
\end{eqnarray}
Definition of $\varepsilon(a,b,c)$ is the same as before $\varepsilon(a,b,c)=\epsilon_{\mu\nu\rho}a^\mu b^\nu c^\rho$, and $f_q^{tri}(\ell,q_1,q_2,q_3)$ is polynomial in momenta. Products of $\varepsilon(a,b,c)$ in (\ref{trifq}) shall further be replaced with the relevant contractions and computed by computer using FORM, like we did for (\ref{boxfq})-(\ref{Nb}).

Summing up four diagrams from figure \ref{fig:Fig31} we have:
\begin{eqnarray}
&&\sum_{r=1}^4 S^{tri}_r(q_1,q_2,q_3;\theta)=(+4)\delta^{A_1}_{A_2}\delta^{A_4}_{A_3}
\int \frac{d^D \ell}{(2\pi)^D}\frac{f_q^{tri}(\ell,q_1,q_2,q_3;\theta)}{\ell^2(\ell+q_1)^2(\ell+q_2)^2}\bigg[
2\cos{\frac{(q_1+q_3)\theta q_2}{2}}
\nonumber\\
&&+e^{-\frac{i}{2}(q_1+q_3)\theta q_2}\Big(e^{-i\ell\theta q_2}-2e^{i\ell\theta(q_1-q_2+q_3)}\Big)+
e^{\frac{i}{2}(q_1+ q_3)\theta q_2}\Big(e^{i\ell\theta q_2}-2e^{-i\ell\theta(q_1-q_2+q_3)}\Big)\bigg].
\nonumber\\
\label{TRI1}
\end{eqnarray}
In triangles from figure \ref{fig:Fig31} we encounter both, the upper--lower asymmetries and decomposition of unordered into different ordered diagrams.

In the first diagram of figure \ref{fig:Fig31} are indicated two needed internal momenta $(q_1,q_2)$. Thus,  we also give definitions of the external momenta $(p_1,p_2,p_3,p_4)$ in terms of internal momenta and vice-versa, necessary to be used during the computation of $f_q^{tri}$-terms (\ref{trifq}) in FORM:
\begin{eqnarray}
p_1&=&q_1,\;\;p_2=q_2-q_1,\;\;p_3=q_3-q_2,,\;\;p_4=-q_3,\;\;{\rm and}
\nonumber\\
q_3&=&p_1+p_2+p_3,\;\;q_2=p_1+p_2.
\label{momqt}
\end{eqnarray}
From (\ref{momqt}) follows $f_q^{tri}(\ell,q_1,q_2,q_3)\to f_p^{tri}(\ell,p_1,p_2,p_3)$, giving
\begin{eqnarray}
N^{tri}_1(\ell, p_1,p_2,p_3;\theta)&=&+4\cdot\delta^{A_1}_{A_2}\delta^{A_4}_{A_3}
\,e^{-\frac{i}{2}(p_1\theta p_4+p_2\theta p_3)}\Big[1+e^{-i\ell\theta (p_1+p_2)} \Big] f_p^{tri}(\ell,p_1,p_2,p_3),
\nonumber\\
N^{tri}_2(\ell, p_1,p_2,p_3;\theta)&=&-8\cdot\delta^{A_1}_{A_2}\delta^{A_4}_{A_3}
\,e^{\frac{i}{2}(p_1\theta p_4+p_2\theta p_3)}e^{-i\ell\theta (p_1+p_3)}f_p^{tri}(\ell,p_1,p_2,p_3),
\nonumber\\
N^{tri}_3(\ell, p_1,p_2,p_3;\theta)&=&-8\cdot\delta^{A_1}_{A_2}\delta^{A_4}_{A_3}
\,e^{-\frac{i}{2}(p_1\theta p_4+p_2\theta p_3)}e^{i\ell\theta (p_1+p_3)}f_p^{tri}(\ell,p_1,p_2,p_3),
\nonumber\\
N^{tri}_4(\ell, p_1,p_2,p_3;\theta)&=&+4\cdot\delta^{A_1}_{A_2}\delta^{A_4}_{A_3}
\,e^{\frac{i}{2}(p_1\theta p_4+p_2\theta p_3)}\Big[1+e^{i\ell\theta (p_1+p_2)} \Big]f_p^{tri}(\ell,p_1,p_2,p_3).
\label{Nt4’}
\end{eqnarray}
Here again $f_p^{tri}(\ell,p_1,p_2,p_3)$, as a polynomial in momenta with maximal power 2 of loop momentum
$\ell$, is long expression need to be handled by the computer using definition (\ref{trifq}).

Important to note is that once one assigns the above external momenta there are two types of triangle diagrams, the ($I$) and ($II$), respectively:\\
($I$) For any further computation of the sum of four triangle diagrams at $\theta^{\mu\nu}\not=0$ from eqs.(\ref{Nt4’}), we are giving the first compact type of contributions $ S_I^{tri}$ from figure \ref{fig:Fig31}, like a sum (\ref {TRI1}), but in terms of external momenta:
\begin{eqnarray}
S_I^{tri}&=&\sum_{r=1}^4 S^{tri}_r=\int \frac{d^D \ell}{(2\pi)^D} \frac{1}{\ell^2(\ell+p_1)^2(\ell+p_1+p_2)^2}
\sum_{r=1}^4N^{tri}_r(\ell,p_1,p_2,p_3) ,
\label{TRI2}\\
&&\hspace{-2cm}\sum_{r=1}^4 N^{tri}_r(\ell,p_1,p_2,p_3)=4\cdot\delta^{A_1}_{A_2}\delta^{A_4}_{A_3}
\Big[e^{\frac{i}{2}(p_1\theta p_3+p_2\theta p_4)}
\Big(1+e^{i\ell\theta (p_1+p_2)}-2e^{-i\ell\theta (p_1+p_3)}\Big)
\nonumber\\
&+&e^{\frac{-i}{2}(p_1\theta p_3+p_2\theta p_4)}
\Big(1+e^{-i\ell\theta (p_1+p_2)}-2e^{i\ell\theta (p_1+p_3)}\Big)\Big]f_p^{tri}(\ell,p_1,p_2,p_3),
\nonumber
\end{eqnarray}
from where we trivially see vanishing of numerator (\ref{TRI2}) in the ${\theta\to 0}$ limit.\\
($II$) The second compact type $S_{II}^{tri}$ can be obtained from the first one by simple exchange of external momenta. So we shall have altogether sum of eight diagrams, i.e. eight contributions.

Inspecting and comparing carefully this paper expressions (\ref{tria1})-(\ref{TRI2}), and relevant integrals in this paper appendices  B.2.3, B.3.3, C.3.1, C.3.2, with analysis regarding integrals $\hat I_1,I_+,I_+^\mu$, Eqs.(8.18-8.25) from \cite{Martin:2017nhg}, one can see that this paper integrals, being the same types as $\hat I_1,I_+,I_+^\mu$ in \cite{Martin:2017nhg}, are both UV and IR finite by power-counting. Therefore we can apply Lebesque’s theorem \cite{Velo,Rudin} to the each of sums $S_{I}^{tri}$ and $S_{II}^{tri}$ separately, and  due to the same arguments as for
the boxes we conclude that the commutative limit for the full sum holds:
\begin{equation}
\lim_{\theta\to 0}\;S^{tri}=\lim_{\theta\to 0}\;(S_I^{tri}+S_{II}^{tri})=\lim_{\theta\to 0}\;\sum_{r=1}^8 S^{tri}_r (p_1,p_2,p_3;\theta)=0.
\label{limStria}
\end{equation}
$\phantom{XXXXXXXXXXXXXXXXXXXXXXXXXXXXXXXXXXXXXXXX.}$Q.E.D.

Additionally, in the appendix E.3 by direct computations of generic sample diagram {\it(2.)} in figure \ref{4scalar}, we have shown that the commutative limit $\theta^{\mu\nu}\to 0$ exist for each individual triangle diagram in figure \ref{fig:Fig31} and for each of the additional four diagrams with exchanged external momenta, i.e. the commutative limit for the sum of eight contributions (\ref{limStria}), also holds.

To obtain $\theta$-final parts of the 4-point function $S^{tri}$ we need relevant integrals $I_2^0$, $I_3^0$, and $I_3^1$ from appendices B, and C, respectively. Since the full $S^{tri}$ expression for $\theta\not=0$ is too long, cumbersome and non-transparent, it has to be handled by the computer, like the $S^{box}$.

\subsection{Bubble diagram contributions to the 4 scalar field 4-point functions}

\begin{figure}[t]
\begin{center}
\includegraphics[width=15cm,angle=0]{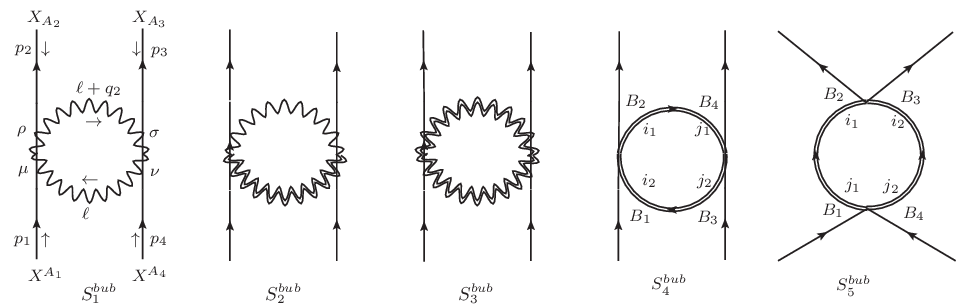}
\end{center}
\caption{Scalar field 4-correlator with gauge and fermion bubble-loops diagrams: $S^{bub}_{1,2,3}$, and $S^{bub}_{4,5}$, respectively. Here arrows on the diagram lines indicate the flow of charge, while separate arrows indicate the flow of incoming momenta on all five diagrams. In $S_2^{bub},S_3^{bub},S_4^{bub}$ diagrams,  momenta are denoted in the same way as in $S_1^{bub}$, while for the $S_5^{bub}$ loop we have the following two propagators momenta: $\ell$, and $-(\ell+p_2+p_3)$, respectively.}
\label{fig:Fig41}
\end{figure}
From figures \ref{fig:Fig41} we have to compute relevant diagrams by using FORM again:
\begin{eqnarray}
S^{bub}=\sum_{r=1}^9 S^{bub}_r&=&\int \frac{d^D \ell}{(2\pi)^D}\bigg(
\sum_{r=1}^4\frac{N^{bub}_r(\ell,p_1,p_2,p_3;\theta)}{\ell^2(\ell+p_1+p_2)^2}
+\sum_{r=5}^8\frac{N^{bub}_r(\ell,p_1,p_2,p_3;\theta)}{\ell^2(\ell+p_1+p_3)^2}\bigg)
\nonumber\\
&+&\int \frac{d^D \ell}{(2\pi)^D}\frac{N^{bub}_9(\ell,p_1,p_2,p_3;\theta)}{\ell^2(\ell+p_2+p_3)^2},
\label{bub9}
\end{eqnarray}
where we have expressed bubble diagram contributions from the very beginning in terms of external momenta, due to the necessity for any further computation of the 4-point functions at general $\theta^{\mu\nu}\not=0$ case.

Here, due to the computations in FORM, we also have to express the external momenta $(p_1,p_2,p_3,p_4)$ as functions of set of the internal momenta  $(q_1,q_2,q_3)$: $p_1=q_1$, $p_2=q_2-q_1$, $p_3=q_3-q_2$, $p_4=-q_3$ and vice-versa $q_2=p_1+p_2$, $q_3=p_1+p_2+p_3$. Note that we indicated, in the first diagram from Figs.\ref{fig:Fig41},  only one needed internal moment in bubble loops as $q_2$.
\begin{eqnarray}
N^{bub}_1(\ell,p_1,p_2,p_3;\theta)&=&-2\cdot\delta^{A_1}_{A_2}\delta^{A_4}_{A_3}
\,e^{-\frac{i}{2}(p_1\theta p_2-p_1\theta p_3-p_2\theta p_3)}
\Big[1+\cos{\big(\ell\theta (p_1+p_2)\big)} \Big]
\nonumber\\
& \cdot& \big(\ell^2+\ell (p_1+p_2)\big),
\label{Nb1}\\
N^{bub}_2(\ell,p_1,p_2,p_3;\theta)&=&+8\cdot\delta^{A_1}_{A_2}\delta^{A_4}_{A_3}
\,e^{-\frac{i}{2}(p_1\theta p_2-p_1\theta p_3-p_2\theta p_3)}e^{i\ell\theta(p_1+p_3)}
\nonumber\\
&\cdot&
\big(\ell^2+\ell(p_1+p_2)\big),
\label{Nb2}\\
N^{bub}_3(\ell,p_1,p_2,p_3;\theta)&=&-2\cdot\delta^{A_1}_{A_2}\delta^{A_4}_{A_3}
\,e^{\frac{i}{2}(p_1\theta p_2-p_1\theta p_3-p_2\theta p_3)}\Big[1+\cos{\big(\ell\theta (p_1+p_2)\big)} \Big]
\nonumber\\
&\cdot&
\big(\ell^2+\ell(p_1+p_2)\big),
\label{Nb3}\\
N^{bub}_4(\ell,p_1,p_2,p_3;\theta)&=&-32\cdot\delta^{A_1}_{A_3}\delta^{A_4}_{A_2}
\,\sin\frac{p_1\theta p_2-\ell\theta (p_1+p_2)}{2}
\nonumber\\
&\cdot&\sin\frac{p_3\theta(p_1+p_2)+\ell\theta(p_1+p_2)}{2}\,
\cdot\big(\ell^2+\ell(p_1+p_2)\big),
\label{Nb4}
\end{eqnarray}
\begin{eqnarray}
N^{bub}_5(\ell,p_1,p_2,p_3;\theta)&=&-2\cdot\delta^{A_1}_{A_3}\delta^{A_4}_{A_2}
\,e^{\frac{i}{2}(p_1\theta p_2-p_1\theta p_3-p_2\theta p_3)}
\Big[1+\cos{\big(\ell\theta (p_1+p_3)\big)} \Big]
\nonumber\\
& \cdot& \big(\ell^2+\ell (p_1+p_3)\big),
\label{Nb5}\\
N^{bub}_6(\ell,p_1,p_2,p_3;\theta)&=&+8\cdot\delta^{A_1}_{A_3}\delta^{A_4}_{A_2}
\,e^{\frac{i}{2}(p_1\theta p_2-p_1\theta p_3-p_2\theta p_3)}e^{i\ell\theta(p_1+p_2)}
\nonumber\\
&\cdot&
\big(\ell^2+\ell(p_1+p_3)\big),
\label{Nb6}\\
N^{bub}_7(\ell,p_1,p_2,p_3;\theta)&=&-2\cdot\delta^{A_1}_{A_3}\delta^{A_4}_{A_2}
\,e^{-\frac{i}{2}(p_1\theta p_2-p_1\theta p_3-p_2\theta p_3)}\Big[1+\cos{\big(\ell\theta (p_1+p_3)\big)} \Big]
\nonumber\\
&\cdot&
\big(\ell^2+\ell(p_1+p_3)\big),
\label{Nb7}\\
N^{bub}_8(\ell,p_1,p_2,p_3;\theta)&=&-32\cdot\delta^{A_1}_{A_2}\delta^{A_4}_{A_3}
\,\sin\frac{p_1\theta p_3-\ell\theta (p_1+p_3)}{2}
\nonumber\\
&\cdot&\sin\frac{p_2\theta(p_1+p_3)+\ell\theta(p_1+p_3)}{2}\,
\cdot\big(\ell^2+\ell(p_1+p_3)\big),
\label{Nb8}\\
N^{bub}_9(\ell,p_1,p_2,p_3;\theta)&=&-8\cdot\Big(\delta^{A_1}_{A_2}\delta^{A_4}_{A_3}-\delta^{A_1}_{A_3}\delta^{A_4}_{A_2}\Big)\big(\ell^2+\ell(p_2+p_3)\big)
\label{Nb9}\\
&\cdot& \bigg\{e^{i\ell\theta (p_1+p_3)}e^{-\frac{i}{2}(p_1\theta p_2+p_1\theta p_3-p_2\theta p_3)}
-e^{-i\ell\theta (p_1+p_2)}e^{\frac{i}{2}(p_1\theta p_2+p_1\theta p_3+p_2\theta p_3)}
\nonumber\\
&-&
e^{i\ell\theta(p_1+p_2)}e^{-\frac{i}{2}(p_1\theta p_2+p_1\theta p_3+p_2\theta p_3)}
+ e^{-i\ell\theta(p_1+p_3)}e^{\frac{i}{2}(p_1\theta p_2+p_1\theta p_3-p_2\theta p_3)}\bigg\}.
\nonumber
\end{eqnarray}
In the above $N^{bub}_5$ to $N^{bub}_8$ contributions are obtained from the first four diagrams in figures \ref{fig:Fig41} with exchange $(A_2,p_2) \longleftrightarrow(A_3,p_3)$. The ninth term $N^{bub}_9$ is obtained from the fifth diagram in figures \ref{fig:Fig41} and by permuting indices and momenta in $N^{bub}_1$, $N^{bub}_4$ and $N^{bub}_5$, $N^{bub}_8$, respectively. The planar parts are self-evident integrals with no $\ell_\mu\theta^{\mu\nu}$-dependent NC phase factor. In figures \ref{fig:Fig41} diagrams with one double-wavy and one wavy lines, like $S_2^{bub}$, are non-planar.

Inspecting each numerator (\ref{Nb1})-(\ref{Nb9}) one could trivially see vanishing of each combination $N^{bub}_4$, $N^{bub}_8$, $N^{bub}_9$, ($N^{bub}_1+N^{bub}_2+N^{bub}_3$) and ($N^{bub}_5+N^{bub}_6+N^{bub}_7$) by itself in the $\theta^{\mu\nu}=0$ point, i.e. that for the numerator sum commutative limit holds:
\begin{equation}
\lim_{\theta\to 0}\; N^{bub}(p_1,p_2,p_3;\theta)=\lim_{\theta\to 0}\;\sum_{r=1}^9 N^{bub}_r (p_1,p_2,p_3;\theta)=0.
\label{limNbub}
\end{equation}
Finally, integrating (\ref{bub9}) in $D=3$ with abbreviations $\tilde p^{\mu}=\theta^{\mu\nu}p_\nu$ gives:

\begin{eqnarray}
S^{bub}_1(p_1,p_2,p_3;\theta)&=&-2\cdot\delta^{A_1}_{A_2}\delta^{A_4}_{A_3}
\,e^{-\frac{i}{2}(p_1+p_3)\theta(p_1+p_2)}
\nonumber\\
& \cdot&
\Big[I_1^0(\tilde p_1+\tilde p_2)-\frac{(p_1+p_2)^2}{2}I^0_2(p_1+p_2;\tilde p_1+\tilde p_2)\Big],
\label{Sb1}\\
S^{bub}_2(p_1,p_2,p_3;\theta)&=&+8\cdot\delta^{A_1}_{A_2}\delta^{A_4}_{A_3}
\Big[\cos{\frac{(p_1+p_3)\theta( p_1+ p_2)}{2}} I_1^0(\tilde p_1+\tilde p_3)
\nonumber\\
&-&\frac{(p_1+p_2)^2}{2}e^{-\frac{i}{2}(p_1+p_3)\theta(p_1+ p_2)}
I^0_2(p_1+p_2;\tilde p_1+\tilde p_3)\Big],
\label{Sb2}\\
S^{bub}_3(p_1,p_2,p_3;\theta)&=&-2\cdot\delta^{A_1}_{A_2}\delta^{A_4}_{A_3}
\,e^{\frac{i}{2}(p_1+p_3)\theta(p_1+p_2)}
\nonumber\\
& \cdot&
\Big[I_1^0(\tilde p_1+\tilde p_2)-\frac{(p_1+p_2)^2}{2}I^0_2(p_1+p_2;\tilde p_1+\tilde p_2)\Big],
\label{Sb3}\\
S^{bub}_4(p_1,p_2,p_3;\theta)&=&-16\cdot\delta^{A_1}_{A_3}\delta^{A_4}_{A_2}
\nonumber\\
&&\hspace{-3.5cm}\cdot
\bigg\{\cos\frac{(p_1-p_3)\theta(p_1+p_2)}{2}
\Big[I_1^0(\tilde p_1+\tilde p_2)-\frac{(p_1+p_2)^2}{2}I^0_2(p_1+p_2;\tilde p_1+\tilde p_2)\Big]
\nonumber\\
&&\hspace{-3.5cm}
-\cos\frac{(p_1+p_3)\theta(p_1+p_2)}{2}
\int \frac{d^3 \ell}{(2\pi)^3}\frac{\ell^2+\ell(p_1+p_2)}{\ell^2 (\ell+p_1+p_2)^2}\bigg\},
\label{Sb4}\\
S^{bub}_5(p_1,p_2,p_3;\theta)&=&-2\cdot\delta^{A_1}_{A_3}\delta^{A_4}_{A_2}
\,e^{\frac{i}{2}(p_1+p_3)\theta(p_1+p_2)}
\nonumber\\
&&\hspace{-3.5cm}\cdot
\bigg\{\Big[ I_1^0(\tilde p_1+\tilde p_3)-\frac{(p_1+p_3)^2}{2}I^0_2(p_1+p_3;\tilde p_1+\tilde p_3) \Big]
+\int \frac{d^3 \ell}{(2\pi)^3}\frac{\ell^2+\ell(p_1+p_3)}{\ell^2 (\ell+p_1+p_3)^2}\bigg\},
\label{Sb5}\\
S^{bub}_6(p_1,p_2,p_3;\theta)&=&+8\cdot\delta^{A_1}_{A_3}\delta^{A_4}_{A_2}
\bigg\{\cos\frac{(p_1+p_3)\theta(p_1+p_2)}{2}I_1^0(\tilde p_1+\tilde p_2)
\nonumber\\
&-&\frac{(p_1+p_3)^2}{2}\,e^{\frac{i}{2}(p_1+p_3)\theta(p_1+p_2)}
I^0_2(p_1+p_3;\tilde p_1+\tilde p_2)\bigg\},
\label{Sb6}\\
S^{bub}_7(p_1,p_2,p_3;\theta)&=&S^{bub}_3(\ell,p_1,p_2,p_3;\theta)\big|_{2\leftrightarrow3}
=-2\cdot\delta^{A_1}_{A_3}\delta^{A_4}_{A_2}\,e^{-\frac{i}{2}(p_1+p_3)\theta(p_1+p_2)}
\nonumber\\
& \cdot&
\Big[I_1^0(\tilde p_1+\tilde p_3)-\frac{(p_1+p_3)^2}{2}I^0_2(p_1+p_3;\tilde p_1+\tilde p_3)\Big],
\label{Sb7}\\
S^{bub}_8(p_1,p_2,p_3;\theta)&=&S^{bub}_4(\ell,p_1,p_2,p_3;\theta)\big|_{2\leftrightarrow3}
=-16\cdot\delta^{A_1}_{A_2}\delta^{A_4}_{A_3}
\nonumber\\
&&\hspace{-3.5cm}\cdot
\bigg\{\cos\frac{(p_1-p_2)\theta(p_1+p_3)}{2}
\Big[I_1^0(\tilde p_1+\tilde p_3)-\frac{(p_1+p_3)^2}{2}I^0_2(p_1+p_3;\tilde p_1+\tilde p_3)\Big]
\nonumber\\
&&\hspace{-3.5cm}\cdot
\cos\frac{(p_1+p_2)\theta(p_1+p_3)}{2}
\int \frac{d^3 \ell}{(2\pi)^3}\frac{\ell^2+\ell(p_1+p_3)}{\ell^2 (\ell+p_1+p_3)^2}\bigg\},
\label{Sb8}\\
S^{bub}_9(p_1,p_2,p_3;\theta)&=&-8\cdot\Big(\delta^{A_1}_{A_2}\delta^{A_4}_{A_3}
-\delta^{A_1}_{A_3}\delta^{A_4}_{A_2}\Big)
\bigg\{2\cos\frac{(p_1-p_2)\theta(p_1+p_3)}{2}I_1^0(\tilde p_1+\tilde p_3)
\nonumber\\
&&\hspace{-3.5cm}
-\frac{(p_2+p_3)^2}{2}\Big[e^{-\frac{i}{2}(p_1-p_2)\theta(p_1+p_3)}I^0_2(p_2+p_3;\tilde p_1+\tilde p_3)
+e^{\frac{i}{2}(p_1-p_2)\theta(p_1+p_3)}I^0_2(p_2+p_3;-\tilde p_1-\tilde p_3)\Big]
\nonumber\\
&&\hspace{-3.5cm}
-2\cos\frac{(p_1-p_3)\theta(p_1+p_2)}{2}I_1^0(\tilde p_1+\tilde p_2)
\nonumber\\
&&\hspace{-3.5cm}
+\frac{(p_2+p_3)^2}{2}\Big[e^{-\frac{i}{2}(p_1-p_3)\theta(p_1+p_2)}I^0_2(p_2+p_3;\tilde p_1+\tilde p_2)
+ e^{\frac{i}{2}(p_1-p_3)\theta(p_1+p_2)}I^0_2(p_2+p_3;-\tilde p_1-\tilde p_2)\Big]\bigg\}.
\nonumber\\
\label{Sb9}
\end{eqnarray}
Above integrals $I^0_{1,2}$ are given in the appendix B, eqs. (\ref{B.5})-(\ref{B.11}), respectively. 

Since the expressions (\ref{bub9}),(\ref{Sb1})-(\ref{Sb9}) from bubble diagrams in figures \ref{fig:Fig41} contain type of integrals which are not UV finite by power-counting rule, the Lebesque’s theorem \cite{Velo,Rudin} does not hold, thus the complete loop integrations and the limits $\theta^{\mu\nu}\to 0$ in {\rm(\ref{bub9})} do not commute. Consequently, after the full $3D$ integrations the above vanishing of numerators (\ref{limNbub}) does not hold anymore for integrated $S^{bub}$ (\ref{bub9}), i.e.
\begin{equation}
\lim_{\theta\to 0}\;S^{bub}=\lim_{\theta\to 0}\;\sum_{r=1}^9 S^{bub}_r (p_1,p_2,p_3;\theta)\not=0.
\label{limSbub}
\end{equation}
Due to its non-planar parts, the $S^{bub}$ is IR unstable.

So, at this point we shall not continue with the computation of the 4-point function $S^{bub}$ (\ref{bub9}) since it contain IR divergent integral $I_1^0(\tilde k)$ (\ref{B.6}). To handle that situation we first isolate the non-planar parts from the sum $S^{bub}$ (\ref{bub9}) and than add it to the non-planar part of tadpole diagram $S^{tad}$, yielding surprisingly good result presented in the next section.

\section{Non-planar $(\cal NP)$ bubble + tadpole contributions and $\theta^{\mu\nu}$ limits}

\subsection{$\cal (NP)$ bubble diagram contributions to the 4 scalar field 4-point functions}

Using results from figures \ref{fig:Fig41} and eqs. (\ref{Sb1})-(\ref{Sb9}), after some lengthy algebra, we can write the following total bubble contribution coming from the sum in dimensional regularisation non-planar diagrams,
\begin{eqnarray}
&&\sum_{r=1}^9 {(\cal NP)}S^{bub}_r(p_1,p_2,p_3;\theta)=
\nonumber\\
&&\delta_{A_2}^{A_1}\delta_{A_3}^{A_4}
\Big\{
\Big[-4\cos{\frac{(p_1+p_3)\theta(p_1+p_2)}{2}}+16\cos{\frac{(p_1-p_3)\theta(p_1+ p_2)}{2}}\Big]
I^0_1(\tilde p_1+\tilde p_2)
\nonumber\\
&&+\Big[8\cos{\frac{(p_1+p_2)\theta(p_1+ p_3)}{2}}-32\cos{\frac{(p_1-p_2)\theta(p_1+ p_3)}{2}}\Big]
I^0_1(\tilde p_1+\tilde p_3)
\nonumber\\
&&+4(p_1+p_2)^2\cos{\frac{(p_1+p_3)\theta(p_1+ p_2)}{2}}
I^0_2(p_1+p_2;\tilde p_1+\tilde p_2)
\nonumber\\
&&-
4(p_2+p_3)^2\Big[
e^{-\frac{i}{2}(p_1-p_3)\theta( p_1+ p_2)}
I^0_2(p_2+p_3;\tilde p_1+\tilde p_2)
\nonumber\\
&&\hspace{2cm}+e^{\frac{i}{2}(p_1-p_3)\theta(p_1+ p_2)}
I^0_2(p_2+p_3;-\tilde p_1-\tilde p_2)\Big]
\nonumber\\
&&-
4(p_1+p_2)^2
e^{-\frac{i}{2}(p_1+p_3)\theta( p_1+ p_2)}
I^0_2(p_1+p_2;\tilde p_1+\tilde p_3)
\nonumber\\
&&+8(p_1+p_3)^2\cos{\frac{(p_1-p_2)\theta( p_1+ p_3)}{2}}
I^0_2(p_1+p_2;\tilde p_1+\tilde p_3)
\nonumber\\
&&+
4(p_2+p_3)^2\Big[
e^{-\frac{i}{2}(p_1-p_2)\theta(p_1+ p_3)}I^0_2(p_2+p_3;\tilde p_1+\tilde p_3)
\nonumber\\
&&\hspace{2cm}+e^{\frac{i}{2}(p_1-p_2)\theta( p_1+ p_3)}I^0_2(p_2+p_3;-\tilde p_1-\tilde p_3)\Big]\Big\}
\nonumber\\
&&+\Big\{same\;\;as\;\;above\;\;  with\;\;(2\leftrightarrow 3)\Big\}.
\label{tnpbub}
\end{eqnarray}
We repeat, the above tadpole scalar type integral $I^0_1(\tilde k)$ shows linear IR divergence  (\ref{B.6}) when $\tilde k\to0$, while bubble scalar type integral $I^0_2(p;\tilde k)$ has well-defined $\theta^{\mu\nu}\to0$ limit with finite value given in (\ref{B.9}).

\subsection{$\cal (NP)$ tadpole diagram contribution to the 4 scalar field 4-point functions}

Using six-scalar fields vertices, figure \ref{fig:FigFRCorr} from the appendix A, with $\pi-,\sigma-$permutations of all fields expressed by Feynman rules (\ref{A.5}) and (\ref{A.6}), we compute all 1-loop diagrams, generically shown in figure \ref{fig:Fig4-6tad} as a single 1-loop tadpole diagram in terms of external momenta,
\begin{equation}
{(\cal NP)}S^{tad}(p_1,p_4,p_2,p_3;\theta)=\int \frac{d^D \ell}{(2\pi)^D}
\frac{V_6(p_1,p_4,\ell|\ell,p_2,p_3;\theta)}{\ell^2}.
\label{Stad}
\end{equation}
Contracted terms from (\ref{A.5}-\ref{A.6}) are as follows:
\begin{figure}[t]
\begin{center}
\includegraphics[width=12cm,angle=0]{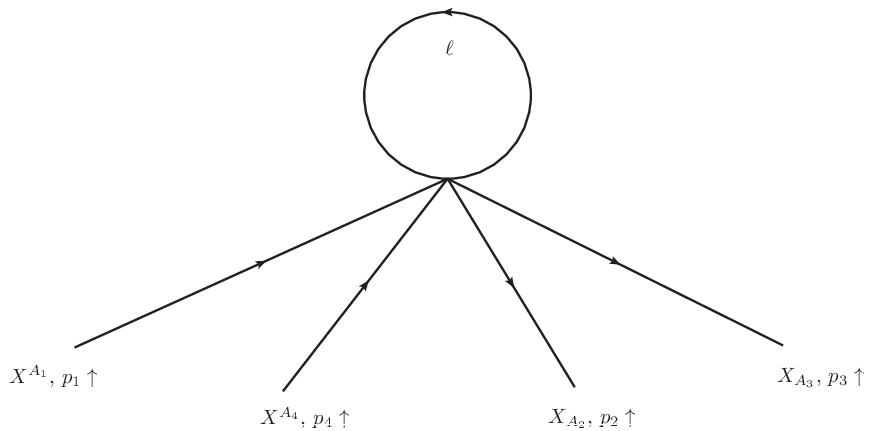}
\end{center}
\caption{Scalar 6-field tadpole diagram contributing to the 4-correlator $S^{tad}$. Arrows on diagram lines indicate flows of charge, while separate up arrows indicate flows of incoming momenta.}
\label{fig:Fig4-6tad}
\end{figure}
\begin{eqnarray}
&&V_6(p_1,p_4,\ell|\ell,p_2,p_3;\theta)=
\nonumber\\
&&\delta_{A_2}^{A_1}\delta_{A_3}^{A_4}
\Big\{-6\;e^{-i\ell\theta(p_1+p_2)}e^{\frac{i}{2}(p_1\theta p_2+p_3\theta p_4)}
+12\;e^{-i\ell\theta(p_1+p_3)}e^{\frac{i}{2}(p_1\theta p_3+p_2\theta p_4)}
\nonumber\\
&&{\;\;\;\;\;\;\;\;\;\;\;\;\,\,}-6\;e^{i\ell\theta(p_1+p_2)}e^{-\frac{i}{2}(p_1\theta p_2+p_3\theta p_4)}
+12\;e^{i\ell\theta(p_1+p_3)}e^{-\frac{i}{2}(p_1\theta p_3+p_2\theta p_4)}
\Big\}
\nonumber\\
&+&\Big\{same\;\;as\;\;above\;\;  with\;\;(2\leftrightarrow 3)\Big\}.
\label{VTadpole}
\end{eqnarray}

Taking into account the $\pi-,\sigma-$permutations of all fields expressed in the Feynman rules (\ref{A.5}-\ref{A.6}) and (\ref{VTadpole}) we found that commutative limit for the numerator holds:
\begin{equation}
\lim_{\theta\to 0}\; V_6 (p_1,p_2,p_3;\theta)=0.
\label{limtad}
\end{equation}
However, since the tadpole type integrals are not UV finite by power-counting rule the loop integration and the limit $\theta^{\mu\nu}\to 0$ do not commute. Finally, after changing variable $\ell\to-\ell^\prime$ in the first line of (\ref{VTadpole}) and integrating (\ref{Stad}), the total tadpole contribution being non-planar in dimensional regularization reads:
\begin{eqnarray}
&&{(\cal NP)}S^{tad}(p_1,p_2,p_3;\theta)=
\delta^{A_1}_{A_2}\delta^{A_4}_{A_3}
\nonumber\\
&\cdot&
\Big\{-12I^0_1(\tilde p_1+\tilde p_2)\cos{\frac{(p_1-p_3)\theta( p_1+ p_2)}{2}}
+24I^0_1(\tilde p_1+\tilde p_3)\cos{\frac{(p_1-p_2)\theta(p_1+ p_3)}{2}}\Big\}
\nonumber\\
&+&\Big\{same\;\;as\;\;above\;\;  with\;\;(2\leftrightarrow3) \Big\}.
\label{Stad1}
\end{eqnarray}

\subsection{The $\theta^{\mu\nu}\to 0$ limits for the $\cal (NP)$ part of bubble plus tadpole diagrams}

Let us repeat. It is plane from previous subsections that the limit $\theta^{\mu\nu}\to 0$ of each box (\ref{B.4}) and each triangle (\ref{B.3}) types of integrals is not divergent, and that this limit agrees with corresponding result in three-dimensional ordinary Minkowski space. Indeed,  if one sets $\theta^{\mu\nu}=0$ in the integrand of the integrals which make up those diagrams one obtains a collection of integrals which are UV finite by power-counting, and therefore the Lebesgue’s theorem \cite{Velo,Rudin} holds, i.e. the limit and the integration do commute, thus the limit $\theta^{\mu\nu}\to 0$ can be taken under the integral sign.

On the other hand, if one sets $\theta^{\mu\nu}= 0$ in the integrand of each bubble diagram, one obtains a Feynman integral which is UV divergent by power-counting, and therefore the non-planar contribution coming from that integral gives rise to the noncommutative IR divergence, since the non-planar part of tadpole diagram for $\theta^{\mu\nu}\to 0$ diverges \footnote{Each bubble diagram contribution (\ref{Sb1})-(\ref{Sb9}) contain integral $I_1^0(\tilde k)$ (\ref{B.5}), linearly divergent for $\tilde k \to0$ (\ref{B.6}), which is source of the UV/IR mixing effect in any Moyal based NCQFT \cite{Matusis:2000jf,Minwalla:1999px,Hayakawa:1999yt,VanRaamsdonk:2000rr,VanRaamsdonk:2001jd,Schupp:2008fs,Horvat:2011bs,Martin:2020ddo}.}. So, the limit $\theta^{\mu\nu}\to 0$ of each bubble diagram does not exist, limits actually diverges.

However, taking into account that all integrals of $I^0_2(p;\tilde k)$ types from (\ref{tnpbub}) have finite value (\ref{B.8}), i.e. they have well-defined limit for $\tilde k^\mu =\theta^{\mu\nu}k_\nu\to 0$ (\ref{B.9}), and since tadpole type integral $I^0_1(\tilde k)$ (\ref{B.6}) in $3D$ has a form:
\begin{equation}
I^0_1(\tilde k)=\frac{Constant}{\sqrt{\tilde k^2}},
\label{1I_1}
\end{equation}
it is clear that the only source which realizes as the noncommutative linear IR divergent contributions to the scalar 4-correlator, is the above integral $I^0_1(\tilde k)$. In eqs. (\ref{tnpbub}) and (\ref{Stad1}) such integrals are $I^0_1({\tilde p}_1 + {\tilde p}_2)$ and $I^0_1({\tilde p}_1 + \tilde p_3)$. And yet, by expanding in power series the cosines in eqs. (\ref{tnpbub}) and (\ref{Stad1}), after summing up all of them and extracting IR divergent integrals only, we obtain the following leading contribution
\begin{eqnarray}
&&\hspace{-1cm}{(\cal NP)}\Big[S^{bub}(p_1,p_2,p_3;\theta)+S^{tad}(p_1,p_2,p_3;\theta)\Big]
\nonumber\\
&=&{\Big\{\delta^{A_1}_{A_2}\delta^{A_4}_{A_3}
\big[-4+16-12\big]+\delta^{A_1}_{A_3}\delta^{A_4}_{A_2}
\big[8-32+24\big]\Big\}I^0_1({\tilde p}_1+{\tilde p}_2)}
\nonumber\\
&+&\Big\{\delta^{A_1}_{A_3}\delta^{A_4}_{A_2}\big[-4+16-12\big]+\delta^{A_1}_{A_2}\delta^{A_4}_{A_3}\big[8-32+24\big]
\Big\}I^0_1({\tilde p}_1+{\tilde p}_3),
\label{I_1+2}
\end{eqnarray}
Sum of all these non-planar divergent contributions obviously vanishes, so it is not divergent when $\theta^{\mu\nu}\to 0$, and what is very important, it has a well-defined finite limit.

The next-to-leading  and higher order contributions from the sum of bubble $S^{bub}$ (\ref{tnpbub}), and tadpole $S^{tad}$ (\ref{Stad1}) contributions, are of the type
\begin{equation}
\frac{{\tilde k}^{\mu_1}}{\sqrt{{\tilde k}^2}} {\tilde k}^{\mu_2}...{\tilde k}
^{\mu_{2n}}T_{\mu_1...\mu_{2n}}(p_1,p_2,p_3),
\label{tildek1}
\end{equation}
with $n\geq 1$, ${\tilde k}^{\mu}=({\tilde p}_1 + {\tilde p}_2)^\mu$ or $({\tilde p}_1 + {\tilde p}_3)^\mu$,
and $T_{\mu_1...\mu_{2n}}(p_1,p_2,p_3)$ being a Lorentz tensor which is a polynomial in terms of external momenta $(p_1,p_2,p_3)$. Each of these contributions vanishes as ${\tilde k}^{\mu}\to 0$.

We conclude that the limit $\theta^{\mu\nu}\to 0$ of the non-planar part obtained after summing bubble and tadpole contributions (\ref{tnpbub})+(\ref{Stad1}) exists, and coincides with the result that is obtained by setting $\theta^{\mu\nu}=0$ in the integrand of relevant integrals arising from the bubble plus tadpole diagrams in figures  \ref{fig:Fig41} and \ref{fig:Fig4-6tad}, respectively. 
$\phantom{So, in this Sections we have shown vanishing of all discussed scalar-gauge and zerosXX}$Q.E.D.\\

Let us stress another fact here, that is that the cancellation of the noncommutative IR divergence upon adding all non-planar contributions does not imply by itself that the limit $\theta^{\mu\nu}\to 0$ exists for a contributions of the type
\begin{equation}
I^0_1(\tilde p_1+\tilde p_2)\sin{\frac{(p_1-p_3)\theta(p_1+ p_2)}{2}},
\label{5.9}
\end{equation}
which are not divergent as $\theta^{\mu\nu}\to 0$, but its value depends on the way $\theta^{\mu\nu}$ approaches to zero. It is the fact that cosines, rather than sines, occur in (\ref{tnpbub}) and (\ref{Stad1}), which is instrumental in having a well-defined limit.

\section{Scalar-guage four-point functions}

\subsection{Scalar-scalar-gauge-hgauge 4-point function: $\big<XXA\hat A\big>$}
\begin{figure}
\begin{center}
\includegraphics[width=15cm,angle=0]{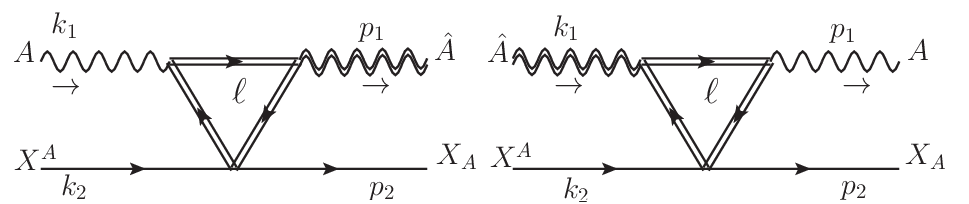}
\end{center}
\caption{2scalar-gauge-hgauge 4-correlator with fermion triangle loop diagrams.
In-out momenta $\{k_1,k_2;p_1,p_2\}$ in this figure correspond to the following set of all incoming momenta $\{p_4,p_1;-p_3,-p_2\}$, in figures \ref{fig:Fig21}-\ref{fig:Fig41}, respectively. Separate arrows indicate momentum flows.}
\label{2scalarghg}
\end{figure}
Scalar-scalar-gauge-hgauge correlator $\big<XXA\hat A\big>$ with the mixed gauges receives only contribution from kinetic terms in the actin (\ref{Akin}) generating fermion triangle-loops in figure \ref{2scalarghg}; see also eqs. (D.8), and (D.9) from appendix D.3 in \cite{Martin:2017nhg}. Instead of three we now have trace of five $\gamma$-matrices:
\begin{equation}
\begin{split}
&\tr \gamma^a\gamma^{\mu}\gamma^b\gamma^{\nu}\gamma^c \ell_a(\ell_b+p_b)(\ell_c+q_c)
\\&=2(g^{a\mu}\epsilon^{b\nu c}-g^{ab}\epsilon^{\mu\nu c}+g^{\mu b}\epsilon^{a\nu c}+g^{\nu c}\epsilon^{a\mu b}) \ell_a(\ell_b+p_b)(\ell_c+q_c),
\end{split}
\label{6.1}
\end{equation}
giving maximal relevant numerator of the $\ell$-power type $\ell^3\sim 2\ell^2\epsilon^{\mu\nu c}\ell_c$. From appendices eqs. (\ref{C.4})-(\ref{C.9}), and eqs. (C.18)-(C.27) in \cite{Martin:2017nhg}, we know that such integrand is in $D=$3 associated with the finite type of integral but with not well-defined commutative limit:
\begin{equation}
\begin{split}
I_2^\mu(p,\tilde k)=\int \frac{d^3 \ell}{(2\pi)^3}\frac{\ell^\mu e^{-i\ell\tilde k}}{\ell^2(\ell-p)^2}
=\frac{1}{8\pi}\,\frac{\tilde k^\mu}{|\tilde k|}+\frac{i}{16}\frac{ p^\mu}{|p|}\,+\mathcal O^{\mu}(\theta^1),
\end{split}
\label{6.2}
\end{equation}
see computation and solution in the appendix B, eqs.(\ref{B.39})-(\ref{B.41}). Namely one can notice that eq. (\ref{6.2}) is bounded as $\tilde k\to0$. However integral is not divergent but it’s limit depends on the way one approaches to the $\tilde k^{\mu}= 0$ point, i.e. leading order depends on the $\tilde k^\mu$ only. The above structure suggests two possible ways of cancellations: \\
($\mathbf{i}$) either two integrals with identical function of $\tilde k^{\mu}$ dependence and opposite relative sign, \\
($\mathbf{ii}$) two integrals with identical strength, i.e. both proportional to the $\pm\tilde k^{\mu}/|\tilde k|$.

In the case of 4-correlator $\big<XXA\hat A\big>$ (figures \ref{2scalarghg}) it is important to notice that exchanging two gauge bosons give opposite relative sign from the Levi-Civita symbol. Explicit computation shows that ($\tilde k^{\mu},p^{\mu}$) dependence remains the same after this permutation, therefore each pair of integrals has not
well-defined $\tilde k^{\mu}\to0$ limit. However, there are two identical terms with opposite signs canceling  each other, which indeed correspond to the ($\mathbf{i}$) case, making total result safely zero.


\subsection{Scalar-scalar-(h)gauge-(h)gauge 4-point functions: $\big<XXAA\big>$, $\big<XX\hat A\hat A\big>$}

\begin{figure}
\begin{center}
\includegraphics[width=15cm,angle=0]{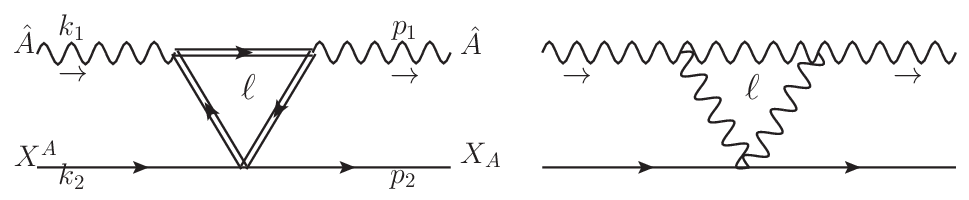}
\end{center}
\caption{2scalar-2gauge 4-correlator with triangle loop diagrams.
Incoming-outgoing momenta $\{k_1,k_2;p_1,p_2\}$ in this figure correspond to the following
set of all incoming momenta $\{p_4,p_1;-p_3,-p_2\}$, in figure \ref{fig:Fig21}-\ref{fig:Fig41}, respectively.
Separate arrows indicate momentum flows.}
\label{2scalar2g}
\end{figure}
\begin{figure}
\begin{center}
\includegraphics[width=15cm,angle=0]{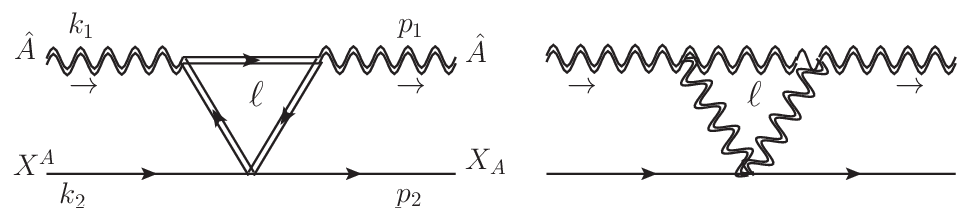}
\end{center}
\caption{2scalar-2hgauge 4-correlator with triangle loop diagrams. Incoming-outgoing momenta $\{k_1,k_2;p_1,p_2\}$ in this figure correspond to the following set of all incoming momenta $\{p_4,p_1;-p_3,-p_2\}$, in figure \ref{fig:Fig21}-\ref{fig:Fig41}, respectively. Separate arrows indicate momentum flows.}
\label{2scalar2hg}
\end{figure}

Two 4-correlators, the $\big<XXAA\big>$ (figure \ref{2scalar2g}) and the $\big<XX\hat A\hat A\big>$ (figure \ref{2scalar2hg}), receive contributions from both, fermion and (h)gauge boson triangle loops, respectively. Results of computed polynomial numerators is equal to the one in (\ref{E.10}), producing integrals evaluated in the appendix E.3. This shows that each of the above correlators vanishes in the limit $\theta^{\mu\nu}\to0$.

In the above subsections we have shown vanishing of all three discussed 2scalar-2gauge 4-correlators: $\big<XXA\hat A\big>$, and $\big<XXAA\big>, \big<XX\hat A\hat A\big>$, in the $\theta^{\mu\nu}\to0$ limit, respectively.\\
$\phantom{So, in this Section we have shown vanishing of all discussed scalar-gauge and is zerX.}$Q.E.D.

\section{Scalar-fermion four-point functions: $\big<XX\bar\Psi\Psi \big>$,$\big<XX\Psi\Psi \big>$,$\big<XX\bar\Psi\bar\Psi \big>$}

\begin{figure}[t]
\begin{center}
\includegraphics[width=15cm,angle=0]{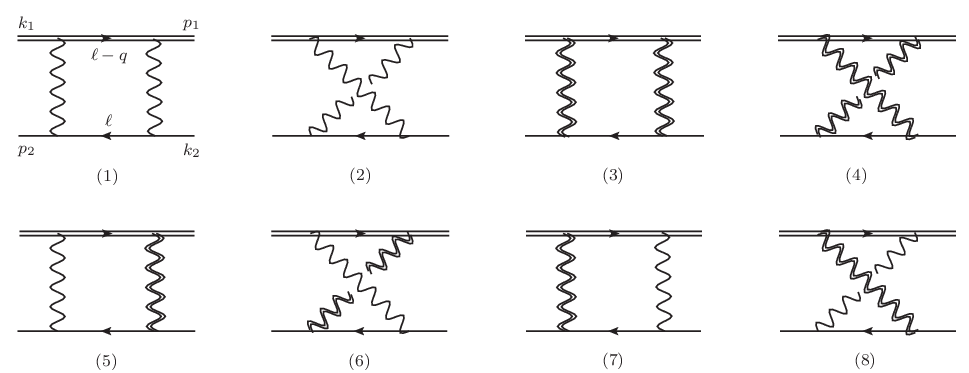}
\end{center}
\caption{Gauge bosons box-loop diagrams contributing to the 2scalar--2fermion 4-correlator. Due to fermions we use the in-out momentum flow $\{k_1,k_2;p_1,p_2\}$ in this figure, respectively.}
\label{2scalar2fermionbox}
\end{figure}
\begin{figure}[t]
\begin{center}
\includegraphics[width=15cm,angle=0]{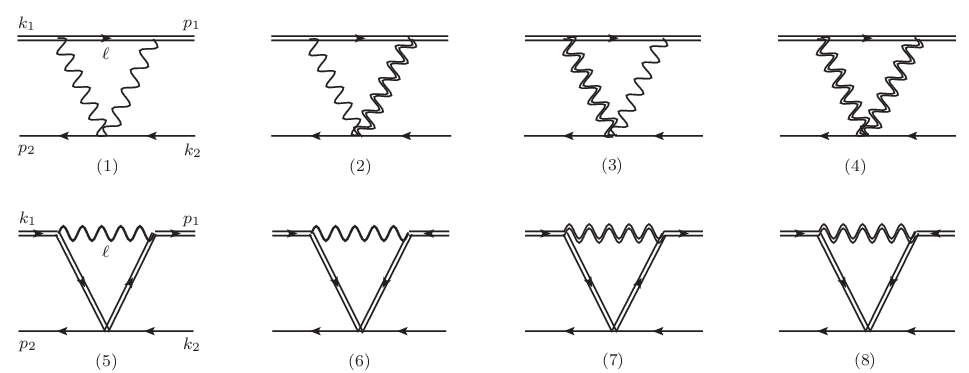}
\end{center}
\caption{Gauge boson (1)-(4) and fermion (5)-(8) triangle-loop diagrams contributing to the 2scalar--2fermion 4-correlator. Again note the in-out momentum flows $\{k_1,k_2;p_1,p_2\}$.}
\label{2scalar2fermiontria}
\end{figure}
Here we employe numerator reductions of the 1-loop 2scalar-2fermion 4-correlators arising from the box-, and triangle-loops in figure \ref{2scalar2fermionbox} and \ref{2scalar2fermiontria}, and they are obtained by analysis of sample generic diagrams from figure \ref{Sample2scalar2fermion} respectively, and given in the appendix E, eqs. (\ref{E.30}) - (\ref{E.32}).

We have for the correlator $\big<XX\bar\Psi\Psi \big>$, both triangles with two gauge and one fermion internal lines, given in the first line of figure \ref{2scalar2fermiontria}, as well as two fermion and one gauge internal lines, in the second line in figure \ref{2scalar2fermiontria}. Notice that the exchange between gauge and hgauge boson internal lines produces phase reversion while keeping the polynomial part identical, therefore generates the second type of cancellation ($\mathbf{ii}$), as discussed in the subsection 6.1.

Other two correlators, $\big<XX\bar\Psi\bar\Psi \big>$ and $\big<XX\Psi\Psi \big>$, have only one gauge boson and two fermion internal triangle lines associated with them, given as the second line of figure \ref{2scalar2fermiontria}. The pairing goes with the exchange of gauge boson line with hgauge boson line, as well as a permutation of the two external scalar fields, producing together zero sum. Thus we have shown vanishing of all three 2scalar-2fermion 4-correlators when $\theta^{\mu\nu}\to0$.

Altogether in the above sections 6, and 7 we have completed our proof for vanishing in the $\theta^{\mu\nu}\to0$ limit of both the 2scalar-2gauge and 2scalar-2fermion 4-point functions, respectively.

$\phantom{In Sections 6 and 7 we have completed our proof for vanishing in limit of all above as}$Q.E.D.\\

\section{Scalar field six-point functions}

In this section we shall show that the limit $\theta^{\mu\nu}\to 0$ of the one-loop 1PI scalar 6-point function exists and that it is equal to the 1-loop 1PI scalar 6-point function of the ordinary U(1) ABJM theory; in other words, that in the IR the 6--point function in the noncommutative ABJM theory flows to the ordinary one.

The 1-loop diagrams that make the 1PI 6-point function can be classified into two broad categories, namely, those involving one triple vertex with two scalars and one gauge field and those which have no triple vertex of this type.  Generic diagrams in the first category are depicted in figure \ref{fig:Fig6-pointfirstclass}, where the dashed lines stand for either of type (h)gauge field propagator as suitable and the continuous lines denote scalar lines. Second category of generic diagrams are given in figure \ref{fig:FigCS6-p-loop}.

\subsection{Six scalar fields 6-point function: $\big<XXXXXX\big>$}

Using the same arguments as before we conclude that the only diagrams contributing to the scalar field 6-correlator
$\big<XXXXXX\big>$ from gauge-scalar sector are gauge-fermion-loops of the first category in figure \ref{fig:Fig6-pointfirstclass}. Since their integrands share the following numerator structure up to constant pre-factors:
\begin{equation}
\epsilon_{abc}\epsilon^{cde}\epsilon_{efa}\ell^b(\ell_d+p_d)(\ell^f+k^f)\:\propto\;\epsilon_{mnr}\ell^m p^n k^r,
\end{equation}
there is no momentum function without commutative limit from these diagrams.

Second category contributions to the 6-correlator $\big<XXXXXX\big>$ comes from fermion triangle-loop diagrams, figure \ref{fig:FigCS6-p-loop}. They, in general, also share a common integrand structure because:
\begin{equation}
\tr \;\gamma^a\gamma^b\gamma^c \ell_a(\ell_b+p_b)(\ell_c+k_c)\:\propto\: \epsilon^{mnr}\ell_m p_n k_r,
\end{equation}
which can be easily derived from the three dimensional $\gamma$-matrices relation
\begin{equation}
\gamma^\mu\gamma^\nu\gamma^\rho=g^{\mu\nu}\gamma^{\rho}-g^{\mu\rho}\gamma^{\nu}+g^{\nu\rho}\gamma^{\mu}+\epsilon^{\mu\nu\rho}.
\label{identity}
\end{equation}
So they do not yield numerator momentum function without commutative limit either.

\subsection{The 6-point function $\big<XXXXXX\big>$ in the commutative limit $\theta^{\mu\nu}\to 0$}

\begin{figure}[t]
\begin{center}
\includegraphics[width=14cm]{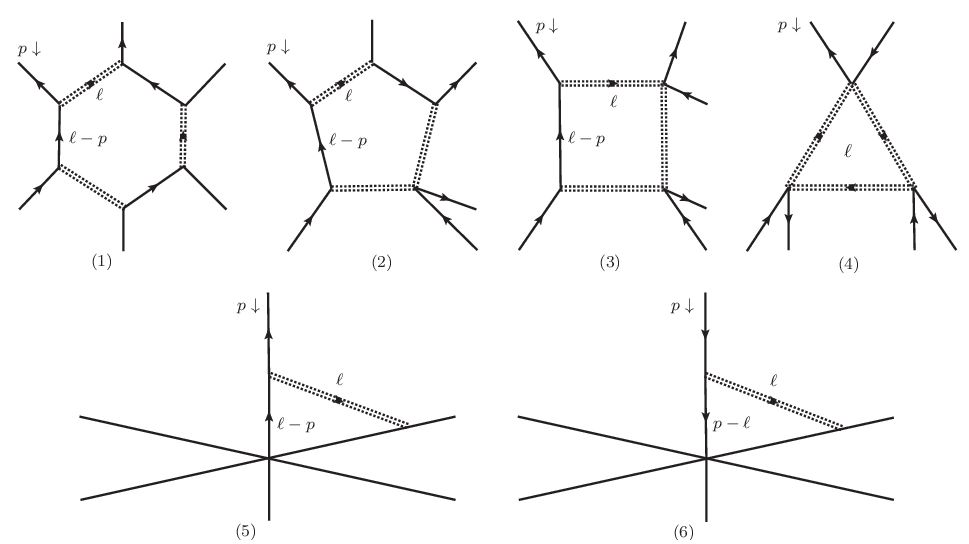}
\end{center}
\caption{First category contributions to the 1-loop scalar field 6-correlator depicted by the generic diagrams with all triple field vertices (1), than with one (2), two (3) and with three (4), four field couplings from kinetic term action (\ref{Akin}).  From 6-scalar fields couplings (\ref{AS6}) we obtain diagrams (5) and (6) given in the second line of the above figure. Double dotted lines stand generically for (h)gauge field propagators as suitable for either type (see figure \ref{fig:notationprop1}), and the continuous lines denote scalar fields. Here the arrow on diagram lines indicate the flow of charge, while separate down arrows indicate the flow of momenta.}
\label{fig:Fig6-pointfirstclass}
\end{figure}
\begin{figure}[t]
\begin{center}
\includegraphics[width=12cm]{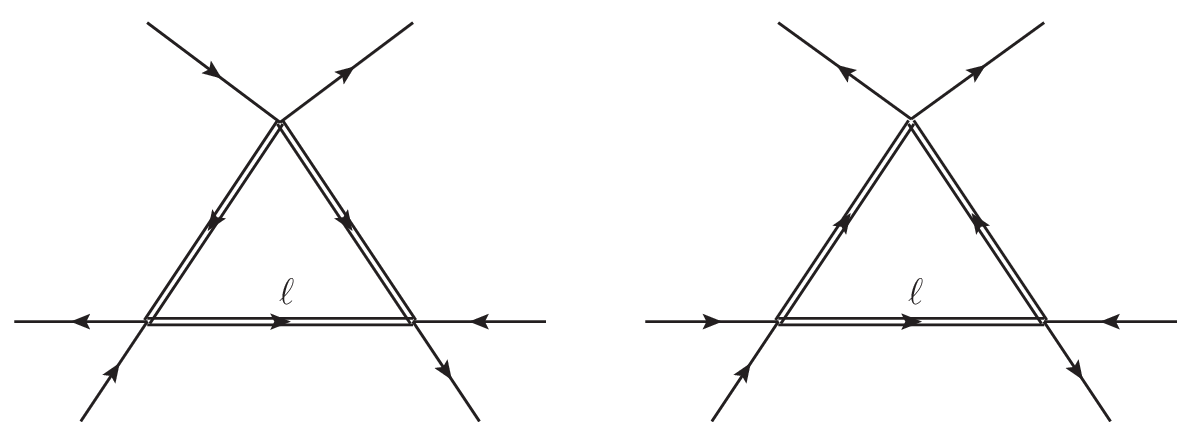}
\end{center}
\caption{Second category of the scalar field 1-loop 6-correlator, from 2fermion-2scalar terms in the four field couplings actions (\ref{AS4a})-(\ref{AS4c}). Double triangle loop lines represent fermions.}
\label{fig:FigCS6-p-loop}
\end{figure}

The UV degree of divergence $\cal D$ of each diagram in figure \ref{fig:Fig6-pointfirstclass}, from power-counting formula\footnote{See relevant explanation after eq. (4.1) in section 4 of ref. \cite{Martin:2017nhg}.}, is zero. This degree is obtained by taking into account polynomial coming from $\epsilon^{\mu\nu\rho}(\ell-2 p)_\nu\ell_\rho$, which has quadratic power in the loop momenta $\ell$. But this quadratic polynomial vanishes due to the antisymmetry of the Levi-Civita symbol. So the integrals with a non-vanishing integrand  contributing to each diagram in figure \ref{fig:Fig6-pointfirstclass} are UV convergent by power-counting and therefore absolutely convergent for the non-exceptional momenta. Hence, the Lebesgue's theorem can be applied and one can take the limit $\theta^{\mu\nu}\to 0$ under the loop integral sign, thus each diagram in figure \ref{fig:Fig6-pointfirstclass} converges to it’s commutative counterpart.

Type of Feynman diagrams that belong to the second category are shown in figures \ref{fig:FigCS6-p-loop}, and they have UV degree of divergence ${\cal D}$ also equal to zero. However this zero value comes from a bit of numerator of the integrand which, as a factor, has the following polynomial in the loop momentum $\ell$:
$\ell^a\epsilon_{\mu a\nu}\ell^b\epsilon_{\nu b\rho}\ell^c\epsilon_{\rho c\mu}.$
This degree is due to the contribution to the numerators of their integrands which have the common structure,
$\tr\big[\slashed{\ell}\slashed{\ell}\slashed{\ell}\big]$,
as a factor. Since that $\tr$ in the previous expression stands for the fermi-statistic trace, we conclude that
previous polynomial in $\ell$ vanishes. So, actually the integrals with a non-vanishing integrand generated by diagrams in Fig.\ref{fig:FigCS6-p-loop} are UV finite by power-counting and, hence, they are absolutely convergent for the non-exceptional momenta. Again, here the Lebesgue's theorem can be applied to each diagram and one can take the limit $\theta^{\mu\nu}\to 0$ under the loop integrals, producing its commutative  counterpart.

In summary we have proven in this section 8 that all integrals with a non-vanishing integrand from figures \ref{fig:Fig6-pointfirstclass}, and \ref{fig:FigCS6-p-loop} which contribute to the 1PI scalar 6-point functions are UV finite by power-counting and hence absolutely convergent for the non-exceptional momenta. The Lebesgue's dominated convergence theorem \cite{Velo,Rudin} than guarantees, for each individual diagram, that taking limits $\theta^{\mu\nu}\to 0$, and integrating over the loop momenta, do commute. This implies that the scalar 6-point function of the NCABJM theory in the $\theta^{\mu\nu}\to 0$ limit are given by the scalar 6-point function of the corresponding ordinary ABJM theory.
$\phantom{XXXXXXXXXXXXXXXXXXXXXXXXXXXXXXXXXXXXXXXX}$Q.E.D.

\section{Discussions}
We summarise proof of the commutative limits existence for the 1-loop higher point ($\ge 4$) 1PI functions in the $\rm U(1)$ NCABJM. Using UV divergence power-counting formula
\begin{equation}
{\cal D}=3-E_G-E_F-\frac{1}{2}\,E_X,
\label{calD}
\end{equation}
with number of: $E_G$ external gauge fields, $E_F$ external fermions, $E_X$ external scalars, and no external ghosts, we did identify the following 1PI higher point correlation functions with degree ${\cal D}\ge 0$, which always shows the presence of UV divergence:

\begin{itemize}

\item $\bullet$ 4 scalar fields 4-correlator $\big<XXXX\big>$; figures \ref{fig:Fig21}, \ref{fig:Fig22}, \ref{fig:Fig31}, \ref{fig:Fig41}, \ref{fig:Fig4-6tad}, and \ref{4scalar}, respectively,

\item $\bullet$ 2scalar-guage-hgauge field 4-correlator $\big<XXA\hat A\big>$; figure \ref{2scalarghg},

\item $\bullet$ 2scalar-2gauge$|$2hgauge field 4-correlators $\big<XXAA\big>|\big<XX\hat A\hat A\big>$; figures \ref{2scalar2g}, \ref{2scalar2hg},

\item $\bullet$ 2scalar-2fermion 4-correlators $\big<XX\bar\Psi\Psi \big>$ and $\big<XX\Psi\Psi\big>, \big<XX\bar\Psi\bar\Psi\big>$; figures \ref{2scalar2fermionbox}, \ref{2scalar2fermiontria}, \ref{Sample2scalar2fermion},

\item $\bullet$ first and second category scalar field 6-correlator $\big<XXXXXX\big>$ from figures \ref{fig:Fig6-pointfirstclass}, \ref{fig:FigCS6-p-loop}.

\end{itemize}
\noindent

Following detailed explicit computation of the 4-scalar correlators $\big<XXXX\big>$ we notice that, in the gauge-scalar sector, only diagrams with least number of internal lines (bubble diagrams in this case) yield integrals without well-defined commutative limits and require cancellation upon summation. This phenomenon is not accidental. The reason is that diagrams with more internal lines are built by inserting 2scalar-2gauge sub-diagrams with one scalar propagator line and two 3-leg vertices in lieu of 2scalar-2gauge field vertices. Such insertion seems to keep the same superficial divergence order at first glance, since it brings in two numerator types, of $\ell^1$ and $\ell^2$ powers, respectively. However a more careful look tells us that new numerators must be attached to the Levi-Civita tensors of the gauge boson propagators in the Landau gauge, which then prevent the loop momenta to co-exist in the same monomial in the numerator. Thus the formal power of loop momenta $\ell$ remains the same in all these diagrams, while denominator power is increasing by two in each insertion. And consequently only the diagrams with least number of internal lines give integrals without well-defined commutative limits. Finally we provide arguments without explicit computation on how correlators acquire well-defined commutative limits.

Through prior sections 4-8 including appendices, we have learned the following properties of 4- \& 6-point functions, and the 1-loop integrals involved:

\begin{itemize}

\item $\bullet$ Landau gauge greatly simplifies calculation, for it is IR safe.

\item $\bullet$ Number of diagrams allowed by the Lorentz structures and topologically, are actually zeros due to the absence of relevant contributing terms in the action $S$ (\ref{Action}).

\item $\bullet$ For the UV degree of divergence (\ref{calD}) $\mathcal D=1$, the 4-correlators $\big<XXXX\big>$ requires cancellation among all sectors to cancel all contributions without commutative limit.

\item $\bullet$ For the rest of correlators the degree of divergence $\mathcal D=0$, and they do achieve commutative limits sector by sector for various reasons listed below:

\begin{itemize}

\item $\ast$ Fermionic triangle-loop contributions to $\big<XXA\hat A\big>, \big<XXAA\big>, \big<XX\hat A\hat A\big>$,
figures \ref{2scalarghg}, \ref{2scalar2g}, \ref{2scalar2hg}, cancel out due to the permutation symmetry; subsections 6.1, and 6.2.

\item $\ast$ Bosonic triangle-loop contributions to $\big<XXAA\big>, \big<XX\hat A\hat A\big>$, figures \ref{2scalar2g}, \ref{2scalar2hg}, disappear because in the Landau gauge integrands of loop integrals vanish; subsection 6.2.

\item $\ast$ Considering $\big<\bar\Psi\Psi XX\big>, \big<\bar\Psi\bar\Psi XX\big>, \big<\Psi\Psi XX\big>$ correlators,  contributions from figure \ref{2scalar2fermionbox}, \ref{2scalar2fermiontria}, \ref{Sample2scalar2fermion}, cancel generally due to the permutation symmetry, similar as for the $\big<XXA\hat A\big>$ case:\\
-- First, contributions to the $\big<\bar\Psi\Psi XX\big>$ cancel each other via gauge/hgauge internal line permutation pairs, which produces phase reversion; section 7. \\
-- Second, contributions to the correlators $\big<\bar\Psi\bar\Psi XX\big>, \big<\Psi\Psi XX\big>$ cancel each other via a pairing by permuting gauge and hgauge internal line with external scalar field lines simultaneously; section 7.

\item $\ast$ The 6-correlators $\big<XXXXXX\big>$ have safe integrands for all suspicious diagrams, after tensor structures get simplified; figures \ref{fig:Fig6-pointfirstclass}, \ref{fig:FigCS6-p-loop} in section 8.

\end{itemize}

\begin{table}
\begin{center}
\begin{tabular}{|c|c|c|c|}
\hline
Integrals ; Equation & limit $\theta^{\mu\nu}\to0$ ; Equation & UV/IR
\\
\hline
$I_1^0(\tilde k)$; (\ref{B.5}) & linear IR divergent; (\ref{B.6})  & Yes
\\
$I_2^0(p;\tilde k)$; (\ref{B.8}) & finite; (\ref{B.9}) & No
\\
$I_3^0(p_1,p_2;\tilde k)$; (\ref{B.16}-\ref{B.32}) & finite; (\ref{B.33}) & No
\\
$I_4^0(q_1,q_2,q_3;\tilde k)$; (\ref{D.3}-\ref{D.4}) & finite; (\ref{D.4}) & No
\\
$I_1^\mu(\tilde k)$; (\ref{B.37}) & quadratic IR divergent; (\ref{B.38}) & Yes
\\
$I_2^\mu(p;\tilde k)$; (\ref{B.39}-\ref{B.41}) & not well-defined; (\ref{B.41}) & \#
\\
$I_3^\mu(q_1,q_2;\tilde k)$; (\ref{B.42})& finite; (\ref{B.42}) & No
\\
$I_3^\mu(q_1,q_2;a(q_1,q_2),\tilde k)$; (\ref{C.12})& finite; (\ref{C.12}-\ref{C.15}) & No
\\
$I_4^\mu(q_1,q_2,q_3;\tilde k)$; (\ref{D.5}) & finite; (\ref{D.5}) & No
\\
$I_2^{\mu\nu}(p;\tilde k)$; (\ref{C.6}) & not well-defined; see \cite{Martin:2017nhg} & \#
\\
$I_3^{\mu\nu}(q_1,q_2;\tilde k)$; (\ref{C.22}-\ref{C.26})& not well-defined; (\ref{C.22}-\ref{C.26}) & \#
\\
$I_4^{\mu\nu}(q_1,q_2,q_3;\tilde k)$; (\ref{D.6}-\ref{D.12}) & finite; (\ref{D.6}-\ref{D.12}) & No
\\
$I_{2_{1,2}}^1(q_1,q_2,q_3;\tilde k)$; (\ref{C.5})&  not well-defined; (\ref{C.7}) & \#
\\
$I_{2_{1}}^2(q_1,q_2,q_3;\tilde k)$; (\ref{C.5})& not well-defined; (\ref{C.8}) &  \#
\\
$I_{2_{2}}^2(q_1,q_2,q_3;\tilde k)$; (\ref{C.5})&  not well-defined; (\ref{C.9})  & \#
\\
$I_3^1(p_1,p_2;a(p_1,p_2),\tilde k)$; (\ref{C.16}-\ref{C.20}) & 0; (\ref{C.21})  & No
\\
\hline
\end{tabular}
\caption{Properties of $3D$ integrals contributing to the 4- \& 6-point functions evaluated and analysed in this work. Label $\#$ in the UV/IR column means that in the column limit $\theta^{\mu\nu}\to0$ expression “not well-defined” denotes that final values of integrals depend on the way how $\theta^{\mu\nu}$ approach to the zero point. However those integrals are not IR divergent. For the additional explanation of the above expression “not well-defined”, see also  sections 5.-6., eqs.(6.3-6.7), and the appendices C.14-C.27, all in reference \cite{Martin:2017nhg}. }
\end{center}
\label{table1}
\end{table}

\item $\bullet$ In table 1 we listed all needed integrals together with their $\theta^{\mu\nu}\to0$ limit properties. Indicated IR divergent integrals are exactly the source of celebrated UV/IR mixing effect, which on top to the Moyal based NCQFT, \cite{Matusis:2000jf,Seiberg:2000gc,Seiberg:2000ms,Minwalla:1999px,Hayakawa:1999yt,VanRaamsdonk:2000rr,VanRaamsdonk:2001jd,Schupp:2008fs,Horvat:2011bs,Horvat:2013rga,Horvat:2015aca,Martin:2020ddo}, appears in other NCQFT spanned on different noncommutative spaces and within various physical environments, as well \cite{Grosse:2005iz,Magnen:2008pd,Blaschke:2009aw,Meljanac:2011cs,Meljanac:2017grw,Meljanac:2017jyk,Meier:2023kzt,Meier:2023lku,ST,Lust:2017wrl,Craig:2019zbn}.

\item By inspecting eqs. (\ref{B.6}) and (\ref{B.38}) one can trivially see that tadpole type of integrals (\ref{B.1})
which are all non-planar, shows that the latter goes to zero when $\tilde k\to\infty$:
\begin{equation}
\lim_{\tilde{k}\to\infty}I^{0}_1(\tilde{k})=\lim_{\tilde{k}\to\infty}I^\mu_1(\tilde{k})=0
\Longrightarrow \lim_{\theta\to\infty} {(\cal NP)}S^{tad}(p_1,p_2,p_3;\theta)=0.
\label{9.2}
\end{equation}

\item $\bullet$ Star product ordered (ribbon) diagrams can help searching for the non-planar integrands. For NCABJM, the reduction of computation load is not so high because there exist various (unordered) diagrams which are sums of non-planar ordered diagrams only.

\item $\bullet$ In this article scalar tadpole, bubble and triangle type of integrals $I^0_{1,2,3}$ including properties are evaluated and given in the appendix B.2. Vector tadpole, bubble and triangle integrals, including their properties, are evaluated in the appendix B.3.

\item $\bullet$ Explicit reductions and evaluations of vector and tensor triangle integrals in terms of relevant scalar integrals are performed by the van Neevern-Vermaseren method and given in the appendx C. In the appendix D we give explicitly the scalar box master integral $I_4^0$ in terms of scalar bubble and triangle integrals (\ref{D.4}). Reductions of vector $I_4^\mu$ and tensor $I_4^{\mu\nu}$ box integrals using again the van Neevern-Vermaseren method are given in terms of scalar bubble and triangle integrals in the rest of the appendix D.

\item $\bullet$ Simplified/reduced integrand numerators, for bubble, triangle and box loops needed to  evaluate scalar-scalar and scalar-fermion 4-point functions, as well as for direct proof of vanishing of triangle-loop contributions to the 4-correlator $\big<XXXX\big>$--diagram {\it(2.)} in figure \ref{4scalar}--, are given in the appendix E.

\item $\bullet$ Finally we have to note that by inspecting length of all 4- \& 6-correlators and the length of loop integrals $I_{1,2,3,4}^{(0;\mu;\mu\nu)}$ in appendices, it is clear that the length of the sums of all contributions are going to be extremely lengthy/massive formulas, very much cumbersome and non-transparent, thus certainly out of scope to write it down explicitly in this paper.  After we have detected and extracted loop integral IR singular parts, to perform calculations of the remaining finite ($\theta^{\mu\nu}\not=0$) parts of 4- \& 6-correlator, it is obviously necessary to use the state of the art packages to execute the computer work.

\end{itemize}

\section{Conclusions}
In our first steps paper \cite{Martin:2017nhg} we have introduced the ABJM  quantum field theory on the noncommutative Moyal manifold and, by using the component formalism, shown that it is ${\cal N}=6$ supersymmetric. For the $\rm U(1)_\kappa\times U(1)_{-\kappa}$ case, we have computed the 1-loop 1PI 2- \& 3-correlators in the Landau gauge and show that they are UV finite, and have well-defined commutative limit $\theta^{\mu\nu}\to 0$, corresponding to the 1PI correlators of the ordinary ABJM theory.

In this paper we continue with the $\rm U(1)_\kappa\times U(1)_{-\kappa}$ case by using again the component formalism in the Landau gauge. We compute the 1-loop 1PI 4-correlators and show that they are UV finite having well-defined commutative limits $\theta^{\mu\nu}\to 0$, which again corresponds exactly to the 1PI correlators of the ordinary ABJM field theory. This result also holds for one-loop correlators which are UV finite by power-counting.

Now note that the effect of UV/IR mixing is present in each tadpole and bubble contributions containing tadpole type of integrals (\ref{B.6}),(\ref{B.38}),(\ref{C.3}), separately. However we have shown that in the sum of non-planar bubble+tadpole contributions (\ref{I_1+2}) the UV/IR mixing effect cancels out. Also scalar 1-loop triangle, and box diagram contributions to the 4-point functions, which are UV finite by the power-counting rule, in $\theta^{\mu\nu}\to 0$ limit vanish. So considering contributions to the scalar field 4-point functions (\ref{S4X}) in $3D$ from our analysis we see that vanishing of sums in the limit $\theta^{\mu\nu}\to 0$ holds. Thus, from sections 4 and 5 we can write:
\begin{equation}
\lim_{\theta\to 0}\; S^{box}\big|_{(\ref{limSbox})}=\lim_{\theta\to 0}S^{tri}\big|_{(\ref{limStria})}=0 \;\; \&\;\;\lim_{\theta\to 0}\; {(\cal NP)}\big[S^{bub}+S^{tad}\big]_{(\ref{I_1+2})}=0,
\label{S'slimit}
\end{equation}
proving this way vanishing of the sums of the scalar 4-point functions in the commutative limit $\theta^{\mu\nu}\to 0$, for the NCABJM theory action (\ref{Action}), i.e.:
\begin{equation}
\lim_{\theta\to 0}\; S_{4X}\big|_{(\ref{S4X})}=0.
\label{S4Xlimit}
\end{equation}

Next, in sections 6 and 7 we analyse and conclude that vanishing of 2scalar-2(h)gauge and 2scalar-2fermion 4-point functions in the limit $\theta^{\mu\nu}\to 0$ also holds. Thus, we may claim that the NCABJM theory is free from the NC IR instabilities on the level of the entire set of the 1-loop 4-point functions.

In section 8 we discussed limiting properties of the scalar 6-point functions too. Namely we have proven that the limit $\theta^{\mu\nu}\to 0$ of the 1-loop 1PI scalar 6-point functions exists, and that it is equal to the 1-loop 1PI scalar 6-point functions of the ordinary U(1) ABJM theory.

Summing altogether up by taking into account results from previous \cite{Martin:2017nhg} and this paper, by applying the Lebesgue’s dominated convergence theorem to the Fourier transformed NCABJM theory \cite{Velo,Rudin}, and by direct computations of the UV divergent by power-counting 1-loop integrals, we have shown that in the limit $\theta^{\mu\nu}\to0$ all IR divergencies of the NCABJM theory disappear, and conclude that noncommutative ABJM theory,  up to the 1-loop scalar 6-point functions order, flows smoothly to the ordinary commutative ABJM theory. \\
$\phantom{XXXXXXXXXXXXXXXXXXXXXXXXXXXXXXXXXXXXXXXX}$Q.E.D.\\

If necessary one could explicitly execute all of the 4- \& 6-point functions by computer program, using our full sets of equations (including all relevant integrals from appendices), respectively. Sum of $\theta$-nonvanishing, UV and IR finite, contributions to the 4-correlators $\big<XXXX\big>$, $\big<XXA\hat A\big>$, $\big<XXAA\big>$, $\big<XX\hat A\hat A\big>$, $\big<XX\bar\Psi\Psi\big>$, $\big<XX\bar\Psi\bar\Psi\big>$, $\big<XX\Psi\Psi \big>$, and the 6-correlator $\big<XXXXXX\big>$, could be presented graphically as a functions of incoming energies and the scale of noncommutativity, which shall certainly show good IR behaviour with respect to the scale of noncommutativity limit $\Lambda_{\rm NC}\to\infty$.

\appendix

\section{Feynman rules}

First to repeat, in the following Feynman rules we are using clockwise circles for the star-product ordering. Second, Feynman rules for the triple fields vertices; (h)gauge, scalar-(h)gauge, fermion-(h)gauge, and for the four fields vertices; 2scalars-2(h)gauge, are already given in the appendix D of our first steps paper \cite{Martin:2017nhg}. Third, in this article we are giving the following couplings  Feynman rules needed: 2scalar-gauge-hgauge fields vertex (\ref{A.1}), three 2scalar-2fermion vertices (\ref{A.2})-(\ref{A.4}), and the six scalar fields vertex (\ref{A.5}), according to figure \ref{fig:FigFRCorr}.

\subsection{Scalar-scalar-gauge-hgauge fields vertex}

The action $S_{\rm kin}$ (\ref{Akin}) in accord with diagram (A.1) figure \ref{fig:FigFRCorr} gives the following Feynman rule:
\begin{figure}[t]
\begin{center}
\includegraphics[width=15cm]{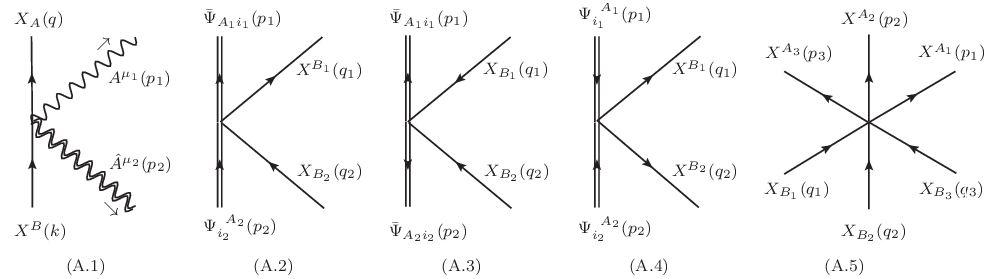}
\end{center}
\caption{Feynman rules for the 2scalar-gauge-hgauge fields vertex $\big(\hat V^{\mu_1\mu_2}\big)^{\;\;B}_{A}$ (A.1); 2scalar-2fermion field vertices, $\big(V_{i_1i_2}\big)^{A_2B_1}_{A_1B_2}$ (A.2), $\big(V_{i_1i_2}\big)^{B_1B_2}_{A_1A_2}$ (A.3) and $\big(V_{i_1i_2}\big)^{A_2A_2}_{B_1B_2}$ (A.4); and the six scalar fields vertex $V_6$ (A.5) with 3 fields incoming and 3 outgoing, respectively.}
\label{fig:FigFRCorr}
\end{figure}
\begin{equation}
\big(\hat V^{\mu_1\mu_2}\big)_{\;A}^B=2i\frac{\kappa}{2\pi}\eta^{\mu_1\mu_2}
\Big[e^{\frac{i}{2}q \theta(k-p_1)}e^{\frac{i}{2}p_1\theta k}\Big]\delta_{\;A}^B,\;\;k=q+p_1+p_2,
\label{A.1}
\end{equation}
correcting also typos in (D.6) from \cite{Martin:2017nhg}.
We recall that $k\theta p=k_\mu\theta^{\mu\nu}p_\nu=-p\theta k$, and $q\theta q=0$.

\subsection{Scalar-scalar-2fermion fields vertices}

From  the action $S_4$  (\ref{AS4}) in accord with diagrams (A.2), (A.3), and (A.4) from figure \ref{fig:FigFRCorr}, we have the following respected Feynman rules:
\begin{eqnarray}
\Big(V_{i_1i_2}\Big)^{A_2B_1}_{A_1B_2}
&=&i\frac{\kappa}{\pi}\delta_{i_1i_2}\Big(\delta^{A_2}_{A_1}\delta^{B_1}_{B_2}
-2\delta^{A_2}_{B_2}\delta^{B_1}_{A_1}
\Big)
\sin\frac{p_1\theta p_2 + q_1\theta q_2}{2},
\label{A.2}\\
\Big(V_{i_1i_2}\Big)^{B_1B_2}_{A_1A_2}
&=&-\frac{\kappa}{2\pi}2\gamma^0_{i_1i_2}
\epsilon^{A_1A_2B_1B_2}\Big[e^{\frac{i}{2}(p_1\theta q_1+p_2\theta q_2)}-e^{\frac{i}{2}(p_1\theta q_2+p_2\theta q_1)}\Big],
\label{A.3}\\
\Big(V_{i_1i_2}\Big)^{A_1A_2}_{B_1B_2}&=&-\Big(V_{i_1i_2}\Big)^{B_1B_2}_{A_1A_2},
\label{A.4}
\end{eqnarray}
were (\ref{A.2}) is repeating the Feynman rule (D.10) in \cite{Martin:2017nhg}.


\subsection{Six scalar fields vertex}

From  the action $S_6$  (\ref{AS6}) in accord with diagram (A.5) in figure \ref{fig:FigFRCorr}, we have found the following Feynman rule
\begin{eqnarray}
V_6(p_1,p_2,p_3|q_1,q_2,q_3;\theta)&=&i\frac{\kappa}{6\pi}\sum_{i=1}^4\sum_{\pi,\sigma}
V^{(i)}\Big[\big(p_{\pi(1)},A_{\pi(1)}\big),\big(p_{\pi(2)},A_{\pi(2)}\big),\big(p_{\pi(3)},A_{\pi(3)}\big)\big|
\nonumber\\
&&\phantom{i\sum_{i=1}^4\sum_\pi XX}\big(q_{\sigma(1)},B_{\sigma(1)}\big),\big(q_{\sigma(2)},B_{\sigma(2)}\big),\big(q_{\sigma(3)},B_{\sigma(3)}\big)\Big],
\label{A.5}
\end{eqnarray}
\begin{eqnarray}
&&V^{(1)}\Big[(p_1,A_1),(p_2,A_2),(p_3,A_3)\big|(q_1,B_1),(q_2,B_2),(q_3,B_3)\Big]
\nonumber\\
&&\phantom{XXXXXXXXXX}=
e^{-\frac{i}{2}(p_1-q_1)\theta(p_2-q_2)}e^{\frac{i}{2}(p_1\theta q_1+p_2\theta q_2+p_3\theta q_3)}
\delta_{A_1}^{B_1}\delta_{A_2}^{B_2}\delta_{A_3}^{B_3},
\nonumber\\
&&V^{(2)}\Big[(p_1,A_1),(p_2,A_2),(p_3,A_3)\big|(q_1,B_1),(q_2,B_2),(q_3,B_3)\Big]
\nonumber\\
&&\phantom{XXXXXXXXXX}=
e^{\frac{i}{2}(p_1-q_1)\theta(p_2-q_2)}e^{-\frac{i}{2}(p_1\theta q_1+p_2\theta q_2+p_3\theta q_3)}
\delta_{A_1}^{B_1}\delta_{A_2}^{B_2}\delta_{A_3}^{B_3}
=\big(V^{(1)}\big)^{*},
\nonumber\\
&&V^{(3)}\Big[(p_1,A_1),(p_2,A_2),(p_3,A_3)\big|(q_1,B_1),(q_2,B_2),(q_3,B_3)\Big]
\nonumber\\
&&\phantom{XXXXXXXXXX}=
4e^{-\frac{i}{2}(q_2-p_1)\theta(q_3-p_2)}e^{\frac{i}{2}(q_2\theta p_1+q_3\theta p_2+q_1\theta p_3)}
\delta_{A_1}^{B_1}\delta_{A_2}^{B_2}\delta_{A_3}^{B_3},
\nonumber\\
&&V^{(4)}\Big[(p_1,A_1),(p_2,A_2),(p_3,A_3)\big|(q_1,B_1),(q_2,B_2),(q_3,B_3)\Big]
\nonumber\\
&&\phantom{XXXXXXXXXX}=
-6e^{-\frac{i}{2}(q_2-p_1)\theta(q_1-p_2)}e^{\frac{i}{2}(p_1\theta q_2+p_2\theta q_1+p_3\theta q_3)}
\delta_{A_1}^{B_1}\delta_{A_2}^{B_2}\delta_{A_3}^{B_3}.
\label{A.6}
\end{eqnarray}

\section{Integrals}

We evaluate a set of 3D non-planar 1-loop integrals which emerge from an OPP \cite{Ossola:2006us} type integrand reduction in NCABJM theory.

The 1-loop structure of the NCQFT allows us to decompose the integrand into two parts: The rational fraction part which is independent from the NC parameter $\theta^{\mu\nu}$ and the NC phase factor part which bears an universal form $e^{-i\ell \theta k}$, in which $\ell_\mu$ is the loop moment, while $(\theta k)^\mu=\theta^{\mu\nu} k_\nu$ does not depend on $\ell^\mu$. We call a loop integral non-planar if the NC phase factor is nonzero.

As the commutative OPP-type integrand reduction \cite{Ossola:2006us} involves only the algebraic structure of rational fractions, we can employ it on the rational fractional part of the NC one loop integrals as well. The general commutative result says that the outcome of such reduction involves two types of integrands: Scalar loop integrand with no numerator and loop integrands which are vanishing in the commutative setting. The latter are therefore called ``spurious terms''.
Once NC phase factors are included, we notice that spurious terms no longer vanish, instead they bear specific NC dependent structure, so we have to evaluate both scalar integrals and the spurious integrals.

We choose to work with the integrals emerging from the 3-dimensional integrand reduction, instead of the so called $D$-dimensional integrand reduction. However, for number of integrals we first perform integrations in $D$-dimensions, and than as the last step we take the limit $D\to3$, respectively.

\subsection{Types of integrals needed}

Inspecting (\ref{box1})--(\ref{Sb9}) we see that, with respect to all integrals used in previous paper \cite{Martin:2017nhg}, there are few of additional integrals needed to be evaluated at $D=3$ for completing this paper.
Next we listed a complete set of integrals $I^{(0;\mu;\mu\nu)}_{1,2,3,4}$ arising from tadpole, bubble, triangle, and box diagrams with the $\ell$-power types $\ell^0, \ell^1$, and $\ell^2$, respectively:
\begin{eqnarray}
I^{(0;\mu;\mu\nu)}_1(\tilde k)&=&\int \frac{d^D \ell}{(2\pi)^D}
\frac{(\ell^0;\ell^\mu;\ell^\mu\ell^\nu)\cdot e^{i\ell\tilde k}}{\ell^2},
\label{B.1}\\
I^{(0;\mu;\mu\nu)}_2(p_1;\tilde k)&=&\int \frac{d^D \ell}{(2\pi)^D}
\frac{(\ell^0;\ell^\mu;\ell^\mu\ell^\nu)\cdot e^{i\ell\tilde k}}{\ell^2(\ell+p_1)^2},
\label{B.2}\\
I_3^{(0;\mu;\mu\nu)}(p_1,p_2;\tilde k)&=&\int \frac{d^D \ell}{(2\pi)^D}
\frac{(\ell^0;\ell^\mu;\ell^\mu\ell^\nu)\cdot e^{i\ell\tilde k}}{\ell^2(\ell+p_1)^2(\ell+p_2)^2},
\label{B.3}\\
I_4^{(0;\mu;\mu\nu)}(p_1,p_2,p_3;\tilde k)&=&\int \frac{d^D \ell}{(2\pi)^D}
\frac{(\ell^0;\ell^\mu;\ell^\mu\ell^\nu)\cdot e^{i\ell\tilde k}}{\ell^2(\ell+p_1)^2(\ell+p_2)^2(\ell+p_3)^2},
\label{B.4}
\end{eqnarray}
with abbreviations $\tilde k^{\mu}=\theta^{\mu\nu}k_\nu$. Here $\ell$ is the loop momenta, while $k$ and $p_i$ are some various momenta, according to Feynman rules used to evaluate certain diagrams. In other figures in this manuscript $p_i$ are corresponding momenta, respectively. Notations for tadpole, bubble, triple and box type of integrals for scalar, vector and tensor cases are self-evident. In the above, integrals $I^{(0,\mu)}_{1}$ belong to the class of the non-planar one, which we handle first. Also, while $I_1^0$ is almost trivial, the integral $I_2^0 $ should be evaluated with help of
eqs. (C.22)-(C.27), and $I_3^0$ should be evaluated with help of eqs. (8.17)-(8.25) and (C.22)-(C.27) both from  \cite{Martin:2017nhg} respectively, while for other $I_3^{(\mu;\mu\nu)}$ and $I_4^{(0;\mu;\mu\nu)}$ types we shall employe the van Neevern-Vermaseren method.

We would like to remind the reader that not all the integrals that we shall deal with in the sequel are UV finite by power-counting; so to define them and manipulate them properly, we shall use Dimensional Regularization — this is why they are defined in $D$ dimensions. Only after we have made sure that the UV divergences cancel out upon adding up relevant contributions, we shall take the limit $D\to3$.

\subsection{Evaluating scalar tadpole, bubble and triangle integrals}

\subsubsection{Scalar tadpole integral}

Computation of tadpole scalar integral $I_1^0(\tilde k)$ (\ref{B.1}) is straightforward. Introducing Feynman and $\alpha$ parametrisations, in Eucleadian metrics using Wick rotation ($\ell^0=-i\ell_E^0$), and integrate over the loop momenta, we get the following $D$ dimensional result:
\begin{eqnarray}
I_1^0(\tilde k)&=&\int\frac{d^D\ell}{(2\pi)^D}\,\frac{e^{i\ell\cdot(\pm\tilde k)}}{\ell^2}
=\int\frac{d^D\ell}{(2\pi)^D}\int\limits_0^\infty d\alpha\,e^{-\alpha\ell^2+i\ell\cdot(\pm\tilde k)}
=-i\int\frac{d^D\ell_E}{(2\pi)^D}\int\limits_0^\infty d\alpha\,e^{-\alpha\ell_E^2-\frac{\tilde k^2}{4\alpha}}
\nonumber\\
&=&-i(4\pi)^{-\frac{D}{2}}\int\limits_0^\infty d\alpha\,\alpha^{-\frac{D}{2}}e^{-\frac{\tilde k^2}{4\alpha}}
=-i(4\pi)^{-\frac{D}{2}}\int\limits_0^\infty d\lambda\,\lambda^{\frac{D}{2}-2}e^{-\lambda\frac{\tilde k^2}{4}},
\label{B.5}
\end{eqnarray}
\begin{eqnarray}
I_1^0(\tilde k)&=&I_1^0(-\tilde k)=-i(4\pi)^{-\frac{D}{2}}\left(\frac{\tilde k^2}{4}\right)^{1-\frac{D}{2}}\Gamma\left(\frac{D}{2}-1\right)\Bigg|^{D=3}=\frac{-i}{4\pi \sqrt{\tilde k^2}},
\label{B.6}
\end{eqnarray}
where in $D=3$ case we have found linear IR-divergence for $\tilde k^\mu\to0$.

\subsubsection{Scalar bubble integral}

Computation of $I_2^0(p;\tilde k)$ in $D$ dimensions from (\ref{B.2}) is also straightforward. Introducing again Feynman,
and $\alpha$ parametrisations, and using Wick rotation gives the following expression:
\begin{eqnarray}
I_2^0(p;\tilde k)&=&i\int\frac{d^D\ell}{(2\pi)^D}\,\frac{e^{i\ell\cdot\tilde k}}{\ell^2(\ell+p)^2}
=\int\frac{d^D\ell}{(2\pi)^D}\,\int\limits_0^1 dx\,\frac{e^{i\ell\cdot\tilde k}}{\big((\ell+xp)^2+x(1-x)p^2\big)^2}
\label{B.7}\\
&=&\int\frac{d^D\ell}{(2\pi)^D}\int\limits_0^1 dx\int\limits_0^\infty d\alpha\;\alpha\;e^{-\alpha((\ell+xp)^2+x(1-x)p^2)+i\ell\cdot\tilde k}
\nonumber\\
&=&-i\int\frac{d^D\ell_E}{(2\pi)^D}\int\limits_0^1 dx e^{-ixp\cdot\tilde k}\int\limits_0^\infty d\alpha\;\alpha\,e^{-\alpha\ell_E^2-\alpha x(1-x)p^2-\frac{\tilde k^2}{4\alpha}}
\nonumber\\
&=&-i(4\pi)^{-\frac{D}{2}}\int\limits_0^1 dx e^{-ixp\cdot\tilde k}\int\limits_0^\infty d\alpha\;\alpha^{1-\frac{D}{2}}\,e^{-\alpha x(1-x)p^2-\frac{\tilde k^2}{4\alpha}}
\nonumber\\
&=&-i(4\pi)^{-\frac{D}{2}}\int\limits_0^1 dx e^{-ixp\cdot\tilde k} 2\big(x(1-x)p^2\big)^{\frac{D}{4}-1}\left(\frac{\tilde k^2}{4}\right)^{1-\frac{D}{4}}K_{\frac{D}{2}-2}\left[\sqrt{x(1-x)p^2\tilde k^2}\right],
\nonumber
\end{eqnarray}
where $K_\nu[z]$ is the modified Bessel function of the second kind. Setting $D\to 3$ and taking into account
$K_{\pm\frac{1}{2}}[z]=\sqrt{\frac{\pi}{2}}\frac{e^{-z}}{\sqrt{z}}$, we have found
\begin{equation}
{I_2^0}(p;\tilde k)\Big|^{D=3}=
\frac{i}{8\pi^2}\int\limits_0^1 dx\,\frac{e^{-\sqrt{x(1-x)p^2\tilde k^2}-ixp\cdot\tilde k}}{\sqrt{x(1-x)p^2}}.
\label{B.8}
\end{equation}
While it is complicated and cumbersome to give a closed formula for the $x$-integration, it is not hard to see that one could expand the exponential function as power series. Then each term has a well-defined integration over x and no divergence occurs when $\theta^{\mu\nu}\to 0$, thus
\begin{equation}
\lim_{\theta\to 0}{I_2^0}(p;\tilde k)\Big|^{D=3}\equiv \lim_{\tilde k\to 0}{I_2^0}(p;\tilde k)\Big|^{D=3}
=\frac{-i}{8\pi}\int\limits_0^1 dx\,\frac{1}{\sqrt{x(1-x)p^2}}=\frac{-i}{8\sqrt{p^2}},
\label{B.9}
\end{equation}
and clearly this integral has well-defined finite commutative limit.\\

Now, in $3D$ we list some properties  of scalar bubble integrals:
\begin{eqnarray}
\int\frac{d^3\ell}{(2\pi)^3}\,\frac{e^{i\ell\tilde k}}{(\ell+q_1)^2(\ell+q_2)^2}
&=&e^{-iq_1\tilde k}I_2^0(q_2-q_1;\tilde k)=e^{-iq_2\tilde k}I_2^0(q_1-q_2;\tilde k),
\label{B.10}\\
\int\frac{d^3\ell}{(2\pi)^3}\,\frac{e^{i\ell\tilde k}}{(\ell+q_1)^2(\ell+q_3)^2}
&=&e^{-iq_1\tilde k}I_2^0(q_3-q_1;\tilde k)=e^{-iq_3\tilde k}I_2^0(q_1-q_3;\tilde k),
\label{B.11}\\
\int\frac{d^3\ell}{(2\pi)^3}\,\frac{e^{i\ell\tilde k}}{(\ell+q_2)^2(\ell+q_3)^2}
&=&e^{-iq_2\tilde k}I_2^0(q_3-q_2;\tilde k)=e^{-iq_3\tilde k}I_2^0(q_2-q_3;\tilde k),
\label{B.12}
\end{eqnarray}
and a  special bubble type integral
\begin{equation}
I_2^1(k;\tilde k)=\int\frac{d^3\ell}{(2\pi)^3}\,\frac{(\ell\cdot k)e^{i\ell\tilde k}}{\ell^2(\ell+k)^2}
=\frac{1}{2}\int\frac{d^3\ell}{(2\pi)^3}\,e^{i\ell\tilde k}\Big(\frac{1}{\ell^2}-\frac{1}{(\ell+k)^2}-\frac{k^2}{\ell^2(\ell+k)^2}\Big)
=-\frac{k^2}{2}{I_2^0}(k;\tilde k),
\label{B.13}
\end{equation}
proportional to (\ref{B.9}), which has well defined, finite commutative limit.\\

Next we work on the bubble type integral $I_2^1(p;\tilde k)$, starting with the following integral property of the typical integral in the $\theta\to 0$ and $D\to3$ limits
\begin{equation}
I_2^1(p;\tilde k)=\int \frac{d^3 \ell}{(2\pi)^3}\frac{(\ell\cdot p)e^{\pm i\ell\theta k}}{\ell^2(\ell+p)^2}
=\frac{1}{2}\Big(1-e^{\mp ip\theta k}\Big)\int \frac{d^3 \ell}{(2\pi)^3}\frac{e^{\pm i\ell\theta k}}{\ell^2}
-\frac{p^2}{2}\int \frac{d^3 \ell}{(2\pi)^3}\frac{e^{\pm i\ell\theta k}}{\ell^2(\ell+p)^2}.
\label{B.14}
\end{equation}
Here we have used $2(\ell\cdot p)=(\ell+p)^2-\ell^2-p^2$ and changed the variable $\ell\to\ell-p$ in term with
$\ell^2$ in nominator. From the above eqs. (\ref{B.13})-(\ref{B.14}) we observe interesting property, as a kind of interplay with the two limits: $\theta\to0$ and $D\to3$, as follows:
\begin{equation}
\lim_{\theta\to 0}\int \frac{d^3 \ell}{(2\pi)^3}\frac{(\ell\cdot p)e^{\pm i\ell\theta k}}{\ell^2(\ell+p)^2}
=\lim_{D\to 3}\int \frac{d^D \ell}{(2\pi)^D}\frac{\ell\cdot p}{\ell^2(\ell+p)^2}
=-\frac{p^2}{2}\int \frac{d^3 \ell}{(2\pi)^3}\frac{1}{\ell^2(\ell+p)^2}.
\label{B.15}
\end{equation}

\subsubsection{Scalar triangle integral}

We evaluate the scalar triangle (\ref{B.3}) by using the standard Feynman parametrisation twice:
\begin{equation}
\begin{split}
I_3^0(p_1,p_2;\tilde k)=&\int\frac{d^D\ell}{(2\pi)^D}\,\frac{e^{i\ell\cdot\tilde k}}{\ell^2(\ell+p_1)^2(\ell+p_2)^2}
\\=&
\Gamma(3)e^{-ip_2\cdot\tilde k}\int\limits_0^1dx\,\int\limits_0^1dy\,(1-y)\,e^{i(1-y)(p_2-xp_1)\cdot\tilde k}\frac{e^{i\ell\cdot\tilde k}}{(\ell^2+\Delta)^3},
\end{split}
\label{B.16}
\end{equation}
where
\begin{equation}
\begin{split}
\Delta=&y(1-y)p_2^2+\big((1-y)x(1-x)+y(1-y)x^2\big)p_1^2-2(1-y)xy(p_1\cdot p_2)
\\=&(1-y)\big(x(1-x)(1-y)p_1^2+(1-x)y p_2^2+xy(p_1-p_2)^2\big).
\end{split}
\label{B.17}
\end{equation}
Next, introducing $\alpha$-parametrization, performing a Wick rotation ($\ell^0=-i\ell_E^0$), and integrating over the loop momenta, gives
\begin{equation}
\begin{split}
I_3^0(p_1,p_2;\tilde k)=&e^{-i p_2\cdot\tilde k}\int\frac{d^D\ell}{(2\pi)^D}\int\limits_0^1dx\int\limits_0^1dy(1-y)
\int\limits_0^\infty d\alpha\alpha^2 e^{i(1-y)(p_2-xp_1)\cdot\tilde k}e^{-\alpha(\ell^2+\Delta)+i\ell\cdot\tilde k}
\\=&-ie^{-i p_2\cdot\tilde k}\int\frac{d^D\ell_E}{(2\pi)^D}\int\limits_0^1dx\int\limits_0^1dy(1-y)
\int\limits_0^\infty d\alpha\alpha^2 e^{i(1-y)(p_2-xp_1)\cdot\tilde k}e^{-\alpha\ell_E^2}e^{-\alpha \Delta-\frac{\tilde k^2}{4\alpha}}
\\=&-i(4\pi)^{-\frac{D}{2}}e^{-i p_2\cdot\tilde k}\int\limits_0^1dx\int\limits_0^1dy(1-y)
e^{i(1-y)(p_2-xp_1)\cdot\tilde k}\int\limits_0^\infty d\alpha\alpha^{2-\frac{D}{2}}e^{-\alpha \Delta-\frac{\tilde k^2}{4\alpha}} .
\end{split}
\label{B.18}
\end{equation}
Than integration over $\alpha$ leads to the familiar Bessel $K$-function
\begin{equation}
\begin{split}
I_3^0=&-i(4\pi)^{-\frac{D}{2}}e^{-i p_2\cdot\tilde k}\int\limits_0^1dx\int\limits_0^1dy(1-y)e^{i(1-y)(p_2-xp_1)\cdot\tilde k}
2\Delta^{\frac{D}{4}-\frac{3}{2}}\left(\frac{\tilde k^2}{4}\right)^{\frac{3}{2}-\frac{D}{4}}K_{\frac{D}{2}-3}\left[\sqrt{\Delta\tilde k^2}\right].
\end{split}
\label{B.19}
\end{equation}
Now we take the limit $D\to 3$ and get
\begin{equation}
\begin{split}
{I_3^0}\Big|^{D=3}=&-i(4\pi)^{-\frac{3}{2}}e^{-i p_2\cdot\tilde k}\int\limits_0^1dx\int\limits_0^1 dy(1-y)
e^{i(1-y)(p_2-xp_1)\cdot\tilde k} 2\Delta^{-\frac{3}{4}}\left(\frac{\tilde k^2}{4}\right)^{\frac{3}{4}}K_{-\frac{3}{2}}\left[\sqrt{\Delta\tilde k^2}\right],
\end{split}
\label{B.20}
\end{equation}
which after using
\begin{equation}
K_{\pm\frac{3}{2}}[z]=\sqrt{\frac{\pi}{2}}e^{-z}(z^{-\frac{1}{2}}+z^{-\frac{3}{2}}),
\label{B.21}
\end{equation}
produces
\begin{equation}
\begin{split}
&{I_3^0}(p_1,p_2;\tilde k)\Big|^{D=3}=-\frac{ie^{-i p_2\tilde k}}{16\pi}\int\limits_0^1dx\int\limits_0^1dy(1-y)
e^{i(1-y)(p_2-xp_1)\cdot\tilde k} e^{-\sqrt{\Delta\tilde k^2}}\left(\Delta^{-1}(\tilde k^2)^\frac{1}{2}+\Delta^{-\frac{3}{2}}\right)
\\&=-\frac{ie^{-i p_2\cdot\tilde k}}{16\pi}\int\limits_0^1dx\int\limits_0^1dy(1-y)
e^{i(1-y)(p_2-xp_1)\tilde k}\sum\limits_{n=0}^\infty\frac{(-)^n}{n!} \left(\Delta^{\frac{n}{2}-1}(\tilde k^2)^\frac{n+1}{2}+\Delta^{\frac{n-3}{2}}(\tilde k^2)^\frac{n}{2}\right),
\\&{I_3^0}(p_1,p_2;\tilde k)\Big|^{D=3}=-\frac{ie^{-i p_2\cdot\tilde k}}{16\pi}\int\limits_0^1dx\sum\limits_{m=0}^\infty\frac{i^m}{m!}((p_2-xp_1)\cdot\tilde k)^m\int\limits_0^1dy(1-y)^{m+1}
\\&
\cdot\sum\limits_{n=0}^\infty\frac{(-)^n}{n!} \left(\Delta^{\frac{n}{2}-1}(\tilde k^2)^\frac{n+1}{2}+\Delta^{\frac{n-3}{2}}(\tilde k^2)^\frac{n}{2}\right).
\end{split}
\label{B.22}
\end{equation}
In order to analyze the commutative limit we evaluate the relevant term-by-term $y$-integrals
\begin{gather}
A(m,n)=\int\limits_0^1 dy\, (1-y)^{m+1}\cdot\Delta^{\frac{n}{2}-1},\; n\neq 0,
\label{B.23}\\
B(m,n)=\int\limits_0^1 dy\, (1-y)^{m+1}\cdot\Delta^{\frac{n-3}{2}},\; n\neq 1.
\label{B.24}
\end{gather}
The $A(m,0)$ and $B(m,1)$ integrals are excluded because they cancel each other for each equal $m$ in the series.
Putting back the explicit expression for $\Delta$, we get
\begin{gather}
\begin{split}
A(m,n)=&\big(x(1-x)p_1^2\big)^{\frac{n}{2}-1}\int\limits_0^1 dy\, (1-y)^{m+1}
\left(1-y\Big(1-\frac{(1-x)p_2^2+x(p_1-p_2)^2}{x(1-x)p_1^2}\Big)\right)^{\frac{n}{2}-1},
\end{split}
\label{B.25}\\
\begin{split}
B(m,n)=&\big(x(1-x)p_1^2\big)^{\frac{n-3}{2}}\int\limits_0^1 dy\, (1-y)^{m+1}
\left(1-y\Big(1-\frac{(1-x)p_2^2+x(p_1-p_2)^2}{x(1-x)p_1^2}\Big)\right)^{\frac{n-3}{2}}.
\end{split}
\label{B.26}
\end{gather}
Priorly we have learned that such integral can be expressed as hypergeometric functions:
\begin{equation}
\int\limits_0^1 x^a(1-x)^b(cx+d)^e=d^e\frac{\Gamma(a+1)\Gamma(b+1)}{\Gamma(a+b+2)} \,_2F_1\left(a+1,-e;a+b+2;-\frac{c}{d}\right),
\label{B.27}
\end{equation}
so the coefficients $A(m,n)$ and $B(m,n)$ are:
\begin{gather}
\begin{split}
A(m,n)=&\frac{\big(x(1-x)p_1^2\big)^{\frac{n}{2}-1}}{\frac{n}{2}+m+1}\,
_2F_1\left(1,1-\frac{n}{2};\frac{n}{2}+m+2;1-\frac{(1-x)p_2^2+x(p_1-p_2)^2}{x(1-x)p_1^2}\right),
\end{split}
\label{B.28}\\
\begin{split}
B(m,n)=&\frac{\big(x(1-x)p_1^2\big)^{\frac{n-3}{2}}}{\frac{n}{2}+m+\frac{1}{2}}\,
_2F_1\left(1,\frac{3-n}{2};\frac{n}{2}+m+\frac{3}{2};1-\frac{(1-x)p_2^2+x(p_1-p_2)^2}{x(1-x)p_1^2}\right).
\end{split}
\label{B.29}
\end{gather}
Using the hypergeometric analytical continuation formula
\begin{equation}
\begin{split}
\,_2F_1\left(a,b;c;1-z^{-1}\right)=&\frac{\Gamma(c)\Gamma(b-a)}{\Gamma(b)\Gamma(c-a)}z^a\,_2F_1\left(a,c-b;a-b+1;z\right)\\&+\frac{\Gamma(c)\Gamma(a-b)}{\Gamma(a)\Gamma(c-b)}\,_2F_1\left(b,c-a;b-a+1;z\right),
\end{split}
\label{B.30}
\end{equation}
we obtain
\begin{gather}
\begin{split}
A(m,n)=&-\frac{2}{n}\frac{\big(x(1-x)p_1^2\big)^{\frac{n}{2}}}{(1-x)p_2^2+x(p_1-p_2)^2}
\,_2F_1\left(1,n+m+1;\frac{n}{2}+1;\frac{x(1-x)p_1^2}{(1-x)p_2^2+x(p_1-p_2)^2}\right)
\\&+\frac{\Gamma\left(\frac{n}{2}+m+1\right)\Gamma\left(\frac{n}{2}\right)}{\Gamma\left(n+m+1\right)}\big((1-x)p_2^2+x(p_1-p_2)^2\big)^{\frac{n}{2}-1}\\&\cdot\,_1F_0\left(\frac{n}{2}+m+1;;\frac{x(1-x)p_1^2}{(1-x)p_2^2+x(p_1-p_2)^2}\right),
\end{split}
\label{B.31}\\
\begin{split}
B(m,n)=&\frac{2}{1-n}\frac{\big(x(1-x)p_1^2\big)^{\frac{n-1}{2}}}{(1-x)p_2^2+x(p_1-p_2)^2}
\,_2F_1\left(1,n+m;\frac{n+1}{2};\frac{x(1-x)p_1^2}{(1-x)p_2^2+x(p_1-p_2)^2}\right)
\\&+\frac{\Gamma\left(\frac{n}{2}+m+\frac{1}{2}\right)\Gamma\left(\frac{n-1}{2}\right)}{\Gamma\left(n+m\right)}\big((1-x)p_2^2+x(p_1-p_2)^2\big)^{\frac{n-3}{2}}\\&\cdot\,_1F_0\left(\frac{n}{2}+m+\frac{1}{2};;\frac{x(1-x)p_1^2}{(1-x)p_2^2+x(p_1-p_2)^2}\right).
\end{split}
\label{B.32}
\end{gather}
Till here we see that $A(m,n)$ and $B(m,n)$ can be expressed as power series with respect to the factor
$\frac{x(1-x)p_1^2}{(1-x)p_2^2+x(p_1-p_2)^2}$, once we expand all hypergeometric functions. The resulting $x$-integrals are well-defined term by term, thus we conclude that the commutative limit exists. Explicit calculation gives its value to be
\begin{equation}
\lim_{\theta\to 0}{I_3^0}(p_1,p_2;\tilde k)\Big|^{D=3}=\frac{-i}{8\sqrt{p_1^2 p_2^2 (p_1-p_2)^2}}.
\label{B.33}
\end{equation}

In summary, of the three master integrals, only ${I_1^0}\big|^{D=3}$ carries an IR divergence and therefore discontinuity at the commutative limit, while both ${I_2^0}\big|^{D=3}$ and ${I_3^0}\big|^{D=3}$ converge to a nonzero, finite commutative value.


We also list some properties of scalar triangle and/or trivial box integral needed further:
\begin{eqnarray}
\int\frac{d^3\ell}{(2\pi)^3}\,\frac{\ell^2\,e^{i\ell\tilde k}}{\ell^2(\ell+q_1)^2(\ell+q_2)^2(\ell+q_3)^2}
&=&e^{-iq_1\tilde k}I_3^0(q_2-q_1,q_3-q_1;\tilde k),
\label{B.34}\\
&=&e^{-iq_2\tilde k}I_3^0(q_1-q_2,q_3-q_2;\tilde k),
\label{B.35}\\
&=&e^{-iq_3\tilde k}I_3^0(q_1-q_3,q_2-q_3;\tilde k).
\label{B.36}
\end{eqnarray}

\subsection{Evaluating vector tadpole, bubble and triangle integrals}

\subsubsection{Vector tadpole integral}

We evaluate vector tadpole integral from (\ref{B.1}) by performing the following trick:
\begin{equation}
\begin{split}
I_1^\mu(\tilde k)=&\int\frac{d^D\ell}{(2\pi)^D}\,\frac{\ell^\mu e^{i\ell\cdot\tilde k}}{\ell^2}
=-i\frac{d}{dz_\mu}\Bigg|_{z_\mu=0}\int\frac{d^D\ell}{(2\pi)^D}\,\frac{e^{i\ell^\mu{\tilde K_\mu}}}{\ell^2};\;\;\tilde K_\mu=\tilde k_\mu+z_\mu,
\\
I_1^\mu(\tilde k)=&(-i)^2\frac{d}{dz_\mu}\Bigg|_{z_\mu=0}\int\limits_0^\infty d\alpha\int\frac{d^D\ell}{(2\pi)^D}
e^{-\alpha\ell^2+i\ell^\mu \tilde K_\mu}
=\frac{-d}{dz_\mu}\Bigg|_{z_\mu=0}\int\limits_0^\infty d\alpha\,\int\frac{d^D\ell_E}{(2\pi)^D}e^{-\alpha\ell_E^2-\frac{\tilde K^2}{4\alpha}}
\\
=&-\frac{d}{dz_\mu}\Bigg|_{z_\mu=0}(4\pi)^{-\frac{D}{2}}
\int\limits_0^\infty d\alpha\,\alpha^{-\frac{D}{2}}e^{-\frac{\tilde K^2}{4\alpha}}
=\frac{-1}{(4\pi)^{\frac{D}{2}}}\frac{d}{dz_\mu}\Bigg|_{z_\mu=0}\left(\frac{\tilde K^2}{4}\right)^{1-\frac{D}{2}}\Gamma\left(\frac{D}{2}-1\right)
\\
=&\frac{-1}{(4\pi)^{\frac{D}{2}}}4^{\frac{D}{2}-1}\Gamma\left(\frac{D}{2}-1\right)\frac{d}{dz^\mu}\Bigg|_{z^\mu=0}
\Big(\big(\tilde k^\mu+z^\mu\big)^2\Big)^{1-\frac{D}{2}}
\\
=&-\frac{\pi^{-\frac{D}{2}}}{4}\Gamma\left(\frac{D}{2}-1\right)\big(2-D\big)\big(\tilde k^\mu\big)^{1-D},
\end{split}
\label{B.37}
\end{equation}
and for $D\to3$ we have solution:
\begin{equation}
I_1^\mu(\tilde k)\Big|^{D=3}=\frac{\Gamma\big(\frac{1}{2}\big)}{4\pi^\frac{3}{2}}\big(\tilde k^\mu\big)^{-2}.
\label{B.38}
\end{equation}
which is quadratically (fairly badly) IR divergent, and the source of celebrated quadratic UV/IR mixing effect
\cite{Matusis:2000jf,Minwalla:1999px,Hayakawa:1999yt,VanRaamsdonk:2000rr,VanRaamsdonk:2001jd,Schupp:2008fs,Horvat:2011bs,Martin:2020ddo}.

\subsubsection{Vector bubble integral}

Using the same trick as in the computation of vector tadpole integral (\ref{B.37}) we have
\begin{equation}
\begin{split}
I_2^\mu(p;\tilde k)=&i\int\frac{d^3\ell}{(2\pi)^3}\,\frac{\ell^\mu\,e^{i\ell\cdot\tilde k}}{\ell^2(\ell+p)^2}
=\int\limits_0^1 dx\,\int\frac{d^3\ell}{(2\pi)^3}\,\frac{\ell^\mu\,e^{i\ell\cdot\tilde k}}{\big[(\ell+xp)^2+x(1-x)p^2\big]^2}
\\=&
i\frac{d}{dz_\mu}\Bigg|_{z_\mu=0}\int\limits_0^1 dx\,e^{-ixp\cdot(\tilde k+z)}\int\frac{d^3\ell}{(2\pi)^3}\,\frac{e^{i\ell\cdot(\tilde k+z)}}{\big[(\ell+xp)^2+x(1-x)p^2\big]^2}\Bigg|_{\ell\to\ell-xp}
\\=&
i\frac{d}{dz_\mu}\Bigg|_{z_\mu=0}\int\limits_0^1 dx\,\,e^{-ixp\cdot(\tilde k+z)}\int\frac{d^3\ell}{(2\pi)^3}\,\frac{e^{i\ell\cdot(\tilde k+z)}}{\big[\ell^2+x(1-x)p^2\big]^2}
\\=&
i\frac{d}{dz_\mu}\Bigg|_{z_\mu=0}\int\limits_0^1 dx\,\,e^{-ixp\cdot(\tilde k+z)}\int\limits_0^\infty\frac{dt}{(2\sqrt\pi)^3}\,t^{\frac{1}{2}-1}e^{-\frac{(\tilde k+z)^2}{4t}}\,e^{-tx(1-x)p^2}
\\=&\frac{p^\mu}{(2\sqrt\pi)^3}
\int\limits_0^1 dx\,xe^{-ixp\cdot\tilde k}\int\limits_0^\infty dt\,t^{\frac{1}{2}-1}e^{-\frac{\tilde k^2}{4t}-tx(1-x)p^2}
\\&
+\frac{i}{2}\frac{\tilde k^\mu}{(2\sqrt\pi)^3}\int\limits_0^1 dxe^{-ixp\cdot\tilde k}\int\limits_0^\infty dt\,t^{\frac{1}{2}-2}
e^{-\frac{\tilde k^2}{4t}-tx(1-x)p^2},
\end{split}
\label{B.39}
\end{equation}
\begin{equation}
\begin{split}
I_2^\mu(p;\tilde k)=&
\frac{2p^\mu}{(2\sqrt\pi)^3}\int\limits_0^1 dx\,x\,e^{-ixp\cdot\tilde k}\Big(\frac{\tilde k^2}{4x(1-x)p^2}\Big)^{1/4}
K_{1/2}\Big[\sqrt{x(1-x)p^2\tilde k^2}\Big]
\\&
+\frac{\tilde k^\mu}{(2\sqrt\pi)^3}\int\limits_0^1 dx\,e^{-ixp\cdot\tilde k}\Big(\frac{\tilde k^2}{4x(1-x)p^2}\Big)^{3/4}
K_{3/2}\Big[\sqrt{x(1-x)p^2\tilde k^2}\Big],
\end{split}
\label{B.40}
\end{equation}
where after expansion and x-integration we obtain the following leading terms:
\begin{equation}
\begin{split}
I_2^\mu(p,\tilde k)=\int \frac{d^3 \ell}{(2\pi)^3}\frac{\ell^\mu e^{-i\ell\tilde k}}{\ell^2(\ell-p)^2}
=\frac{1}{8\pi}\,\frac{\tilde k^\mu}{\sqrt{\tilde k^2}}+\frac{i}{16}\frac{ p^\mu}{\sqrt{ p^2}}\,+\mathcal O^{\mu}(\tilde k^1),
\end{split}
\label{B.41}
\end{equation}
confirming earlier result, see i.e. (C.21,C.27) in \cite{Martin:2017nhg}. Here in the limit $\tilde k\to0$ term $\mathcal O^{\mu}(\tilde k^1)$ vanishes, while the above integral $I_2^\mu(p,\tilde k)$ is not well defined, though not divergent. By claiming not well-defined integral we mean that integral’s value in that limit depends on the way how $\tilde k$ approach to the zero point, however integral is not divergent.

\subsubsection{Vector triangle integral}

After some lengthy computations in Euclidian metrics, using Wick rotations, from (\ref{B.3}) in $3D$ we obtain
\begin{eqnarray}
I_3^\mu(p_1,p_2;\tilde k)&=&\int\frac{d^3\ell}{(2\pi)^3}\frac{\ell^\mu e^{i\ell\tilde k}}{\ell^2(\ell+p_1)^2(\ell+p_2)^2}
=\frac{1}{(2\sqrt\pi)^3\sqrt2}\int_0^1 xdx\int_0^1dy\; e^{-iQ\tilde k}
\nonumber\\
&\cdot&\bigg[-\tilde k^\mu
\bigg(\frac{\tilde k^2}{M}\bigg)^{\frac{1}{4}}K_{\frac{1}{2}}\big(\sqrt{M\tilde k^2}\big)
-iQ^\mu\bigg(\frac{\tilde k^2}{M}\bigg)^{\frac{3}{4}}K_{\frac{3}{2}}\big(\sqrt{M\tilde k^2}\big)\bigg],
\nonumber\\
M&=&x(1-y)p_1^2 +(1-y)p_2^2-Q^2,\;\;\;\;p_1^\mu{\bot}p_2^\mu,
\nonumber\\
Q^\mu&=&x(1-y)p_1^\mu+(1-y)p_2^\mu.
\label{B.42}
\end{eqnarray}
If some vector $a_\mu$ is perpendicular on $\{p_1^\mu,p_2^\mu\}$, than $ap_1=ap_2=0\Rightarrow a\cdot Q=0$, is giving
\begin{eqnarray}
I_3^1(p_1,p_2;a(p_1,p_2),\tilde k)&=&\int\frac{d^3\ell}{(2\pi)^3}\frac{(\ell\cdot a) e^{i\ell\tilde k}} {\ell^2(\ell+p_1)^2(\ell+p_2)^2}
\label{B.43}\\
&=&\frac{-(\tilde k a)(\tilde k^2)^{\frac{1}{4}}}{(2\sqrt\pi)^3\sqrt2}\int_0^1 xdx\int_0^1dy\;
\frac{e^{-iQ\tilde k}K_{\frac{1}{2}}\big(\sqrt{M\tilde k^2}\big)}{M^{\frac{1}{4}}}.
\nonumber
\end{eqnarray}
Explicit and detailed calculation of the above integrals is given in the following appendices C.3.1. and C.3.2, respectively.

Nevertheless, we shall next reduce/compute vector and tensor integrals with the NC phase factor included. We call them further on the spurious integrals of bubble, triangle and box diagrams by applying the van Neerven-Vermaseren method.

\section{NC deformation of the spurious integrals}

Commutative spurious integrals (terms with loop momenta powers $\ell^1$ and $\ell^2$ in numerator, respectively) vanish in loop integration because of the Lorentz structure, which is no longer the case for the NC non-planar integrals. In $3D$, there should be one type of triangle spurious term, four types of bubble and three types of tadpole spurious terms if we start with a triangle tensor integral and presume the renormalisable gauge condition that says the formal power of loop momenta in the numerator should not exceed the number of loop momenta in denominators. Short expressions of the bubble and triangle numerators should be $(\ell\cdot \omega_1)$ and \{$(\ell \omega_1)$, $(\ell \omega_2)$, $(\ell \omega_1)(\omega_2)$, $(\ell \omega_1)^2-(\ell \omega_2)^2$\}, where
\begin{equation}
\omega_1^\mu=\epsilon^{\mu jk}p_{1_j}p_{2_k},\;\;
\omega_2^\mu=\frac{p_1^\mu(p_1 p_2)-p_2^\mu p_1^2}{\sqrt{p_1^2}}.
\label{C.1}
\end{equation}
Vectors $\omega_1^\mu$ and $\omega_2^\mu$ satisfy the following conditions:
\begin{equation}
p_1\omega_1=0,\;p_2\omega_1=0,\; p_1\omega_2=0,\;\omega_1\omega_2=0,\;|\omega_1|=|\omega_2|.
\label{C.2}
\end{equation}

\subsection{Tadpole vector term}

Tadpole spurious terms may be realized with a simple vector numerator $\ell^\mu$. So, the three tadpole spurious terms can be expressed as a single vector integral already computed, as quadratically IR divergent (\ref{B.38}), which for $D=3$  we can rewrite as:
\begin{equation}
{I}_1^\mu(\tilde k)=\int\frac{d^D\ell}{(2\pi)^D}\,\frac{\ell^\mu}{\ell^2}e^{i\ell\tilde k}\;\;
\longrightarrow\;\;
{I_1^\mu}(\tilde k)\Big|^{D=3}=\frac{1}{4\pi}\frac{\tilde k^\mu}{(\tilde k^2)^{\frac{3}{2}}}.
\label{C.3}
\end{equation}

\subsection{Bubble terms}

Next we consider the bubble spurious terms. While superficially divergent, the planar integral converges in dimensional regularization because massless tadpoles vanish in this prescription. Non-planar integral gives a value which diverges at commutative limit. Eventually we could still reach the right value by work in $D=3$. In this case we need two auxiliary vectors $\omega_1^\mu$ and $\omega_2^\mu$ to decompose $\ell^\mu$, they should satisfy the conditions $\omega_i\cdot p=0$ and $\omega_i\cdot \omega_j=\delta_{ij}$. Then
\begin{equation}
\ell^\mu=(\ell\cdot\omega_1)\omega_1^\mu+(\ell\cdot\omega_2)\omega_2^\mu+(\ell\cdot p)\frac{p^\mu}{p^2},
\label{C.4}
\end{equation}
\begin{equation*}
\frac{\ell^\mu}{\ell^2(\ell+p)^2}=\omega_1^\mu\frac{\ell\cdot\omega_1}{\ell^2(\ell+p)^2}+\omega_2^\mu\frac{\ell\cdot\omega_2}{\ell^2(\ell+p)^2}+\frac{p^\mu}{2p^2}\left(\frac{1}{\ell^2}-\frac{1}{(\ell+p)^2}-\frac{p^2}{\ell^2(\ell+p)^2}\right).
\end{equation*}

Next we handle the bubble spurious terms with $\ell^1$ and $\ell^2$ types of numerators. For completeness taking two auxiliary vectors $\omega_1^\mu$, and $\omega_2^\mu$ from eqs.(\ref{C.1}) and (\ref{C.2}) we obtain four combinations for integrals needed $(\ell\cdot\omega_1)$, $(\ell\cdot \omega_2)$, $(\ell\cdot \omega_1)(\ell\cdot \omega_2)$, and $(\ell\cdot\omega_1)^2-(\ell\cdot\omega_2)^2$:
\begin{eqnarray}
I_{2_1}^1&=&\int\frac{d^D\ell}{(2\pi)^D}\,\frac{\ell\cdot\omega_1}{\ell^2(\ell+p)^2}e^{i\ell\cdot\tilde k},
\;\;
I_{2_2}^1=\int\frac{d^D\ell}{(2\pi)^D}\,\frac{\ell\cdot\omega_2}{\ell^2(\ell+p)^2}e^{i\ell\cdot\tilde k},
\nonumber\\
I_{2_1}^2&=&\int\frac{d^D\ell}{(2\pi)^D}\,\frac{(\ell\cdot\omega_1)(\ell\cdot\omega_2)}{\ell^2(\ell+p)^2}e^{i\ell\cdot\tilde k},
\;\;
I_{2_2}^2=\int\frac{d^D\ell}{(2\pi)^D}\,\frac{(\ell\cdot\omega_1)^2-(\ell\cdot\omega_2)^2}{\ell^2(\ell+p)^2}e^{i\ell\cdot\tilde k}.
\label{C.5}
\end{eqnarray}
Using the existing results of bubble diagram computations in sections 5.-6., eqs. (6.3-6.7) and appendix C from our previous paper \cite{Martin:2017nhg}, for vector (we have it also in appendix B) and tensor bubble type of integrals,
\begin{equation}
I_2^{\mu}(p;\tilde k)=\int\frac{d^D\ell}{(2\pi)^D}\,\frac{\ell^\mu e^{i\ell\tilde k}}{\ell^2(\ell+p)^2}\bigg|_{(\ref{B.41})},
\;\;{\rm and}\;\;
I_2^{\mu\nu}(p;\tilde k)=\int\frac{d^D\ell}{(2\pi)^D}\,\frac{\ell^\mu\ell^\nu e^{i\ell\tilde k}}{\ell^2(\ell+p)^2},
\label{C.6}
\end{equation}
it is not difficult to obtain $3D$ case for the above bubble spurious integrals (\ref{C.5}):
\begin{equation}
I_{2_1}^1=\frac{1}{8\pi}\frac{(\tilde k\cdot\omega_1)}{\sqrt{\tilde k^2}}\int\limits_0^1 dx\,e^{-\sqrt{x(1-x)p^2\tilde k^2}-ip\cdot\tilde k},\;\;
I_{2_2}^1=\frac{1}{8\pi}\frac{(\tilde k\cdot\omega_2)}{\sqrt{\tilde k^2}}\int\limits_0^1 dx\,e^{-\sqrt{x(1-x)p^2\tilde k^2}-ip\cdot\tilde k},
\label{C.7}
\end{equation}
\begin{equation}
I_{2_1}^2=\frac{i}{8\pi}\frac{(\tilde k\cdot\omega_1)(\tilde k\cdot\omega_2)}{\tilde k^2}\int\limits_0^1 dx\,e^{-\sqrt{x(1-x)p^2\tilde k^2}-ip\cdot\tilde k}\left(\sqrt{x(1-x)p^2}+\frac{1}{\sqrt{\tilde k^2}}\right),
\label{C.8}
\end{equation}
\begin{equation}
I_{2_2}^2=\frac{i}{8\pi}\frac{((\tilde k\cdot\omega_1)^2-(\tilde k\cdot\omega_2)^2)}{\tilde k^2}\int\limits_0^1 dx\,e^{-\sqrt{x(1-x)p^2\tilde k^2}-ip\cdot\tilde k}\left(\sqrt{x(1-x)p^2}+\frac{1}{\sqrt{\tilde k^2}}\right).
\label{C.9}
\end{equation}
Till here we can notice clearly that the above vector bubble spurious integrals do not have well-defined $\theta^{\mu\nu}\to 0$ limits.

\subsection{Reduction of triangle integrals}

We start to compute integrals by introducing decomposition of loop momenta $\ell^\mu$ in the van Neerven-Vermaseren basis $\{w_1^\mu,w_2^\mu,a^\mu\}$ as a functions of two momenta $\{q_1,q_2\}$:
\begin{eqnarray}
\ell^\mu&=&(\ell q_1)w_1^\mu(q_1,q_2) +(\ell q_2)w_2^\mu(q_1,q_2) +(\ell a)\frac{a^\mu(q_1,q_2)}{a^2} ,
\label{C.10}\\
w_1^\mu(q_1,q_2)&=&\frac{\delta^{\mu q_2}_{q_1q_2}}{\Delta(q_1,q_2)},\;
w_2^\mu(q_1,q_2)=\frac{\delta^{ q_1\mu}_{q_1q_2}}{\Delta(q_1,q_2)},\;
a^\mu(q_1,q_2)=\epsilon^{\mu\nu\rho}k_{1_\nu}q_{2_\rho} ,
\label{C.11}\\
\Delta(q_1,q_2)&=&\delta_{q_1q_2}^{q_1q_2}=\det(q_{1_m}\cdot q_{2_n})=q_1^2q_2^2-(q_1q_2)^2,
\;a^2(q_1,q_2)=-\Delta(q_1,q_2),
\nonumber
\end{eqnarray}
with properties $w_i\cdot q_j=\delta_{ij}$, and $a\cdot q_i=0\;\Rightarrow a\cdot w_i=0,\;i,j=1,2$.

\subsubsection{Vector triangle integral  in $D=3$}

First we compute $I_3^\mu(q_1,q_2;a(q_1,q_2),\tilde k)$ integral by using decomposition of $\ell^\mu$ (\ref{C.10}) in basis (\ref{C.11}), and in terms  of scalar master integrals obtain:
\begin{eqnarray}
I_3^\mu(q_1,q_2;a,\tilde k)&=&w_1^\mu\int\frac{d^3\ell}{(2\pi)^3}\,\frac{(\ell q_1)e^{i\ell\tilde k}}{\ell^2(\ell+q_1)^2(\ell+q_2)^2}+w_2^\mu\int\frac{d^3\ell}{(2\pi)^3}\,\frac{(\ell q_2)e^{i\ell\tilde k}}{\ell^2(\ell+q_1)^2(\ell+q_2)^2}
\nonumber\\&+&\frac{a^\mu}{a^2}I_3^1(q_1,q_2;a(q_1,q_2),\tilde k)
\label{C.12}\\
&=&
\frac{w_1^\mu}{2}\left(I^0_2(q_2;\tilde k)-e^{-iq_1\tilde k}I^0_2(q_2-q_1;\tilde k)-q_1^2I^0_3(q_1, q_2;\tilde k)\right)
\nonumber\\
&+&\frac{w_2^\mu}{2}\left(I^0_2(q_1;\tilde k)-e^{-iq_2\tilde k}I^0_2(q_1-q_2;\tilde k)-q_2^2I^0_3(q_1, q_2;\tilde k)\right)+\frac{a^\mu}{a^2}{I_3^1}(q_1,q_2;a,\tilde k).
\nonumber
\end{eqnarray}
where $I^0_2(q;\tilde k)$ [with $q$ being either: $q_1$, $q_2$, $q_1-q_2$, $q_2-q_1$], is given in (\ref{B.7})-(\ref{B.8}), while $I^0_3(q_1, q_2;\tilde k)$ in (\ref{B.16})-(\ref{B.32}), and $I^1_3(q_1, q_2;a(q_1,q_2),\tilde k)$ is just presented in (\ref{B.43}), while it is computed completely and given explicitly in the next appendix C.3.2 as eqs. (\ref{C.16})-(\ref{C.20}).\\

With help of (\ref{C.12}), we obtain more general reduced triangle integrals in basis (\ref{C.11}), with power $\ell^1$ type of  numerators needed further on:
\begin{eqnarray}
&&\int\frac{d^3\ell}{(2\pi)^3}\,\frac{(\ell q_1)e^{i\ell\tilde k}}{\ell^2(\ell+q_2)^2(\ell+q_3)^2}
\nonumber\\
&=&\frac{1}{2}q_1\cdot w_1(q_2,q_3)\Big[I^0_2(q_3;\tilde k)-e^{-iq_2\tilde k}I_2^0(q_3-q_2;\tilde k)-q_2^2I_3^0(q_2,q_3;\tilde k)\Big]
\nonumber\\
&+&\frac{1}{2}q_1\cdot w_2(q_2,q_3)\Big[I^0_2(q_2;\tilde k)-e^{-iq_2\tilde k}I_2^0(q_3-q_2;\tilde k)-q_3^2I_3^0(q_2,q_3;\tilde k)\Big]
\nonumber\\
&-&\frac{q_1\cdot a(q_2,q_3)}{\Delta(q_2,q_3)}I_3^0(q_2,q_3;a(q_2,q_3),\tilde k),
\nonumber\\
&&\int\frac{d^3\ell}{(2\pi)^3}\,\frac{(\ell q_2)e^{i\ell\tilde k}}{\ell^2(\ell+q_1)^2(\ell+q_3)^2}
\nonumber\\
&=&\frac{1}{2}q_2\cdot w_1(q_1,q_3)\Big[I^0_2(q_3;\tilde k)-e^{-iq_1\tilde k}I_2^0(q_3-q_1;\tilde k)-q_1^2I_3^0(q_1,q_3;\tilde k)\Big]
\nonumber\\
&+&\frac{1}{2}q_2\cdot w_2(q_1,q_3)\Big[I^0_2(q_1;\tilde k)-e^{-iq_1\tilde k}I_2^0(q_3-q_1;\tilde k)-q_3^2I_3^0(q_1,q_3;\tilde k)\Big]
\nonumber\\
&-&\frac{q_2\cdot a(q_1,q_3)}{\Delta(q_1,q_3)}I_3^0(q_1,q_3;a(q_1,q_3),\tilde k).
\label{C.13}
\end{eqnarray}

We also list two triangle integrals which we shall need in the reduction/computation of a box spurious integrals further on:
\begin{eqnarray}
&&\int\frac{d^3\ell}{(2\pi)^3}\,\frac{(\ell q_3)e^{i\ell\tilde k}}{\ell^2(\ell+q_1)^2(\ell+q_2)^2}
\nonumber\\
&=&\frac{1}{2}q_3\cdot w_1(q_1,q_2)\Big[I^0_2(q_2;\tilde k)-e^{-iq_1\tilde k}I_2^0(q_2-q_1;\tilde k)-q_1^2I_3^0(q_1,q_2;\tilde k)\Big]
\nonumber\\
&+&\frac{1}{2}q_3\cdot w_2(q_1,q_2)\Big[I^0_2(q_1;\tilde k)-e^{-iq_1\tilde k}I_2^0(q_2-q_1;\tilde k)-q_2^2I_3^0(q_1,q_2;\tilde k)\Big]
\nonumber\\
&-&\frac{q_3\cdot a(q_1,q_2)}{\Delta(q_1,q_2)}I_3^0(q_1,q_2;a(q_1,q_2),\tilde k),
\label{C.14}
\end{eqnarray}
and
\begin{eqnarray}
&&\int\frac{d^3\ell}{(2\pi)^3}\,\frac{(\ell q_1)e^{i\ell\tilde k}}{\ell^2(\ell+q_2-q_1)^2(\ell+q_3-q_1)^2}
\nonumber\\
&=&\frac{1}{2}q_1\cdot w_1(q_2-q_1,q_3-q_1)\Big[I^0_2(q_3-q_1;\tilde k)-e^{-i(q_2-q_1)\tilde k}I_2^0(q_3-q_2;\tilde k)
\nonumber\\
&-&(q_2-q_1)^2I_3^0(q_2-q_1,q_3-q_1;\tilde k)\Big]
\nonumber\\
&+&\frac{1}{2}q_1\cdot w_2(q_2-q_1,q_3-q_1)\Big[I^0_2(q_2-q_1;\tilde k)-e^{-i(q_2-q_1)\tilde k}I_2^0(q_3-q_2;\tilde k)
\nonumber\\
&-&(q_3-q_1)^2I_3^0(q_2-q_1,q_3-q_1;\tilde k)\Big]
\nonumber\\
&-&\frac{q_1\cdot a(q_2-q_1,q_3-q_1)}{\Delta(q_2-q_1,q_3-q_1)}I_3^0(q_2-q_1,q_3-q_1;a(q_2-q_1,q_3-q_1),\tilde k),
\label{C.15}
\end{eqnarray}
from where we easy get other combinations when necessary.

\subsubsection{Evaluating triangle integral with $\ell^1$ power term in numerator}

Next, we explicitly evaluate integral $I_3^1(p_1,p_2;a(p_1,p_2),\tilde k)$,
\begin{equation}
I_3^1(p_1,p_2;a(p_1,p_2),\tilde k)=\int\frac{d^D\ell}{(2\pi)^D}\,\frac{\big(\ell\cdot a(p_1,p_2)\big) e^{i\ell\tilde k}}{\ell^2(\ell+p_1)^2(\ell+p_2)^2},
\label{C.16}
\end{equation}
in which $a^\mu(p_1,p_2)$ is (up to scaling) the third vector in the van Neerven-Vermaseren basis
(\ref{C.11}). The vanishing of the corresponding commutative integral can be observed by realising that tensor reduction of the following integral can only be the linear combination of $p_1^\mu$ and $p_2^\mu$, and therefore orthogonal to $a_{\mu}$. We can now exploit this fact and evaluate $I_3^1$ (and other NC deformed spurious terms) relatively quickly using the conventional Feynman parametrization, as only the NC tensor structure $\tilde k^{\mu}=\theta^{\mu\nu}k_\nu$ based terms ($(\tilde k\cdot a)$,....) are relevant. For this reason from (\ref{C.16}), following the same prescription as in the appendix B.3.2, we have
\begin{eqnarray}
I_3^1(p_1,p_2;a(p_1,p_2),\tilde k)&=&\Gamma(3)e^{-ip_2\tilde k}\int\limits_0^1dx\,
\int\limits_0^1dy\,(1-y)\,e^{i(1-y)(p_2-xp_1)\tilde k}(\ell\cdot a)\frac{e^{i\ell\tilde k}}{(\ell^2+\Delta)^3},
\nonumber\\
\label{C.17}
\end{eqnarray}
where $\Delta$ is the same as defined for scalar triangle (\ref{B.16})-(\ref{B.17}).
Notice that $(\ell\cdot a)$ remains unchanged upon any redefinitions of $\ell$, because all the shifts introduced are linear combinations of $p_1$ and $p_2$. Next one can introduce the $\alpha$-parametrization and first perform the loop integral in $D$ dimensions:
\begin{equation}
\begin{split}
I_3^1=&e^{-i p_2\tilde k}\int\frac{d^D\ell}{(2\pi)^D}\int\limits_0^1dx\int\limits_0^1dy(1-y) e^{i(1-y)(p_2-xp_1)\tilde k}
\int\limits_0^\infty d\alpha\alpha^2 (\ell\cdot a)e^{-\alpha(\ell^2+\Delta)+i\ell\tilde k}
\\=&e^{-i p_2\tilde k}\int\frac{d^D\ell_E}{(2\pi)^D}\int\limits_0^1dx\int\limits_0^1dy(1-y)e^{i(1-y)(p_2-xp_1)\tilde k}
\frac{(\tilde k\cdot a)}{2}\int\limits_0^\infty d\alpha\alpha e^{-\alpha\ell_E^2-\alpha \Delta-\frac{\tilde k^2}{4\alpha}}
\\=&(4\pi)^{-\frac{D}{2}}e^{-i p_2\tilde k}\frac{(\tilde k\cdot a)}{2} \int\limits_0^1dx\int\limits_0^1dy(1-y)
e^{i(1-y)(p_2-xp_1)\tilde k}\int\limits_0^\infty d\alpha\alpha^{1-\frac{D}{2}}e^{-\alpha \Delta-\frac{\tilde k^2}{4\alpha}}.
\end{split}
\label{C.18}
\end{equation}
Evaluating the $\alpha$-integration we than have
\begin{equation}
\begin{split}
I_3^1=&(4\pi)^{-\frac{D}{2}}e^{-i p_2\cdot\tilde k}(\tilde k\cdot a)\int\limits_0^1dx\int\limits_0^1dy(1-y)e^{i(1-y)(p_2-xp_1)\tilde k}
\Delta^{\frac{D}{4}-1}\Big(\frac{\tilde k^2}{4}\Big)^{1-\frac{D}{4}}K_{\frac{D}{2}-2}\left[\sqrt{\Delta\tilde k^2}\right],
\end{split}
\label{C.19}
\end{equation}
and taking the limit $D\to 3$ obtain
\begin{equation}
\begin{split}
I_3^1\Big|^{D=3}=&(4\pi)^{-\frac{3}{2}}e^{-i p_2\tilde k}(\tilde k\cdot a)\int\limits_0^1dx\int\limits_0^1 dy(1-y)
e^{i(1-y)(p_2-xp_1)\tilde k}\cdot \Delta^{-\frac{1}{4}}\left(\frac{\tilde k^2}{4}\right)^{\frac{1}{4}}K_{\frac{1}{2}}\left[\sqrt{\Delta\tilde k^2}\right]
\\=&\frac{1}{16\pi}e^{-i p_2\tilde k}(\tilde k\cdot a)\int\limits_0^1dx\int\limits_0^1 dy(1-y)
e^{i(1-y)(p_2-xp_1)\cdot\tilde k} \Delta^{-\frac{1}{2}}e^{-\sqrt{\Delta\tilde k^2}}
\\=&\frac{1}{16\pi}e^{-i p_2\tilde k}(\tilde k\cdot a)\int\limits_0^1dx\sum\limits_{m=0}^\infty\frac{i^m}{m!}\big((p_2-xp_1)\tilde k\big)^m\int\limits_0^1 dy
\sum\limits_{n=0}^\infty\frac{(-)^n}{n!} (1-y)^{m+1}\Delta^{\frac{n-1}{2}}(\tilde k^2)^\frac{n}{2}.
\end{split}
\label{C.20}
\end{equation}
Thus the $y$-integration could be evaluated using the $B(m,n)$ result, starting at $(m=0,\,n=2)$. So since the
$\theta\to 0$ limit of $I_3^1|^{D=3}$ is controlled by $(\tilde k\cdot a)$, such integrals are regular and
\begin{equation}
\lim_{\theta\to 0}{I_3^1}(p_1,p_2;a(p_1,p_2),\tilde k)\Big|^{D=3}=0,
\label{C.21}
\end{equation}
i.e. the above integral $I_3^1(p_1,p_2;a(p_1,p_2),\tilde k)$ has also well-defined commutative limit.

Inspecting $3D$ integral solutions (\ref{B.8})-(\ref{B.32}) for $I^0_2$, $I^0_3$ and the above integral $I_3^1(p_1,p_2;a,\tilde k)$ (\ref{C.20}), it is clear that spurious $3D$ integral $I_3^\mu(p_1,p_2;a,\tilde k)$ (\ref{C.12}) has also well-defined commutative limit.

\subsubsection{Reduction of tensor triangle integral in $D=3$}

Using again decomposition of vector $\ell^\mu$ in the basis $\{w_1^\mu,w_2^\mu,a^\mu\}$ (\ref{C.10})-(\ref{C.11}) we have:
\begin{eqnarray}
&&I_3^{\mu\nu}(q_1,q_2;\tilde k)=\int\frac{d^3\ell}{(2\pi)^3}\,\frac{\ell^\mu\ell^\nu \cdot e^{i\ell\tilde k}}{\ell^2(\ell+q_1)^2(\ell+q_2)^2}
=\sum_{i,j=1}^3w_i^\mu w_j^\nu\int\frac{d^3\ell}{(2\pi)^3}\,\frac{(\ell q_i)(\ell q_j) \cdot e^{i\ell\tilde k}}{\ell^2(\ell+q_1)^2(\ell+q_2)^2}
\nonumber\\
&=&\int\frac{d^3\ell}{(2\pi)^3}e^{i\ell\tilde k}\;\frac{w_1^\mu w_1^\nu(\ell q_1)^2+w_2^\mu w_2^\nu(\ell q_2)^2
+(w_1^\mu w_2^\nu+w_2^\mu w_1^\nu)(\ell q_1)(\ell q_2)}
{\ell^2(\ell+q_1)^2(\ell+q_2)^2}
\nonumber\\
&+&\int\frac{d^3\ell}{(2\pi)^3}e^{i\ell\tilde k}\;\frac{(w_1^\mu a^\nu+a^\mu w_1^\nu)(\ell q_1)(\ell a)+(w_2^\mu a^\nu+a^\mu w_2^\nu)(\ell q_2)(\ell a)+a^\mu a^\nu(\ell a)^2}{\ell^2(\ell+q_1)^2(\ell+q_2)^2}.
\label{C.22}
\end{eqnarray}
Above integrals with numerators $(\ell q_1)^2; (\ell q_2)^2;(\ell q_1)(\ell q_2)$ can be reduced into bubble with power $\ell^1$ numerator. Next two with $(\ell q_1)(\ell a),(\ell q_2)(\ell a)$ we reduced into bubble spurious terms in the appendix C.2. The last term $(\ell a)^2$ requires some extra work. Here we notice that
\begin{equation}
\begin{split}
a^\mu a^\nu=&\epsilon^{\mu\rho\sigma}\epsilon^{\nu\delta\xi}{q_1}_\rho{q_2}_\sigma{q_1}_\delta{q_2}_\xi
\\=&g^{\mu\nu}\big((q_1 q_2)^2-q_1^2q_2^2\big)-(q_1 q_2)\big(q_1^\mu q_2^\nu+q_1^\nu q_2^\mu\big)+q_1^\mu q_1^\nu q_2^2+q_2^\mu q_2^\nu q_1^2,
\end{split}
\nonumber
\end{equation}
therefore
\begin{equation}
\frac{(\ell a)^2}{\ell^2(\ell+q_1)^2(\ell+q_2)^2}
=\frac{\ell^2\big((q_1 q_2)^2-q_1^2q_2^2\big)+q_2^2(\ell q_1)^2-2(q_1 q_2)(\ell q_1)(\ell q_2) + q_1^2(\ell q_2)^2}{\ell^2(\ell+q_1)^2(\ell+q_2)^2}.
\label{C.23}
\end{equation}
One could then classify the reduction by tensor structures and express this integrand in the above basis,
i.e. finally in terms of master bubble integrals.

Explicitly, we first decompose $\ell q_i=\frac{1}{2}[(\ell+q_i)^2-\ell^2-q_i^2]$, compute, and obtain reduction of triangle integrals with power $\ell^2$ into power $\ell^1$ type of bubble and triangle integrals given in appendix B and C, needed to evaluate the complete tensor triangle integral (\ref{C.22}):
\begin{eqnarray}
&&\int\frac{d^3\ell}{(2\pi)^3}\,\frac{(\ell q_1)^2e^{i\ell\tilde k}}{\ell^2(\ell+q_1)^2(\ell+q_2)^2}
=\frac{1}{2}q_1^2 e^{-iq_1\tilde k}I_2^0(q_2-q_1;\tilde k),
\nonumber\\
&+&\frac{1}{2}{q_1}_\mu \Big[I _2^\mu(q_2;\tilde k)-e^{-iq_1\tilde k} I_2^\mu(q_2-q_1;\tilde k)
-q_1^2 I_3^\mu(q_1,q_2;\tilde k)\Big]
\label{C.24}
\end{eqnarray}
\begin{eqnarray}
&&\int\frac{d^3\ell}{(2\pi)^3}\,\frac{(\ell q_1)(\ell q_2)e^{i\ell\tilde k}}{\ell^2(\ell+q_1)^2(\ell+q_2)^2}
=\frac{1}{2}q_1^2e^{-iq_1\tilde k}I_2^0(q_2-q_1;\tilde k)
\nonumber\\
&+&\frac{1}{2}{q_1}_\mu\Big[ I_2^\mu(q_1;\tilde k)-e^{-iq_1\tilde k} I_2^\mu(q_2-q_1;\tilde k)
-q_2^2I_3^\mu(q_1,q_2;\tilde k)\Big],
\label{C.25}
\end{eqnarray}
\begin{eqnarray}
&&\int\frac{d^3\ell}{(2\pi)^3}\,\frac{(\ell q_2)^2e^{i\ell\tilde k}}{\ell^2(\ell+q_2)^2(\ell+q_1)^2}\;\stackrel{q_2\leftrightarrow q_1}{=}\;\;\int\frac{d^3\ell}{(2\pi)^3}\,\frac{(\ell q_1)^2e^{i\ell\tilde k}}{\ell^2(\ell+q_1)^2(\ell+q_2)^2},
\label{C.26}
\end{eqnarray}

\begin{eqnarray}
\int\frac{(\ell q_1)(\ell a)e^{i\ell\tilde k}}{\ell^2(\ell+q_1)^2(\ell+q_2)^2}
&=&\frac{1}{2}\int\Big[\frac{(\ell a)e^{i\ell\tilde k}}{\ell^2(\ell+q_2)^2}-
\frac{(\ell a)e^{i\ell\tilde k}}{(\ell+q_1)^2(\ell+q_2)^2}
-q_1^2\frac{(\ell a)e^{i\ell\tilde k}}{\ell^2(\ell+q_1)^2(\ell+q_2)^2}\Big],
\nonumber\\
\int\frac{(\ell q_2)(\ell a)e^{i\ell\tilde k}}{\ell^2(\ell+q_1)^2(\ell+q_2)^2}
&=&\frac{1}{2}\int\Big[\frac{(\ell a)e^{i\ell\tilde k}}{\ell^2(\ell+q_1)^2}-
\frac{(\ell a)e^{i\ell\tilde k}}{(\ell+q_1)^2(\ell+q_2)^2}
-q_2^2\frac{(\ell a)e^{i\ell\tilde k}}{\ell^2(\ell+q_1)^2(\ell+q_2)^2}\Big],
\nonumber \\
\label{C.27}
\end{eqnarray}
with shorthand notation $\int\frac{d^3\ell}{(2\pi)^3}\to\int$. In the above (\ref{C.24})-(\ref{C.26}) necessary bubble $I^\mu_2$ and triangle $I^\mu_3$ types of integrals with $\ell^\mu$ in numerator can be evaluated by using results for vector bubble (being not well-defined but certainly not divergent (\ref{B.41})), and by using vector triangle integrals (\ref{C.12}),(\ref{C.20}), which all have well-defined commutative limit, respectively.

Above scalar bubble integrals $I^0_2(p;\tilde k)$, are computed and presented in the appendix B, as (\ref{B.7})-(\ref{B.15}). Scalar triangle integrals $I_3^0(q_1,q_2;\tilde k)$ and $I_3^1(q_1,q_2;a(q_1,q_2),\tilde k)$ are given in (\ref{B.16})-(\ref{B.32}) and (\ref{C.16})-(\ref{C.20}), respectively. Other integrals in (\ref{C.27}) one can evaluate directly with help of vector bubble (\ref{B.40}), and vector triangle (\ref{C.12}).

\section{Reduction of box spurious integrals}

Using again the van Neevern-Vermaseren method and notation, where loop moment $\ell^\mu$ being decomposed in new basis $\{v_1^\mu,v_2^\mu,v_3^\mu\}$, as a functions of three momenta
$(q_1,q_2,q_3)$:
\begin{eqnarray}
\ell^{\mu}&=&\sum\limits_{i=1}^3(\ell q_i)v_i^{\mu}(q_1,q_2,q_3),
\label{D.1}\\
v_1^\mu(q_1,q_2,q_3)&=&\frac{\delta^{\mu q_2,q_3}_{q_1q_2q_3}}{\Delta(q_1,q_2,q_3)},\;
v_2^\mu(q_1,q_2,q_3)=\frac{\delta^{ q_1\mu q_3}_{q_1q_2q_3}}{\Delta(q_1,q_2,q_3)},\;
v_3^\mu(q_1,q_2,q_3)=\frac{\delta^{q_1q_2\mu }_{q_1q_2q_3}}{\Delta(q_1,q_2,q_3)},
\nonumber\\
\Delta(q_1,q_2,q_3)&=&\delta_{q_1,q_2,q_3}^{q_1,q_2,q_3}=\det(q_i q_j)
=-\epsilon^{i_1i_2i_3}\epsilon_{j_1j_2j_3}q_{1i_1}q_{2i_2}q_{3i_3}q^{j_1}_1q^{j_2}_2q^{j_3}_3,\;v_i\cdot q_j=\delta_{ij},
\nonumber\\
\label{D.2}
\end{eqnarray}
we next reduce and compute the box spurious integrals.

\subsection{Scalar box master integral}

To compute full integral $I_4^0(q_1,q_2,q_3;\tilde k)$ from (\ref{B.4}), on top of the basis
$\{v_1^\mu,v_2^\mu,v_3^\mu\}$ (\ref{D.2}) being  functions of three momenta $(q_1,q_2,q_3)$ necessary to reduce scalar integral with box type of denominators into linear combinations of integrals with triangle and bubble type of denominators, we also need previous basis $\{w_1^\mu,w_2^\mu,a^\mu\}$ (\ref{C.11}), again as a functions of the same three momenta $(q_1,q_2,q_3)$  permuted by three pairs $(1,2);(1,3);(2,3)$, respectively. After some lengthly algebra we obtain:
\begin{eqnarray}
I_4^0(q_1,q_2,q_3;\tilde k)&=&\int\frac{d^3\ell}{(2\pi)^3}\,\frac{e^{i\ell\tilde k}}{\ell^2(\ell+q_1)^2(\ell+q_2)^2(\ell+q_3)^2}
\nonumber\\
&=&\frac{1}{\sum^3_{i,j=1}q_i^2q_j^2v_iv_j}\sum_{n=1}^6 {I_{4_n}^0}(q_1,q_2,q_3;\tilde k),
\label{D.3}
\end{eqnarray}
where
\begin{eqnarray}
I^0_{4_1}&=&-v^2_1\Big[e^{-iq_2\tilde k}I^0_2(q_3-q_2;\tilde k)
\nonumber\\
&+&q_1\cdot w_1(q_2,q_3)\Big(I^0_2(q_3;\tilde k)-e^{-iq_2\tilde k}I_2^0(q_3-q_2;\tilde k)-q_2^2I_3^0(q_2,q_3;\tilde k)\Big)
\nonumber\\
&+&q_1\cdot w_2(q_2,q_3)\Big(I^0_2(q_2;\tilde k)-e^{-iq_2\tilde k}I_2^0(q_3-q_2;\tilde k)-q_3^2I_3^0(q_2,q_3;\tilde k)\Big)
\nonumber\\
&-&2\frac{q_1\cdot a(q_2,q_3)}{\Delta(q_2,q_3)}I_3^0(q_2,q_3;a(q_2,q_3),\tilde k)+q_1^2I_3^0(q_2,q_3;\tilde k)\Big],
\nonumber\\
I^0_{4_2}&=&-v^2_2\Big[e^{-iq_1\tilde k}I^0_2(q_3-q_1;\tilde k)
\nonumber\\
&+&q_2\cdot w_1(q_1,q_3)\Big(I^0_2(q_3;\tilde k)-e^{-q_1\tilde k}I^0_2(q_3-q_1;\tilde k)-q_1^2I_3^0(q_1,q_3;\tilde k)\Big)
\nonumber\\
&+&q_2\cdot w_2(q_1,q_3)\Big(I^0_2(q_1;\tilde k)-e^{-iq_1\tilde k}I_2^0(q_3-q_1;\tilde k)-q_3^2I_3^0(q_1,q_3;\tilde k)\Big)
\nonumber\\
&-&2\frac{q_2\cdot a(q_1,q_3)}{\Delta(q_1,q_3)}I_3^0(q_1,q_3;a(q_1,q_3),\tilde k)+q_2^2I_3^0(q_1,q_3;\tilde k)\Big],
\nonumber
\end{eqnarray}
\begin{eqnarray}
I^0_{4_3}&=&-v^2_3\Big[e^{-iq_1\tilde k}I^0_2(q_2-q_1;\tilde k)
\nonumber\\
&+&q_3\cdot w_1(q_1,q_2)\Big(I^0_2(q_2;\tilde k)-e^{-iq_1\tilde k}I^0_2(q_2-q_1;\tilde k)-q_1^2I_3^0(q_1,q_2;\tilde k)\Big)
\nonumber\\
&+&q_3\cdot w_2(q_1,q_2)\Big(I^0_2(q_1;\tilde k)-e^{-iq_1\tilde k}I_2^0(q_2-q_1;\tilde k)-q_2^2I_3^0(q_1,q_2;\tilde k)\Big)
\nonumber\\
&-&2\frac{q_3\cdot a(q_1,q_2)}{\Delta(q_1,q_2)}I_3^0(q_1,q_2;a(q_1,q_2),\tilde k)+q_3^2I_3^0(q_1,q_2;\tilde k)\Big]
\nonumber\\
I^0_{4_4}&=&2\sum_{j=1}^3\Big[v_1\Big(v_je^{-iq_2\tilde k}I^0_2(q_3-q_2;\tilde k)+v_jq_j^2I^0_3(q_2,q_3;\tilde k)-v_2I^0_2(q_3;\tilde k)\Big)
\nonumber\\
&&\phantom{Xx}+v_2\Big(v_je^{-q_1\tilde k}I^0_2(q_3-q_1;\tilde k)+v_jq_j^2I^0_3(q_1,q_3;\tilde k)-v_3I^0_2(q_1;\tilde k)\Big)
\nonumber\\
&&\phantom{Xx}+v_3\Big(v_je^{-q_1\tilde k}I^0_2(q_2-q_1;\tilde k)+v_jq_j^2I^0_3(q_1,q_2;\tilde k)-v_1I^0_2(q_2;\tilde k)\Big)\Big]
\nonumber\\
I^0_{4_5}&=&-\Big(\sum_{i,j=1}^3v_iv_j\Big)e^{-iq_1\tilde k}\Big[e^{-i(q_2-q_1)\tilde k}I^0_2(q_3-q_2;\tilde k)
\nonumber\\
&-&q_1\cdot w_1(q_2-q_1,q_3-q_1)\Big(I^0_2(q_3-q_1;\tilde k)-e^{-i(q_2-q_1)\tilde k}I_2^0(q_3-q_2;\tilde k)
\nonumber\\
&-&(q_2-q_1)^2I_3^0(q_2-q_1,q_3-q_1;\tilde k)\Big)
\nonumber\\
&-&q_1\cdot w_2(q_2-q_1,q_3-q_1)\Big(I^0_2(q_2-q_1;\tilde k)-e^{-i(q_2-q_1)\tilde k}I_2^0(q_3-q_2;\tilde k)
\nonumber\\
&-&(q_3-q_1)^2I_3^0(q_2-q_1,q_3-q_1;\tilde k)\Big)
\nonumber\\
&+&2\frac{q_1\cdot a(q_2-q_1,q_3-q_1)}{\Delta(q_2-q_1,q_3-q_1)}I_3^0(q_2-q_1,q_3-q_1;a(q_2-q_1,q_3-q_1),\tilde k)
\nonumber\\
&+&q_1^2I_3^0(q_2-q_1,q_3-q_1;\tilde k)\Big],
\nonumber\\
I^0_{4_6}&=&2\Big[2-\sum_{i,j=1}^3v_iv_jq_j^2 \Big]e^{-iq_1\tilde k}I_3^0(q_2-q_1,q_3-q_1;\tilde k),
\label{D.4}
\end{eqnarray}
To fully reduce/compute the above scalar box integral $I_4^0(q_1,q_2,q_3;\tilde k)$ we also need integrals
obtained with help of equation (\ref{C.12}), for reduction of vector triangle spurious integral.

\subsection{Vector box integral}
Using the van Neevern-Vermaseren decomposition (\ref{D.1}), from (\ref{B.4}) we obtain
\begin{eqnarray}
&&I_4^\mu(q_1,q_2,q_3;\tilde k)=\int\frac{d^3\ell}{(2\pi)^3}\,\frac{\ell^\mu \cdot e^{i\ell\tilde k}}{\ell^2(\ell+q_1)^2(\ell+q_2)^2(\ell+q_3)^2}
\label{D.5}\\
&=&\frac{1}{2}v_1^\mu(q_1,q_2,q_3) \Big[I_3^0(q_2,q_3;\tilde k)-e^{-iq_1\tilde k}I_3^0(q_2-q_1,q_3-q_1;\tilde k)-q_1^2I_4^0(q_1,q_2,q_3;\tilde k) \Big]
\nonumber\\
&+&\frac{1}{2}v_2^\mu(q_1,q_2,q_3)\Big[I^0_3(q_1,q_3;\tilde k)-e^{-iq_1\tilde k}I_3^0(q_2-q_1,q_3-q_1;\tilde k)-q_2^2I_4^0(q_1,q_2,q_3;\tilde k)\Big]
\nonumber\\
&+&\frac{1}{2}v_3^\mu(q_1,q_2,q_3)\Big[I^0_3(q_1,q_2;\tilde k)-e^{-iq_1\tilde k}I_3^0(q_2-q_1,q_3-q_1;\tilde k)-q_3^2I_4^0(q_1,q_2,q_3;\tilde k)\Big],
\nonumber
\end{eqnarray}
Above scalar triangle integrals $I_3^0$ are given in appendix B.2.3, while scalar box integral $I^0_4(q_1,q_2,q_3;\tilde k)$ is given in (\ref{D.3})-(\ref{D.4}).

\subsection{Tensor box integral}

Using again the van Neevern-Vermaseren method and (\ref{D.1}), the $I_4^{\mu\nu}(q_1,q_2,q_3;\tilde k)$ we write as
\begin{eqnarray}
I_4^{\mu\nu}(q_1,q_2,q_3;\tilde k)&=&\int\frac{d^3\ell}{(2\pi)^3}\,\frac{\ell^\mu\ell^\nu \cdot e^{i\ell\tilde k}}{\ell^2(\ell+q_1)^2(\ell+q_2)^2(\ell+q_3)^2}
\label{D.6}\\
&=&\sum_{i,j=1}^3v_i^\mu v_j^\nu\int\frac{d^3\ell}{(2\pi)^3}\,\frac{(\ell q_i)(\ell q_j) \cdot e^{i\ell\tilde k}}{\ell^2(\ell+q_1)^2(\ell+q_2)^2(\ell+q_3)^2}.
\nonumber
\end{eqnarray}
Using decomposition $\ell q_i=\frac{1}{2}[(\ell+q_i)^2-\ell^2-q_i^2]$ and computing, we obtain the following six subintegrals  needed to evaluate complete tensor box integral (\ref{D.6}):
\begin{eqnarray}
&&\int\frac{d^3\ell}{(2\pi)^3}\,\frac{(\ell q_1)^2e^{i\ell\tilde k}}{\ell^2(\ell+q_1)^2(\ell+q_2)^2(\ell+q_3)^2}
\label{D.7}\\
&=&\frac{1}{2}\bigg\{\frac{1}{2}q_1\cdot w_1(q_2,q_3)
\Big(I^0_2(q_3;\tilde k)-e^{-iq_2\tilde k}I_2^0(q_3-q_2;\tilde k)-q_2^2I_3^0(q_2,q_3;\tilde k)\Big)
\nonumber\\
&+&\frac{1}{2}q_1\cdot w_2(q_2,q_3)
\Big(I^0_2(q_2;\tilde k)-e^{-iq_2\tilde k}I_2^0(q_3-q_2;\tilde k)-q_3^2I_3^0(q_2,q_3;\tilde k)\Big)
\nonumber\\
&-&\frac{q_1\cdot a(q_2,q_3)}{\Delta(q_2,q_3)}I_3^0(q_2,q_3;a(q_2,q_3),\tilde k)
\nonumber\\
&-&e^{-iq_1\tilde k}\bigg[\frac{1}{2}q_1\cdot w_1(q_2-q_1,q_3-q_1)
\nonumber\\
&\cdot&
\Big(I^0_2(q_3-q_1;\tilde k)-e^{-i(q_2-q_1)\tilde k}I_2^0(q_3-q_2;\tilde k)-(q_2-q_1)^2I_3^0(q_2-q_1,q_3-q_1;\tilde k)\Big)
\nonumber\\
&+&\frac{1}{2}q_1\cdot w_2(q_2-q_1,q_3-q_1)
\nonumber\\
&\cdot&\Big(I^0_2(q_2-q_1;\tilde k)-e^{-i(q_2-q_1)\tilde k}I_2^0(q_3-q_2;\tilde k)-(q_3-q_1)^2I_3^0(q_2-q_1,q_3-q_1;\tilde k)\Big)
\nonumber\\
&-&\frac{q_1\cdot a(q_2-q_1,q_3-q_1)}{\Delta(q_2-q_1,q_3-q_1)}I_3^0(q_2-q_1,q_3-q_1;a(q_2-q_1,q_3-q_1),\tilde k\bigg]
\nonumber\\
&-&
\frac{q_1^2}{2}\Big(I_3^0(q_2,q_3;\tilde k)-3e^{-iq_1\tilde k}I_3^0(q_2-q_1,q_3-q_1;\tilde k)
-q_1^2I_4^0(q_1,q_2,q_3;\tilde k)\Big)\bigg\},
\nonumber
\end{eqnarray}

\begin{eqnarray}
&&\int\frac{d^3\ell}{(2\pi)^3}\,\frac{(\ell q_2)^2e^{i\ell\tilde k}}{\ell^2(\ell+q_1)^2(\ell+q_2)^2(\ell+q_3)^2}
\label{D.8}\\
&=&\frac{1}{2}\bigg\{\frac{1}{2}q_2\cdot w_1(q_1,q_3)
\Big(I^0_2(q_3;\tilde k)-e^{-iq_1\tilde k}I_2^0(q_3-q_1;\tilde k)-q_1^2I_3^0(q_1,q_3;\tilde k)\Big)
\nonumber\\
&+&\frac{1}{2}q_2\cdot w_2(q_1,q_3)
\Big(I^0_2(q_1;\tilde k)-e^{-iq_1\tilde k}I_2^0(q_3-q_1;\tilde k)-q_3^2I_3^0(q_2,q_3;\tilde k)\Big)
\nonumber\\
&-&\frac{q_2\cdot a(q_1,q_3)}{\Delta(q_1,q_3)}I_3^0(q_1,q_3;a(q_1,q_3),\tilde k)
\nonumber\\
&-&e^{-iq_1\tilde k}\bigg[\frac{1}{2}q_2\cdot w_1(q_2-q_1,q_3-q_1)
\nonumber\\
&\cdot&
\Big(I^0_2(q_3-q_1;\tilde k)-e^{-i(q_2-q_1)\tilde k}I_2^0(q_3-q_2;\tilde k)-(q_2-q_1)^2I_3^0(q_2-q_1,q_3-q_1;\tilde k)\Big)
\nonumber\\
&+&\frac{1}{2}q_2\cdot w_2(q_2-q_1,q_3-q_1)
\nonumber\\
&\cdot&\Big(I^0_2(q_2-q_1;\tilde k)-e^{-i(q_2-q_1)\tilde k}I_2^0(q_3-q_2;\tilde k)-(q_3-q_1)^2I_3^0(q_2-q_1,q_3-q_1;\tilde k)\Big)
\nonumber\\
&-&\frac{q_2\cdot a(q_2-q_1,q_3-q_1)}{\Delta(q_2-q_1,q_3-q_1)}I_3^0(q_2-q_1,q_3-q_1;a(q_2-q_1,q_3-q_1),\tilde k)
\nonumber\\
&-&(q_1q_2)I_3^0(q_2-q_1,q_3-q_1;\tilde k)\bigg]
\nonumber\\
&-&\frac{q_2^2}{2}\Big(I_3^0(q_1,q_3;\tilde k)-e^{-iq_1\tilde k}I_3^0(q_2-q_1,q_3-q_1;\tilde k)
-q_2^2I_4^0(q_1,q_2,q_3;\tilde k)\Big)\bigg\},
\nonumber
\end{eqnarray}

\begin{eqnarray}
&&\int\frac{d^3\ell}{(2\pi)^3}\,\frac{(\ell q_3)^2e^{i\ell\tilde k}}{\ell^2(\ell+q_1)^2(\ell+q_2)^2(\ell+q_3)^2}
\label{D.9}\\
&=&\frac{1}{2}\bigg\{\frac{1}{2}q_3\cdot w_1(q_1,q_2)
\Big(I^0_2(q_2;\tilde k)-e^{-iq_1\tilde k}I_2^0(q_2-q_1;\tilde k)-q_1^2I_3^0(q_1,q_2;\tilde k)\Big)
\nonumber\\
&+&\frac{1}{2}q_3\cdot w_2(q_1,q_2)
\Big(I^0_2(q_1;\tilde k)-e^{-iq_1\tilde k}I_2^0(q_2-q_1;\tilde k)-q_2^2I_3^0(q_1,q_2;\tilde k)\Big)
\nonumber\\
&-&\frac{q_3\cdot a(q_1,q_2)}{\Delta(q_1,q_2)}I_3^0(q_1,q_2;a(q_1,q_2),\tilde k)
\nonumber\\
&-&e^{-iq_1\tilde k}\bigg[\frac{1}{2}q_3\cdot w_1(q_2-q_1,q_3-q_1)
\nonumber\\
&\cdot&
\Big(I^0_2(q_3-q_1;\tilde k)-e^{-i(q_2-q_1)\tilde k}I_2^0(q_3-q_2;\tilde k)-(q_2-q_1)^2I_3^0(q_2-q_1,q_3-q_1;\tilde k)\Big)
\nonumber\\
&+&\frac{1}{2}q_3\cdot w_2(q_2-q_1,q_3-q_1)
\nonumber\\
&\cdot&\Big(I^0_2(q_2-q_1;\tilde k)-e^{-i(q_2-q_1)\tilde k}I_2^0(q_3-q_2;\tilde k)-(q_3-q_1)^2I_3^0(q_2-q_1,q_3-q_1;\tilde k)\Big)
\nonumber\\
&-&\frac{q_3\cdot a(q_2-q_1,q_3-q_1)}{\Delta(q_2-q_1,q_3-q_1)}I_3^0(q_2-q_1,q_3-q_1;a(q_2-q_1,q_3-q_1),\tilde k)
\nonumber\\
&-&(q_1q_3)I_3^0(q_2-q_1,q_3-q_1;\tilde k)\bigg]
\nonumber\\
&-&
\frac{q_3^2}{2}\Big(I_3^0(q_1,q_2;\tilde k)-e^{-iq_1\tilde k}I_3^0(q_2-q_1,q_3-q_1;\tilde k)
-q_3^2I_4^0(q_1,q_2,q_3;\tilde k)\Big)\bigg\},
\nonumber
\end{eqnarray}

\begin{eqnarray}
&&\int\frac{d^3\ell}{(2\pi)^3}\,\frac{(\ell q_1)(\ell q_2)e^{i\ell\tilde k}}{\ell^2(\ell+q_1)^2(\ell+q_2)^2(\ell+q_3)^2}
\label{D.10}\\
&=&\frac{1}{4}\bigg\{I^0_2(q_3;\tilde k)-e^{-iq_2\tilde k}I_2^0(q_3-q_1;\tilde k)-q_1^2I_3^0(q_1,q_3;\tilde k)
\nonumber\\
&-&e^{-iq_1\tilde k}\bigg[q_1\cdot w_1(q_2-q_1,q_3-q_1)
\nonumber\\
&\cdot&
\Big(I^0_2(q_3-q_1;\tilde k)-e^{-i(q_2-q_1)\tilde k}I_2^0(q_3-q_2;\tilde k)-(q_2-q_1)^2I_3^0(q_2-q_1,q_3-q_1;\tilde k)\Big)
\nonumber\\
&+&q_1\cdot w_2(q_2-q_1,q_3-q_1)
\nonumber\\
&\cdot&\Big(I^0_2(q_2-q_1;\tilde k)-e^{-i(q_2-q_1)\tilde k}I_2^0(q_3-q_2;\tilde k)-(q_3-q_1)^2I_3^0(q_2-q_1,q_3-q_1;\tilde k)\Big)
\nonumber\\
&-&\frac{2q_1\cdot a(q_2-q_1,q_3-q_1)}{\Delta(q_2-q_1,q_3-q_1)}I_3^0(q_2-q_1,q_3-q_1;a(q_2-q_1,q_3-q_1),\tilde k)\bigg]
\nonumber\\
&+&
\Big(2q_1^2+q_2^2\Big)e^{-iq_1\tilde k}I_3^0(q_2-q_1,q_3-q_1;\tilde k)-q_2^2\Big(I_3^0(q_2,q_3;\tilde k)
-q_1^2I_4^0(q_1,q_2,q_3;\tilde k)\Big)\bigg\},
\nonumber
\end{eqnarray}

\begin{eqnarray}
&&\int\frac{d^3\ell}{(2\pi)^3}\,\frac{(\ell q_1)(\ell q_3)e^{i\ell\tilde k}}{\ell^2(\ell+q_1)^2(\ell+q_2)^2(\ell+q_3)^2}
\label{D.11}\\
&=&\frac{1}{4}\bigg\{
I^0_2(q_2;\tilde k)-e^{-iq_1\tilde k}I_2^0(q_2-q_1;\tilde k)-q_1^2I_3^0(q_1,q_2;\tilde k)
\nonumber\\
&-&e^{-iq_1\tilde k}\bigg[q_1\cdot w_1(q_2-q_1,q_3-q_1)
\nonumber\\
&\cdot&
\Big(I^0_2(q_3-q_1;\tilde k)-e^{-i(q_2-q_1)\tilde k}I_2^0(q_3-q_2;\tilde k)-(q_2-q_1)^2I_3^0(q_2-q_1,q_3-q_1;\tilde k)\Big)
\nonumber\\
&+&q_1\cdot w_2(q_2-q_1,q_3-q_1)
\nonumber\\
&\cdot&\Big(I^0_2(q_2-q_1;\tilde k)-e^{-i(q_2-q_1)\tilde k}I_2^0(q_3-q_2;\tilde k)-(q_3-q_1)^2I_3^0(q_2-q_1,q_3-q_1;\tilde k)\Big)
\nonumber\\
&-&\frac{2q_1\cdot a(q_2-q_1,q_3-q_1)}{\Delta(q_2-q_1,q_3-q_1)}I_3^0(q_2-q_1,q_3-q_1;a(q_2-q_1,q_3-q_1),\tilde k)\bigg]
\nonumber\\
&+&
\Big(2q_1^2+q_3^2\Big)e^{-iq_1\tilde k}I_3^0(q_2-q_1,q_3-q_1;\tilde k)-q_3^2\Big(I_3^0(q_2,q_3;\tilde k)
-q_1^2I_4^0(q_1,q_2,q_3;\tilde k)\Big)\bigg\},
\nonumber
\end{eqnarray}

\begin{eqnarray}
&&\int\frac{d^3\ell}{(2\pi)^3}\,\frac{(\ell q_2)(\ell q_3)e^{i\ell\tilde k}}{\ell^2(\ell+q_1)^2(\ell+q_2)^2(\ell+q_3)^2}
\label{D.12}\\
&=&\frac{1}{4}\bigg\{
I^0_2(q_1;\tilde k)-e^{-iq_1\tilde k}I_2^0(q_2-q_1;\tilde k)-q_2^2I_3^0(q_1,q_2;\tilde k)
\nonumber\\
&-&e^{-iq_1\tilde k}\bigg[q_2\cdot w_1(q_2-q_1,q_3-q_1)
\nonumber\\
&\cdot&
\Big(I^0_2(q_3-q_1;\tilde k)-e^{-i(q_2-q_1)\tilde k}I_2^0(q_3-q_2;\tilde k)-(q_2-q_1)^2I_3^0(q_2-q_1,q_3-q_1;\tilde k)\Big)
\nonumber\\
&+&q_2\cdot w_2(q_2-q_1,q_3-q_1)
\nonumber\\
&\cdot&\Big(I^0_2(q_2-q_1;\tilde k)-e^{-i(q_2-q_1)\tilde k}I_2^0(q_3-q_2;\tilde k)-(q_3-q_1)^2I_3^0(q_2-q_1,q_3-q_1;\tilde k)\Big)
\nonumber\\
&-&\frac{2q_2\cdot a(q_2-q_1,q_3-q_1)}{\Delta(q_2-q_1,q_3-q_1)}I_3^0(q_2-q_1,q_3-q_1;a(q_2-q_1,q_3-q_1),\tilde k)\bigg]
\nonumber\\
&+&
\Big(2q_1q_2+q_3^2\Big)e^{-iq_1\tilde k}I_3^0(q_2-q_1,q_3-q_1;\tilde k)-q_3^2\Big(I_3^0(q_1,q_3;\tilde k)
-q_2^2I_4^0(q_1,q_2,q_3;\tilde k)\Big)\bigg\}.
\nonumber
\end{eqnarray}
Again all above scalar bubble integrals $I^0_2(p;\tilde k)$, etc. are computed and presented in the appendix (\ref{B.7})-(\ref{B.12}). Scalar triangle integrals $I_3^0(q_1,q_2;\tilde k)$ and $I_3^0(q_1,q_2;a(q_1,q_2),\tilde k)$ are like before given in (\ref{B.16})-(\ref{B.32}) and (\ref{C.10})-(\ref{C.20}), respectively. Scalar box $I_4^0(q_1,q_2,q_3;\tilde k)$ as a master integral is given in (\ref{D.4}).\\

This way we have completed the full set of integrals from (\ref{B.1})-(\ref{B.4}) needed to perform this research on 4- \& 6-point functions properties of our quantum NCABJM theory.

\section{Simplified 2scalar-2scalar/2fermion 4-point functions}

In this section we discuss and present integrand numerators, somewhat simplified relative to bubble (\ref{Nb1})-(\ref{Nb9}), triangle (\ref{trifq}), and box (\ref{boxfq}) numerators, i.e with specifically rearranged momenta with respect to the main body of this manuscript.  Diagrams of generic samples, figures \ref{4scalar} and \ref{Sample2scalar2fermion}, have only loop and momenta in accord to the incoming-outgoing notations indicated, while Lorentz and other quantum number indices of fields were skipped.

\subsection{Numerators and integrals of 4-point function from bubble-loops}

Similar to the bubble integrals in subsection 4.2, eqs.(\ref{bub9})-(\ref{Sb9}), we have minimally three different bubble type of integrals which can be expressed as follows:
\begin{eqnarray}
 I^{bub}_1&=&-e^{\frac{i}{2}(p_1\theta p_2-p_3\theta p_4)}\int \frac{d^3 \ell}{(2\pi)^3}
\frac{2(\ell+p_2)\cdot(\ell-p_1)}{(\ell+p_2)^2(\ell-p_1)^2}=({\hat I}_1^{bub})^\ast,
\label{E.1}\\
 I^{bub}_2&=&-e^{\frac{i}{2}(p_1\theta p_2+p_3\theta p_4)}\int \frac{d^3 \ell}{(2\pi)^3}
\frac{e^{-i\ell\theta(p_1+p_2)}2(\ell+p_2)\cdot(\ell-p_1)}{(\ell+p_2)^2(\ell-p_1)^2}=({\hat I}_2^{bub})^\ast,
\label{E.2}\\
I^{bub}_3&=&4e^{-\frac{i}{2}(p_1\theta p_3+p_2\theta p_4)}\int \frac{d^3 \ell}{(2\pi)^3}
\frac{e^{i\ell\theta(p_1+p_3)}2(\ell+p_2)\cdot(\ell-p_1)
}{(\ell+p_2)^2(\ell-p_1)^2}=({\hat I}_3^{bub})^\ast,
\label{E.3}
\end{eqnarray}
where the case of one wavy and one double-wavy lines diagrams give integrals denoted  as $\hat I^{bub}_i$, and they are non-planar. The above common integrand could be decomposed as follows:
 \begin{eqnarray}
\frac{2(\ell+p_2)(\ell-p_1)}{(\ell+p_2)^2(\ell-p_1)^2}&=&\frac{1}{(\ell+p_2)^2}+\frac{1}{(\ell-p_1)^2}
-\frac{(p_1+p_2)^2}{(\ell+p_2)^2(\ell-p_1)^2}.
\label{E.4}
\end{eqnarray}

There are two more NC factor related properties of other integrals, i.e. to $I^{bub}_2$ (\ref{E.2}) we apply variable change, $\ell\to\ell+p_1$, yielding:
\begin{eqnarray}
 I^{bub}_2&=&-e^{-\frac{i}{2}(p_1\theta p_2-p_3\theta p_4)}\int \frac{d^3 \ell}{(2\pi)^3}
\frac{e^{-i\ell\theta(p_1+p_2)}\;2\ell\cdot(\ell+p_1+p_2)}{{\ell}^2(\ell+p_1+p_2)^2}
\label{E.5}\\
 &=&-e^{-\frac{i}{2}(p_1\theta p_2-p_3\theta p_4)}\int \frac{d^3 \ell}{(2\pi)^3}
e^{-i\ell\theta(p_1+p_2)}\bigg[\frac{1}{(\ell+p_1+p_2)^2}+\frac{1}{\ell^2}-\frac{(p_1+p_2)^2}{\ell^2(\ell+p_1+p_2)^2}\bigg].
\nonumber
\end{eqnarray}

To find properties of integrals $I^{bub}_{1,2,3}$ at the integrand level, we apply the following change of variable
$\ell\to-\ell-p_2+p_1$ in the above integrals. For $I^{bub}_3$ (\ref{E.3}) we have:
\begin{eqnarray}
\nonumber\\
I^{bub}_3&=&4e^{\frac{i}{2}(p_1\theta p_3+p_2\theta p_4)}\int \frac{d^3 \ell}{(2\pi)^3}
\frac{e^{-i\ell\theta(p_1+p_3)}2(\ell-p_1)\cdot(\ell+p_2)}{(\ell-p_1)^2(\ell+p_2)^2}.
\label{E.6}
\end{eqnarray}

Now using decomposition into three terms (\ref{E.4}) of common integrand, from the above eqs. (\ref{E.1})-(\ref{E.3})  we obtain the following results:
\begin{eqnarray}
 I^{bub}_1&=&e^{\frac{i}{2}(p_1\theta p_2-p_3\theta p_4)}(p_1+p_2)^2\int \frac{d^3 \ell}{(2\pi)^3}
\frac{1}{(\ell+p_2)^2(\ell-p_1)^2},
\label{E.7}\\
 I^{bub}_2&=&e^{-\frac{i}{2}(p_1\theta p_2-p_3\theta p_4)}\Big[-2I_1^0(\tilde p_1+\tilde p_2)
+(p_1+p_2)^2 I_2^0(p_1+p_2;\tilde p_1+\tilde p_2)\Big],
\label{E.8}\\
I^{bub}_3&=&4\Big[2\cos{\frac{(p_1\theta p_3-p_2\theta p_4)}{2}}I_1^0(\tilde p_1+\tilde p_3)
-(p_1+p_2)^2e^{\frac{i}{2}(p_1\theta p_3-p_2\theta p_4)}
I_2^0(p_1+p_2;\tilde p_1+\tilde p_3)\Big],
\nonumber\\
\label{E.9}
\end{eqnarray}
where integrals $I^0_1$ and $I^0_2$,  are given in (\ref{B.5}) and (\ref{B.7}), respectively. Now we list properties of bubble diagrams (figure \ref{fig:Fig41}) represented by the above typical integrals $I^{bub}_{1,2,3}$:
\begin{itemize}
\item $\bullet$ trivial existence of commutative limits for integral $I^{bub}_1$ (\ref{E.7}),
\item $\bullet$ integral $I^0_1$ has no commutative limit, showing linear UV/IR mixing (\ref{B.6}),
\item $\bullet$ integral $I^0_2$ has finite commutative limit (\ref{B.9}), and
\item $\bullet$ from above and subsection 4.3 we have $I^{bub}_2=({\hat I}^{bub}_2)^\ast = S^{bub}_1$,
${\hat I}^{bub}_2=(I^{bub}_2)^\ast = S^{bub}_3$, and {\phantom{X}}${\hat I}^{bub}_3=(I^{bub}_3)^\ast = S^{bub}_2$.
\end{itemize}

\subsection{Numerator of the 4 scalar 4-point function with triangle-loop}

\begin{figure}
\begin{center}
\includegraphics[width=15cm,angle=0]{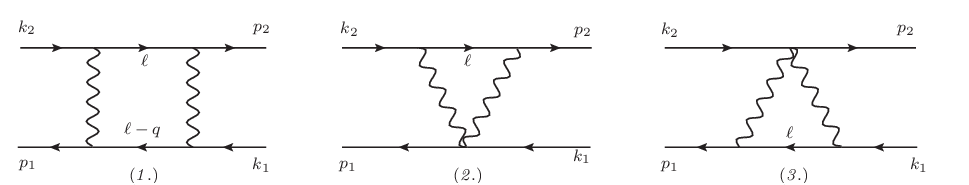}
\end{center}
\caption{Generic samples, with specified fields and moments only, of 4-scalar diagrams from figure \ref{fig:Fig21} and figure \ref{fig:Fig31}.  Note that the incoming-outgoing momenta $\{k_1,k_2;p_1,p_2\}$ in this figure correspond to the following set of all incoming momenta $\{p_4,p_1;-p_3,-p_2\}$, in figures \ref{fig:Fig21}-\ref{fig:Fig41}, respectively. Diagram $\it(2.)\to \it(3.)$, after exchanging  momenta $2\leftrightarrow1$ in diagram $\it(2.)$.  }
\label{4scalar}
\end{figure}

Computing numerator $f_{(\it 2.)}^{triangle}$ from diagram $(\it 2.)$ in figure \ref{4scalar} of generic triangle-loop pointing down we obtain:
\begin{equation}
f_{(\it 2.)}^{triangle}=4\big(\ell^2(k_2 p_2)-(\ell k_2)(\ell p_2)\big).
\label{E.10}
\end{equation}
After changing in-out momenta in accord with all incoming momenta from figures \ref{fig:Fig21} and \ref{fig:Fig22} $p_2\to -p_2,k_2\to p_1$, we have
\begin{equation}
f_{(\it 2.)}^{triangle}=-4\big(\ell^2(p_1 p_2)-(\ell p_1)(\ell p_2)\big).
\label{E.11}
\end{equation}

Decomposition of the common part of simplified relevant integrals with triangle denominator
and above numerator (\ref{E.11}) gives:
\begin{eqnarray}
&&\frac{4\big(\ell^2(p_1p_2)-(\ell p_1)(\ell p_2)\big)}{\ell^2(\ell+p_2)^2(\ell-p_1)^2}=-\frac{p_2^2p_1^2}{\ell^2(\ell+p_2)^2(\ell-p_1)^2}+\frac{p_1^2 }{\ell^2(\ell-p_1)^2}+\frac{p_2^2 }{\ell^2(\ell+p_2)^2}-\frac{1}{\ell^2}
\nonumber\\
&&+\frac{2p_2^2p_1^2-p^2 _2}{(\ell+p_2)^2(\ell-p_1)^2}+\frac{1}{(\ell-p_1)^2}
-\frac{2(\ell+p_2)p_1}{(\ell+p_2)^2(\ell-p_1)^2}
\nonumber\\
&&=-\frac{p_2^2p_1^2}{\ell^2(\ell+p_2)^2(\ell-p_1)^2}+\frac{p_1^2 }{\ell^2(\ell-p_1)^2}+\frac{p_2^2 }{\ell^2(\ell+p_2)^2}
+\frac{-3p_1p_2-p_1^2-p_2^2}{(\ell+p_2)^2(\ell-p_1)^2}
\nonumber\\
&&+\frac{p_2^2+p_1p_2}{(p_1+p_2)^2}\Big(\frac{1}{(\ell-p_1)^2}-\frac{1}{\ell^2}\Big)
+\frac{p_1^2+p_1p_2}{(p_1+p_2)^2}\Big(\frac{1}{(\ell+p_2)^2}-\frac{1}{\ell^2}\Big)
\nonumber\\
&&-\frac{1}{(p_1+p_2)^2}\frac{2(\ell+p_2)\omega}{(\ell+p_2)^2(\ell-p_1)^2},
\label{E.12}
\end{eqnarray}
where auxiliary vector $\omega$ is defined after we introduced the following momentum decomposition:
$\omega^\mu=p_1^\mu(p_2^2+p_1p_2)-p_2^\mu(p_1^2+p_1p_2)$, with condition
$p_1^\mu=c_1(p_1+p_2)^\mu+c_2\omega^\mu$, where $c_1=\frac{p_1(p_1+p_2)}{(p_1+p_2)^2}$, and $c_2=\frac{1}{(p_1+p_2)^2}$, respectively.

\subsection{Vanishing of 4 scalar 4-point function with triangle-loop in $\theta^{\mu\nu}\to0$ limit}

To prove (\ref{limStria}) by direct computations we use somewhat simplified expression from figure \ref{4scalar} $(\it 2.)$ of generic triangle-loop diagram, where only loop and the incoming-outgoing momenta are indicated, which represents any diagram in figure \ref{fig:Fig31}. Thus, the nominator (\ref{E.11}) with relevant NC-phases generate maximally three possible  integrals $I^{tri}_{1,2,3}$:
\begin{eqnarray}
 I^{tri}_1&=&-e^{-\frac{i}{2}(p_1\theta p_2-p_3\theta p_4)}\int \frac{d^3 \ell}{(2\pi)^3}
\frac{4\big(\ell^2(p_1p_2)-(\ell p_1)(\ell p_2)\big)
}{\ell^2(\ell+p_2)^2(\ell-p_1)^2}=({\hat I}_1^{tri})^\ast, 
\label{E.13}\\
 I^{tri}_2&=&-e^{\frac{i}{2}(p_1\theta p_2+p_3\theta p_4)}\int \frac{d^3 \ell}{(2\pi)^3}
\frac{e^{-i\ell\theta(p_1+p_2)}4\big(\ell^2(p_1p_2)-(\ell p_1)(\ell p_2)\big)}
{\ell^2(\ell+p_2)^2(\ell-p_1)^2}=({\hat I}_2^{tri})^\ast.
\label{E.14}
\end{eqnarray}
Identical hgauge boson triangles can be generated using the gauge boson line flipping relation.
Next we consider mixed two upper triangle diagrams, which are connected by flipping gauge boson lines, giving
\begin{eqnarray}
 I^{tri}_3&=&2e^{-\frac{i}{2}(p_1\theta p_3+p_2\theta p_4)}\int \frac{d^3 \ell}{(2\pi)^3}
\frac{e^{i\ell\theta(p_1+p_3)}4\big(\ell^2(p_1p_2)-(\ell p_1)(\ell p_2)\big)}
{\ell^2(\ell+p_2)^2(\ell-p_1)^2}=({\hat I}_3^{tri})^\ast,
\label{E.15}
\end{eqnarray}
where again one wavy and one double-wavy lines diagrams denoted as $\hat I^{tri}_i$ are non-planar integrals.
Note that the $\ell$-dependent NC phase factors in $I^{tri}_{1,2,3}$ are equal to the $\ell$-dependent NC factors in $I^{bub}_{1,2,3}$, (\ref{E.1})-(\ref{E.3}), respectively.

Expressions  $S^{tri}_r$ (\ref{tria1}) from figure \ref{fig:Fig31} can be decomposed in terms of the above three integrals (\ref{E.13})-(\ref{E.15}) as follows:  $S^{tri}_1\sim I^{tri}_1+I^{tri}_2$, $S^{tri}_2\sim {\hat I}^{tri}_3$,
$S^{tri}_3\sim I^{tri}_3$, and $S^{tri}_4\sim {\hat I}^{tri}_1+{\hat I}^{tri}_2$. Here we notice that $I^{tri}_1$ and $I^{tri}_2$ do not show the same external NC factor, while the planar and non-planar integrals in $S^{tri}_1$ do. This is due to the changing variable $\ell\to \ell+p_1$, which provides an additional phase factor $(-p_1\theta p_2)$ for $I^{tri}_2$ while no change for $I^{tri}_1$. 

Decomposition of the common part of integrand from integrals in (\ref{E.13})-(\ref{E.15}) is given in the above as
(\ref{E.12}), leading to the following three integrals:
\begin{eqnarray}
 I^{tri}_1&=&-e^{-\frac{i}{2}(p_1\theta p_2-p_3\theta p_4)}\int \frac{d^3 \ell}{(2\pi)^3}
\bigg(-\frac{p_2^2p_1^2}{\ell^2(\ell+p_2)^2(\ell-p_1)^2}+\frac{p_1^2 }{\ell^2(\ell-p_1)^2}+\frac{p_2^2 }{\ell^2(\ell+p_2)^2}
\nonumber\\
&+&\frac{-3p_1p_2-p_1^2-p_2^2}{(\ell+p_2)^2(\ell-p_1)^2}\bigg),
\label{E.16}\\
 I^{tri}_2&=&-e^{\frac{i}{2}(p_1\theta p_2+p_3\theta p_4)}\int \frac{d^3 \ell}{(2\pi)^3}e^{-i\ell\theta(p_1+p_2)}
\bigg(-\frac{p_2^2p_1^2}{\ell^2(\ell+p_2)^2(\ell-p_1)^2}+\frac{p_1^2 }{\ell^2(\ell-p_1)^2}
\nonumber\\
&+&\frac{p_2^2 }{\ell^2(\ell+p_2)^2}
+\frac{-3p_1p_2-p_1^2-p_2^2}{(\ell+p_2)^2(\ell-p_1)^2}
+\frac{p_2^2+p_1p_2}{(p_1+p_2)^2}\Big(\frac{1}{(\ell-p_1)^2}-\frac{1}{\ell^2}\Big)
\nonumber\\
&+&\frac{p_1^2+p_1p_2}{(p_1+p_2)^2}\Big(\frac{1}{(\ell+p_2)^2}-\frac{1}{\ell^2}\Big)
-\frac{1}{(p_1+p_2)^2}\frac{2(\ell+p_2)\omega}{(\ell+p_2)^2(\ell-p_1)^2}\bigg),
\label{E.17}
\end{eqnarray}
\begin{eqnarray}
I^{tri}_3&=&2e^{-\frac{i}{2}(p_1\theta p_3+p_2\theta p_4)}\int \frac{d^3 \ell}{(2\pi)^3}e^{i\ell\theta(p_1+p_3)}
\bigg(-\frac{p_2^2p_1^2}{\ell^2(\ell+p_2)^2(\ell-p_1)^2}+\frac{p_1^2 }{\ell^2(\ell-p_1)^2}
\nonumber\\
&+&\frac{p_2^2 }{\ell^2(\ell+p_2)^2}
+\frac{-3p_1p_2-p_1^2-p_2^2}{(\ell+p_2)^2(\ell-p_1)^2}
+\frac{p_2^2+p_1p_2}{(p_1+p_2)^2}\Big(\frac{1}{(\ell-p_1)^2}-\frac{1}{\ell^2}\Big)
\nonumber\\
&+&\frac{p_1^2+p_1p_2}{(p_1+p_2)^2}\Big(\frac{1}{(\ell+p_2)^2}-\frac{1}{\ell^2}\Big)
-\frac{1}{(p_1+p_2)^2}\frac{2(\ell+p_2)\omega}{(\ell+p_2)^2(\ell-p_1)^2}\bigg).
\label{E.18}
\end{eqnarray}

Above one can trivially see the existence of commutative limit for integral $I^{tri}_1$ in (\ref{E.16}), since there is no $\ell$-dependent NC phase part, i.e. that integral is planar. Yet we have to show explicitly the existence of commutative limits in integrals (\ref{E.17}) and (\ref{E.18}) respectively. To do that we first split integrals $I^{tri}_2$ and $I^{tri}_3$ into the following forms
\begin{eqnarray}
I_2^{tri}&=&-e^{\frac{i}{2}(p_1\theta p_2+p_3\theta p_4)}
\Big(\widehat I^{tri}(-p_1-p_2)+{\tilde I}^{tri}(p_1+p_2)\Big), 
\label{E.19}\\
I_3^{tri}&=&2e^{-\frac{i}{2}(p_1\theta p_3+p_4\theta p_2)}
\Big(\widehat I^{tri}(p_1+p_3)+{\tilde I}^{tri}(-p_1-p_3)\Big), 
\label{E.20}
\end{eqnarray}
where:
\begin{eqnarray}
{\widehat I^{tri}}(\kappa)&=&\int\frac{d^3 \ell}{(2\pi)^3}e^{i\ell\theta\kappa}
\bigg(-\frac{p_2^2p_1^2}{\ell^2(\ell+p_2)^2(\ell-p_1)^2}+\frac{p_1^2 }{\ell^2(\ell-p_1)^2}
+\frac{p_2^2 }{\ell^2(\ell+p_2)^2}
\nonumber\\
&+&
\frac{-3p_1p_2-p_1^2-p_2^2}{(\ell+p_2)^2(\ell-p_1)^2}\bigg),
\label{E.21}\\
{\tilde I}^{tri}(\kappa)&=& \int\frac{d^3 \ell}{(2\pi)^3}e^{-i\ell\theta\kappa}
\bigg(\frac{p_2^2+p_1p_2}{(p_1+p_2)^2}\Big(\frac{1}{(\ell-p_1)^2}-\frac{1}{\ell^2}\Big)
\nonumber\\
&+&\frac{p_1^2+p_1p_2}{(p_1+p_2)^2}\Big(\frac{1}{(\ell+p_2)^2}-\frac{1}{\ell^2}\Big)
-\frac{1}{(p_1+p_2)^2}\frac{2(\ell+p_2)\omega}{(\ell+p_2)^2(\ell-p_1)^2}\bigg),
\label{E.22}
\end{eqnarray}
In $I_2^{tri}$, $\kappa=p_1+p_2$, and in $I_3^{tri}$, $\kappa=p_1+p_3$, being independent on the NC parameter
$\theta^{\mu\nu}$.

Here the absence of tadpole type integrals in $\widehat I^{tri}(\kappa)$ (\ref{E.21}) means that all four terms in $\widehat I^{tri}(\kappa)$ are UV finite by power-counting rule, thus have commutative limits. But four of those in ${\tilde I}^{tri}(\kappa)$  (\ref{E.22}) are tadpole type --not UV finite by power-counting rule--, thus at first look not having the  commutative limit. However explicit calculation of complete ${\tilde I}^{tri}(\kappa)$ from (\ref{E.22}) shows
\begin{eqnarray}
{\tilde I}^{tri}(\kappa)&=&\frac{1}{4\pi}\frac{\big((\theta\kappa)^2\big)^{-\frac{1}{2}}}{(p_1+p_2)^2}
\Big(\big(p_2^2+p_1p_2\big)p_1\theta\kappa-\big(p_1^2+p_1p_2\big)p_2\theta\kappa-
\omega\theta\kappa\Big)
+{\cal O}(\theta^1).
\label{E.23}
\end{eqnarray}
Since $\omega\theta\kappa=\big(p_2^2+p_1p_2\big)p_1\theta\kappa-\big(p_1^2+p_1p_2\big)p_2\theta\kappa$,
potentially IR divergent terms in (\ref{E.23}) cancels out, and we are finally left only with ${\tilde I}^{tri}(\kappa)={\cal O}(\theta^1)$. This way in (\ref{E.23}) we have shown a good commutative limit for ${\tilde I}^{tri}(\kappa)$ too. Thus, the commutative limits exist for all three integrals $I_1^{tri},I_2^{tri}$, and $I_3^{tri}$ (\ref{E.13})-(\ref{E.15}).

$\phantom{XXXXXXXXXXXXXXXXXXXXXXXXXXXXXXXXXXXXXX}$Q.E.D.

\subsection{Numerator of the 4 scalar 4-point function with box-loop}

Considering the parametrization of generic box-loop diagrams we compute them with choice of loop momentum variable  from  individual diagram $(\it 1.)$ in figure \ref{4scalar}, and obtain the following expression for the gauge boson box numerator $f_{(\it 1.)}^{box}(\ell,p_2,k_2,q)$:
\begin{equation}
\begin{split}
f_{(\it 1.)}^{box}(\ell,p_2,k_2,q)=&16\Big(\ell^2\big[(k_2 q)(p_2 q)+q^2(k_2 p_2)\big]+(\ell q)^2(k_2 p_2)
\\&+(\ell p_2)(\ell k_2)q^2-(\ell q)(\ell p_2)(k_2 q)-(\ell q)(\ell k_2)(p_2 q)\Big).
\end{split}
\label{E.24}
\end{equation}
After changing in-out momenta into all incoming momenta in accord with Figs.{\ref{fig:Fig21} and \ref{fig:Fig22}}: {$(p_2\to -p_2,k_2\to p_1,q\to\eta_i)$}, the $f_{(\it 1.)}^{box}(\ell,p_2,k_2,q)$ corresponds to the following numerator:
\begin{eqnarray}
{\cal N}(\ell,-p_2,p_1,\eta_i)&=&16\Big(-\ell^2\big[(p_2\eta_i)(p_1\eta_i)+\eta_i^2(p_1p_2)\big]-(\ell\eta_i)^2(p_1p_2)
\label{E.25}\\
&-&(p_2\ell)(p_1\ell)\eta_i^2+(p_1\ell)(\ell\eta_i)(p_2\eta_i)+(p_2\ell)(\ell\eta_i)(p_1\eta_i)\Big),\,i=1,...,4;
\nonumber
\end{eqnarray}
which is clearly much shorter/simpler with respect to the starting one in Eq. (\ref{boxfq}). Above $\eta_1=-p_2-p_4=p_3+p_1$,
$\eta_2=p_1+p_4=-p_3-p_2$, $\eta_3=-p_2-p_4=p_1+p_3\equiv\eta_1$, $\eta_4=p_1+p_4=-p_3-p_2\equiv\eta_2$.
Note that the above expression (\ref{E.25}) contain only two terms, both with power 2 in the loop momenta $\ell$. First is proportional to $\ell^2$, and second proportional to $\ell^\mu\ell^\nu$, respectively.

Integrands with the above numerators (\ref{E.24}) and/or (\ref{E.25}) can be decomposed into scalar box (\ref{D.3})-(\ref{D.5}), scalar triangle (\ref{B.33}), and vector triangle (\ref{C.12}) integrals using conventional tensor reduction. All these three types of integrals have been shown to have well-defined commutative limits in prior, B, C, and D, sections of the appendix, which ensures that box integrals share the same property.

\subsection{Numerators of 2scalar-2fermion 4-point functions with triangle- \& box-loops}

\begin{figure*}
\begin{center}
\includegraphics[width=10cm,angle=0]{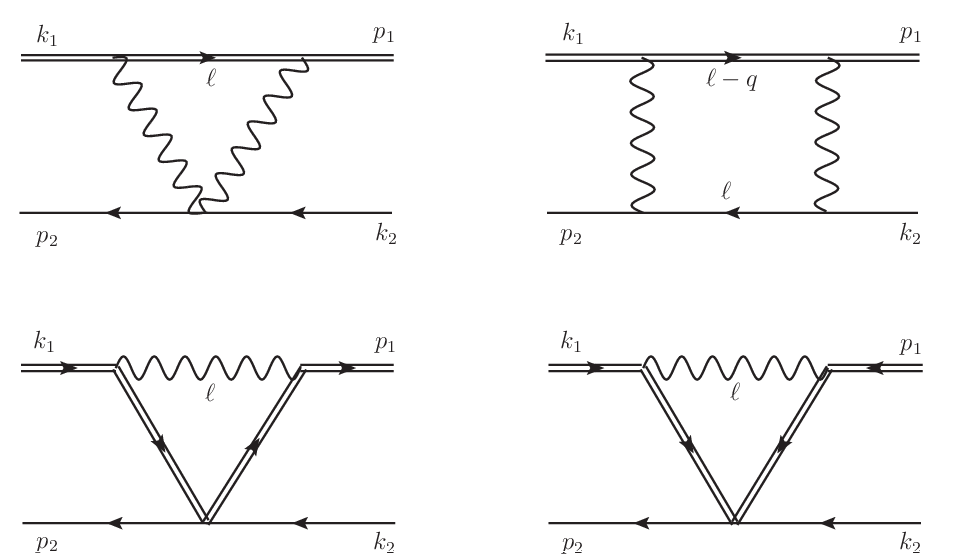}
\end{center}\caption{Generic samples, with specified fields and moment only, of the 2scalar-2fermion diagrams from figure \ref{2scalar2fermionbox} (1), and from figure \ref{2scalar2fermiontria} (1), (5), (6). Here, since the fermion momentum flow has to be respected we keep momentum assignment of incoming-outgoing momenta $\{k_1,k_2;p_1,p_2\}$ like in figures \ref{2scalar2fermionbox}, and \ref{2scalar2fermiontria}, respectively.}
\label{Sample2scalar2fermion}
\end{figure*}

Only triangle and box numerators are discussed in this sector. Sample diagrams are given in figure \ref{Sample2scalar2fermion}, with all additional structures (NC phase factors, R-symmetry indices, etc.)
being  neglected.
\begin{itemize}
\item $\bullet$ Fermion-2gauge bosons triangle-loop numerator is:
\begin{equation}
\begin{split}
{g^{tri}_1}=&\slashed{\ell}\Big(2\ell^2-\ell(p_1+k_1)\Big)-(\slashed{p}_1+\slashed{k}_1)\ell^2+\slashed{p}_1(\ell k_1)+\slashed{k}_1(\ell p_1)-\epsilon_{\alpha\beta\gamma} k_1^\alpha\ell^\beta p_1^\gamma,
\end{split}
\label{E.30}
\end{equation}
while from the 2fermion-gauge boson triangle-loop  we have numerator:
\begin{equation}
{g^{tri}_2}=2\slashed{\ell}(\ell+k_1)(\ell+p_1)+2\epsilon_{\alpha\beta\gamma} k_1^\alpha\ell^\beta p_1^\gamma.
\label{E.31}
\end{equation}
\item $\bullet$ Fermion-2gauge bosons-scalar box-loop numerator is than as follows:
\begin{equation}
\begin{split}
g^{box}&=4(\slashed{\ell}+\slashed{q})\Big((\ell p_2)(\ell k_2)-\ell^2(k_2 p_2)\Big)
+4\slashed{\ell}\Big((\ell p_2)(k_2 q)+(\ell k_2)(p_2 q)-2(\ell q)(k_2 p_2)\Big)
\\&+4\slashed{k}_2\Big((\ell q)(\ell p_2)-\ell^2(p_2 q)\Big)+4\slashed{p}_2\Big((\ell q)(\ell k_2)-\ell^2(k_2 q)\Big)
+4\epsilon_{\alpha\beta\gamma}k_2^\alpha\ell^\beta p_2^\gamma\Big(\ell(\ell-q)\Big).
\end{split}
\label{E.32}
\end{equation}
\end{itemize}
Finally, note that in the above numerators due to fermions we have terms proportional to the maximal power of 3 for the loop $\ell$-momentum dependence. However inspecting figures \ref{2scalar2fermionbox}-\ref{2scalar2fermiontria}  one see that in all diagrams there is at least one massless gauge boson exchange carrying/inducing power of $\ell^2$ peace in denominators reducing thus $\ell^3$ power  of numerator dependence down to the $\ell^1$ power dependence.

\noindent
{\bf Acknowledgments}\\
The work by C.P. Martin has been financially supported in part by the Spanish Ministry of Science, Innovation and Universities through grant PID2023-149834NB-I00. J.Trampetic would like thank Dieter L\"ust for many discussions and to acknowledge support of Max-Planck-Institute for Physics, M\"unchen,  for hospitality. Also J.T. would like to thank S.J. van Tongeren for private communications regarding reference \cite{ST}. The work of J.You has been supported by Croatian Science Foundation.



\begin{thebibliography}{99}

\bibitem{Martin:2017nhg}
  C.~P.~Martin, J.~Trampetic and J.~You,
 {\it Quantum noncommutative ABJM theory: first steps,}
 JHEP {\bf 1804} (2018) 070
  doi:10.1007/JHEP04(2018)070
  [arXiv:1711.09664 [hep-th]].

\bibitem{Aharony:2008ug}
  O.~Aharony, O.~Bergman, D.~L.~Jafferis and J.~Maldacena,
{\it N=6 superconformal Chern-Simons-matter theories, M2-branes and their gravity duals,}
JHEP {\bf 0810} (2008) 091,  doi:10.1088/1126-6708/2008/10/091
  [arXiv:0806.1218 [hep-th]].

\bibitem{Maldacena:1997re}
  J.~M.~Maldacena,
{\it The Large N limit of superconformal field theories and supergravity,}
  Int.\ J.\ Theor.\ Phys.\  {\bf 38} (1999) 1113,   [Adv.\ Theor.\ Math.\ Phys.\  {\bf 2} (1998) 231],  doi:10.1023/A:1026654312961  [hep-th/9711200].

\bibitem{Drukker:2010nc}
  N.~Drukker, M.~Marino and P.~Putrov,
{\it From weak to strong coupling in ABJM theory,}
  Commun.\ Math.\ Phys.\  {\bf 306} (2011) 511,  doi:10.1007/s00220-011-1253-6
  [arXiv:1007.3837 [hep-th]].

\bibitem{Dabholkar:2014wpa}
  A.~Dabholkar, N.~Drukker and J.~Gomes,
{\it Localization in supergravity and quantum $AdS_4/CFT_3$ holography,}
  JHEP {\bf 1410} (2014) 90,  doi:10.1007/JHEP10(2014)090
  [arXiv:1406.0505 [hep-th]].

\bibitem{Asano:2004hy}
E.~A.~Asano, L.~C.~T.~Brito, M.~Gomes, A.~Y.~Petrov and A.~J.~da Silva,
{\it Consistent interactions of the 2+1 dimensional noncommutative Chern-Simons field,}
Phys. Rev. D \textbf{71} (2005), 105005
doi:10.1103/PhysRevD.71.105005
[arXiv:hep-th/0410257 [hep-th]].

\bibitem{Ferrari:2005kx}
A.~F.~Ferrari, M.~Gomes, A.~Y.~Petrov and A.~J.~da Silva,
{\it Supersymmetric non-Abelian noncommutative Chern-Simons theory,}
Phys. Lett. B \textbf{638} (2006), 275-282
doi:10.1016/j.physletb.2006.05.031
[arXiv:hep-th/0511059 [hep-th]].

\bibitem{Ammon:2015wua}
  M.~Ammon and J.~Erdmenger,
 {\it Gauge/gravity duality: Foundations and applications,}
 Cambridge University Press. 2015.

\bibitem{Bea:2014yda}
  Y.~Bea, N.~Jokela, M.~Lippert, A.~V.~Ramallo and D.~Zoakos,
{\it Flux and Hall states in ABJM with dynamical flavors,}
  JHEP {\bf 1503} (2015) 009,  doi:10.1007/JHEP03(2015)009
  [arXiv:1411.3335 [hep-th]].

\bibitem{Bandres:2008ry}
  M.~A.~Bandres, A.~E.~Lipstein and J.~H.~Schwarz,
{\it Studies of the ABJM Theory in a Formulation with Manifest SU(4) R-Symmetry},
JHEP {\bf 0809} (2008) 027,  doi:10.1088/1126-6708/2008/09/027,  [arXiv:0807.0880 [hep-th]].

\bibitem{Kwon:2009ar}
  O.~K.~Kwon, P.~Oh and J.~Sohn,
   {\it Notes on Supersymmetry Enhancement of ABJM Theory,}
  JHEP {\bf 0908}, 093 (2009),  doi:10.1088/1126-6708/2009/08/093
  [arXiv:0906.4333 [hep-th]].

\bibitem{Buchbinder:2008vi}
  I.~L.~Buchbinder, E.~A.~Ivanov, O.~Lechtenfeld, N.~G.~Pletnev, I.~B.~Samsonov and B.~M.~Zupnik,
{\it ABJM models in N=3 harmonic superspace,}
  JHEP {\bf 0903} (2009) 096,  doi:10.1088/1126-6708/2009/03/096
  [arXiv:0811.4774 [hep-th]].

\bibitem{Buchbinder:2009dc}
  I.~L.~Buchbinder, E.~A.~Ivanov, O.~Lechtenfeld, N.~G.~Pletnev, I.~B.~Samsonov and B.~M.~Zupnik,
{\it Quantum N=3, d=3 Chern-Simons Matter Theories in Harmonic Superspace,}
  JHEP {\bf 0910} (2009) 075,  doi:10.1088/1126-6708/2009/10/075
  [arXiv:0909.2970 [hep-th]].

\bibitem{Arkani-Hamed:2017mur}
  N.~Arkani-Hamed, Y.~Bai, S.~He and G.~Yan,
{\it Scattering Forms and the Positive Geometry of Kinematics, Color and the Worldsheet,}
JHEP {\bf 1805} (2018) 096, doi:10.1007/JHEP05(2018)096,  [arXiv:1711.09102 [hep-th]].

\bibitem{Bargheer:2010hn}
  T.~Bargheer, F.~Loebbert and C.~Meneghelli,
{\it Symmetries of Tree-level Scattering Amplitudes in N=6 Superconformal Chern-Simons Theory,}
Phys.\ Rev.\ D {\bf 82} (2010) 045016,  doi:10.1103/PhysRevD.82.045016,  [arXiv:1003.6120 [hep-th]].

\bibitem{Bargheer:2012cp}
  T.~Bargheer, N.~Beisert, F.~Loebbert and T.~McLoughlin,
{\it Conformal Anomaly for Amplitudes in $\mathcal{N}=6$ Superconformal Chern-Simons Theory,}
J.\ Phys.\ A {\bf 45} (2012) 475402,  doi:10.1088/1751-8113/45/47/475402  [arXiv:1204.4406 [hep-th]].

\bibitem{Bianchi:2012cq}
  M.~S.~Bianchi, M.~Leoni, A.~Mauri, S.~Penati and A.~Santambrogio,
{\it One Loop Amplitudes In ABJM,}
JHEP {\bf 1207} (2012) 029,  doi:10.1007/JHEP07(2012)029  [arXiv:1204.4407 [hep-th]].

\bibitem{Lee:2010du}
  S.~Lee,
{\it Yangian Invariant Scattering Amplitudes in Supersymmetric Chern-Simons Theory,}
Phys.\ Rev.\ Lett.\  {\bf 105} (2010) 151603,  doi:10.1103/PhysRevLett.105.151603  [arXiv:1007.4772 [hep-th]].

\bibitem{Elvang:2014fja}
  H.~Elvang, Y.~t.~Huang, C.~Keeler, T.~Lam, T.~M.~Olson, S.~B.~Roland and D.~E.~Speyer,
{\it Grassmannians for scattering amplitudes in 4d $\mathcal{N}=4$ SYM and 3d ABJM,}
JHEP {\bf 1412} (2014) 181, doi:10.1007/JHEP12(2014)181,  [arXiv:1410.0621 [hep-th]].

\bibitem{Beisert:2017pnr}
  N.~Beisert, A.~Garus and M.~Rosso,
{\it Yangian Symmetry and Integrability of Planar N=4 Supersymmetric Yang-Mills Theory,}
Phys.\ Rev.\ Lett.\  {\bf 118} (2017) no.14,  141603,
doi:10.1103/PhysRevLett.118.141603  [arXiv:1701.09162 [hep-th]].

\bibitem{Seiberg:1999vs}
N.~Seiberg and E.~Witten,
{\it String theory and noncommutative geometry,}
JHEP \textbf{09} (1999), 032, doi:10.1088/1126-6708/1999/09/032, [arXiv:hep-th/9908142 [hep-th]].

\bibitem{Gomis:2000zz}
  J.~Gomis and T.~Mehen,
{\it Space-time noncommutative field theories and unitarity,}
  Nucl.\ Phys.\  B {\bf 591}, 265 (2000), [arXiv:hep-th/0005129].

\bibitem{Aharony:2000gz}
  O.~Aharony, J.~Gomis, T.~Mehen,
{\it On theories with lightlike noncommutativity,}
JHEP{\bf 0009}, 023 (2000),  [hep-th/0006236].

\bibitem{Trampetic:2003fu}
J.~Trampetic and J.~Wess,
{\it Particle physics in the new millennium. Proceedings, 8th Adriatic Meeting, Dubrovnik, Croatia, September 4-14, 2001,}
Lect. Notes Phys. \textbf{616} (2003), pp.1-365
doi:10.1007/3-540-36539-7

\bibitem{Szabo:2001kg}
  R.~J.~Szabo,
{\it Quantum field theory on noncommutative spaces,}
  Phys.\ Rept.\  {\bf 378} (2003) 207,  doi:10.1016/S0370-1573(03)00059-0, [hep-th/0109162].

\bibitem{Trampetic:2005ib}
J.~Trampetic and J.~Wess,
{\it Proceedings, 9th Adriatic Meeting on Particle Physics and the Universe: Dubrovnik, Croatia, September 4-14, 2003,}
Springer Proc. Phys. \textbf{98} (2005), pp.1-495, doi:10.1007/b138101

\bibitem{Raju:2009yx}
  S.~Raju,
  {\it The Noncommutative S-Matrix,}
  JHEP {\bf 0906} (2009) 005,  doi:10.1088/1126-6708/2009/06/005,   [arXiv:0903.0380].

\bibitem{Huang:2010fc}
  J.~H.~Huang, R.~Huang and Y.~Jia,
 {\it Tree amplitudes of noncommutative U(N) Yang-Mills Theory,}
  J.\ Phys.\ A {\bf 44} (2011) 425401,  doi:10.1088/1751-8113/44/42/425401,  [arXiv:1009.5073 [hep-th]].

\bibitem{Martin:2016hji}
  C.~P.~Martin, J.~Trampetic and J.~You,
 {\it Equivalence of quantum field theories related by the $\theta$-exact Seiberg-Witten map,}
 Phys.\ Rev.\ D {\bf 94} (2016) no.4,  041703, doi:10.1103/PhysRevD.94.041703, [arXiv:1606.03312 [hep-th]].

\bibitem{Martin:2016saw}
  C.~P.~Martin, J.~Trampetic and J.~You,
  {\it Quantum duality under the $\theta$-exact Seiberg-Witten map,}
  JHEP {\bf 1609} (2016) 052,  doi:10.1007/JHEP09(2016)052, [arXiv:1607.01541 [hep-th]].

\bibitem{Mizera:2019blq}
  S.~Mizera,
  {\it Kinematic Jacobi Identity is a Residue Theorem: Geometry of Color-Kinematics Duality for Gauge and Gravity Amplitudes,}
  Phys.\ Rev.\ Lett.\  {\bf 124} (2020) no.14,  141601,
  doi:10.1103/PhysRevLett.124.141601,  [arXiv:1912.03397 [hep-th]].

\bibitem{Latas:2020nji}
D.~Latas, J.~Trampeti\'c and J.~You,
{\it Seiberg-Witten map invariant scatterings},
Phys. Rev. D \textbf{104} (2021) no.1, 015021, doi:10.1103/PhysRevD.104.015021
[arXiv:2012.07891 [hep-ph]].

\bibitem{Trampetic:2021awu}
J.~Trampeti\'c and J.~You,
{\it Seiberg-Witten maps and scattering amplitudes of NCQED,}
Phys. Rev. D \textbf{105} (2022) no.7, 075016
doi:10.1103/PhysRevD.105.075016, [arXiv:2111.04154 [hep-th]].

\bibitem{Trampetic:2022tij}
J.~Trampeti\'c and J.~You,
{\it Revisiting NCQED and scattering amplitudes},
Eur. Phys. J. Spec. Top. https://doi.org/10.1140/ep js/s11734-023-00837-1,
[arXiv:2209.06636 [hep-th]].

\bibitem{Pisarski:1985yj}
  R.~D.~Pisarski and S.~Rao,
 {\it Topologically Massive Chromodynamics in the Perturbative Regime,}
  Phys.\ Rev.\ D {\bf 32} (1985) 2081,  doi:10.1103/PhysRevD.32.2081.

\bibitem{Martin:1999aq}
C.P.~Martin, D.~Sanchez-Ruiz,
{\it The One-loop UV Divergent Structure of {\rm U(1)} Yang-Mills Theory on Noncommutative $R^4$},
Phys. Rev. Lett.  {\bf 83} (1999) 476--479, [hep-th/9903077].

\bibitem{Bigatti:1999iz}
D.~Bigatti and L.~Susskind,
{\it Magnetic fields, branes and noncommutative geometry,}
Phys. Rev. D \textbf{62} (2000), 066004
doi:10.1103/PhysRevD.62.066004
[arXiv:hep-th/9908056 [hep-th]].

  \bibitem{Matusis:2000jf}
A.~Matusis, L.~Susskind, and N.~Toumbas,
{\it The IR/UV connection in the  non-commutative gauge theories},
 {\em JHEP} {\bf 12} (2000) 002,
[arXiv:hep-th/0002075].

\bibitem{Seiberg:2000gc}
N.~Seiberg, L.~Susskind and N.~Toumbas,
{\it Space-time noncommutativity and causality,}
JHEP \textbf{06} (2000), 044
doi:10.1088/1126-6708/2000/06/044
[arXiv:hep-th/0005015 [hep-th]].

\bibitem{Seiberg:2000ms}
N.~Seiberg, L.~Susskind and N.~Toumbas,
{\it Strings in background electric field, space / time noncommutativity and a new noncritical string theory,}
JHEP \textbf{06} (2000), 021
doi:10.1088/1126-6708/2000/06/021
[arXiv:hep-th/0005040 [hep-th]].

    \bibitem{Minwalla:1999px}
  S.~Minwalla, M.~Van Raamsdonk and N.~Seiberg,
{\it Noncommutative perturbative dynamics,}
  JHEP {\bf 0002}, 020 (2000), [arXiv:hep-th/9912072].

\bibitem{Hayakawa:1999yt}
M.~Hayakawa,
{\it Perturbative analysis on infrared aspects of noncommutative QED on R**4,}
Phys. Lett. B \textbf{478} (2000), 394-400
doi:10.1016/S0370-2693(00)00242-2
[arXiv:hep-th/9912094 [hep-th]].

\bibitem{VanRaamsdonk:2000rr}
  M.~Van Raamsdonk and N.~Seiberg,
{\it Comments on noncommutative perturbative dynamics,} JHEP {\bf 0003} (2000) 035,
doi:10.1088/1126-6708/2000/03/035, [hep-th/0002186].

\bibitem{VanRaamsdonk:2001jd}
M.~Van Raamsdonk,
{\it The Meaning of infrared singularities in noncommutative gauge theories,'}
JHEP \textbf{11} (2001), 006, doi:10.1088/1126-6708/2001/11/006, [arXiv:hep-th/0110093 [hep-th]].

\bibitem{Ferrari:2004ex}
A.~F.~Ferrari, H.~O.~Girotti, M.~Gomes, A.~Y.~Petrov, A.~A.~Ribeiro, V.~O.~Rivelles and A.~J.~da Silva,
{\it Towards a consistent noncommutative supersymmetric Yang-Mills theory: Superfield covariant analysis,}
Phys. Rev. D \textbf{70} (2004), 085012
doi:10.1103/PhysRevD.70.085012
[arXiv:hep-th/0407040 [hep-th]].

    \bibitem{Schupp:2008fs}
P.~Schupp and J.~You,
{\it {UV/IR mixing in noncommutative QED defined by  Seiberg-Witten map}},
  JHEP {\bf 08} (2008) 107,

\bibitem{Horvat:2011bs}
  R.~Horvat, A.~Ilakovac, J.~Trampetic and J.~You,
{\it On UV/IR mixing in noncommutative gauge field theories,}
JHEP {\bf 12} (2011) 081, arXiv:1109.2485 [hep-th].

\bibitem{Horvat:2013rga}
R.~Horvat, A.~Ilakovac, J.~Trampetic and J.~You,
{\it Self-energies on deformed spacetimes,}
JHEP \textbf{11} (2013), 071
doi:10.1007/JHEP11(2013)071
[arXiv:1306.1239 [hep-th]].


\bibitem{Horvat:2015aca}
  R.~Horvat, J.~Trampetic and J.~You,
{\it Photon self-interaction on deformed spacetime,}
  Phys.\ Rev.\ D {\bf 92} (2015) no.12,  125006,  doi:10.1103/PhysRevD.92.125006, [arXiv:1510.08691 [hep-th]].


\bibitem{Martin:2016zon}
  C.~P.~Martin, J.~Trampetic and J.~You,
{\it Super Yang-Mills and $\theta$-exact Seiberg-Witten map: absence of quadratic noncommutative IR divergences,}
JHEP {\bf 1605} (2016) 169,  doi:10.1007/JHEP05(2016)169, [arXiv:1602.01333 [hep-th]].

\bibitem{Martin:2020ddo}
C.~P.~Martin, J.~Trampeti\'c and J.~You,
{\it UV/IR mixing in noncommutative SU(N) Yang\textendash{}Mills theory,}
Eur. Phys. J. C \textbf{81} (2021) no.10, 878, doi:10.1140/epjc/s10052-021-09686-5,
[arXiv:2012.09119 [hep-th]].


\bibitem{Grosse:2005iz}
  H.~Grosse and M.~Wohlgenannt,
 {\it On $\kappa$-deformation and UV/IR mixing,}
  Nucl.\ Phys.\ B {\bf 748} (2006) 473,  doi:10.1016/j.nuclphysb.2006.05.004, [hep-th/0507030].

\bibitem{Magnen:2008pd}
  J.~Magnen, V.~Rivasseau and A.~Tanasa,
{\it Commutative limit of a renormalizable noncommutative model,}
  Europhys.\ Lett.\  {\bf 86} (2009) 11001,  [arXiv:0807.4093 [hep-th]].
  %
\bibitem{Blaschke:2009aw}
  D.~N.~Blaschke, H.~Grosse, E.~Kronberger, M.~Schweda and M.~Wohlgenannt,
{\it Loop Calculations for the Non-Commutative $\rm U^\star(1)$ Gauge Field Model with Oscillator Term,}
  Eur.\ Phys.\ J.\ C {\bf 67} (2010) 575,  [arXiv:0912.3642 [hep-th]].

\bibitem{Meljanac:2011cs}
  S.~Meljanac, A.~Samsarov, J.~Trampetic and M.~Wohlgenannt,
{\it Scalar field propagation in the $\phi^4$ kappa-Minkowski model,}
  JHEP {\bf 12} (2011) 010,  arXiv:1111.5553 [hep-th].

\bibitem{Meljanac:2017grw}
  S.~Meljanac, S.~Mignemi, J.~Trampetic and J.~You,
{\it Nonassociative Snyder $\phi^4$ Quantum Field Theory,}
  Phys.\ Rev.\ D {\bf 96} (2017) no.4,  045021,  doi:10.1103/PhysRevD.96.045021, [arXiv:1703.10851 [hep-th]].

\bibitem{Meljanac:2017jyk}
  S.~Meljanac, S.~Mignemi, J.~Trampetic and J.~You,
{\it UV-IR mixing in nonassociative Snyder $\phi^4$ theory,}
  Phys.\ Rev.\ D {\bf 97} (2018) no.5,  055041,  doi:10.1103/PhysRevD.97.055041, [arXiv:1711.09639 [hep-th]].

\bibitem{Hanada:2014ima}
M.~Hanada and H.~Shimada,
{\it On the continuity of the commutative limit of the 4d N=4 non-commutative super Yang{\textendash}Mills theory,}
Nucl. Phys. B \textbf{892} (2015), 449-474, doi:10.1016/j.nuclphysb.2015.01.016, [arXiv:1410.4503 [hep-th]].

\bibitem{Meier:2023kzt}
T.~Meier and S.~J.~van Tongeren,
{\it Quadratic Twist-Noncommutative Gauge Theory,}
Phys. Rev. Lett. \textbf{131} (2023) no.12, 121603
doi:10.1103/PhysRevLett.131.121603
[arXiv:2301.08757 [hep-th]].

\bibitem{Meier:2023lku}
T.~Meier and S.~J.~van Tongeren,
{\it Gauge theory on twist-noncommutative spaces,}
JHEP \textbf{12} (2023), 045
doi:10.1007/JHEP12(2023)045
[arXiv:2305.15470 [hep-th]].

\bibitem{ST} Stijn J. van Tongeren private communications.

\bibitem{Hewett:2000zp}
  J.~L.~Hewett, F.~J.~Petriello and T.~G.~Rizzo,
 {\it Signals for noncommutative interactions at linear colliders},
  Phys.\ Rev.\ D {\bf 64} (2001) 075012,  [hep-ph/0010354].

\bibitem{Mathews:2000we}
  P.~Mathews,
{\it Compton scattering in noncommutative space-time at the NLC,}
  Phys.\ Rev.\ D {\bf 63} (2001) 075007, [hep-ph/0011332].

\bibitem{Behr:2002wx}
W.~Behr, N.~G.~Deshpande, G.~Duplancic, P.~Schupp, J.~Trampetic and J.~Wess,
{\it The Z ---{\ensuremath{>}} gamma gamma, g g decays in the noncommutative standard model,}
Eur. Phys. J. C \textbf{29} (2003), 441-446
doi:10.1140/epjc/s2003-01207-4
[arXiv:hep-ph/0202121 [hep-ph]].

\bibitem{Schupp:2002up}
P.~Schupp, J.~Trampetic, J.~Wess, and G.~Raffelt,
{\it {The photon neutrino  interaction in noncommutative gauge field theory and astrophysical bounds}},
   Eur. Phys. J. {\bf C36} (2004) 405--410.

\bibitem{Minkowski:2003jg}
P.~Minkowski, P.~Schupp, and J.~Trampetic,
{\it Neutrino dipole moments and  charge radii in noncommutative space-time},
Eur. Phys. J. {\bf C37}  (2004) 123--128.

\bibitem{Abel:2006wj}
  S.~A.~Abel, J.~Jaeckel, V.~V.~Khoze and A.~Ringwald,
{\it Vacuum birefringence as a probe of Planck scale noncommutativity,}
  JHEP {\bf 0609}, 074 (2006), [arXiv:hep-ph/0607188].

\bibitem{Horvat:2010sr}
  R.~Horvat, D.~Kekez, J.~Trampetic,
{\it Spacetime noncommutativity and ultra-high energy cosmic ray experiments,}
 Phys.\ Rev.\  {\bf D83 } (2011)  065013,  [arXiv:1005.3209 [hep-ph]].

\bibitem{Horvat:2010km}
  R.~Horvat, J.~Trampetic,
{\it Constraining noncommutative field theories with holography,}
  JHEP {\bf 1101 } (2011)  112,  [arXiv:1009.2933 [hep-ph]].

\bibitem{Horvat:2011iv}
R.~Horvat, D.~Kekez, P.~Schupp, J.~Trampetic and J.~You,
{\it Photon-neutrino interaction in theta-exact covariant noncommutative field theory,}
Phys. Rev. D \textbf{84} (2011), 045004, doi:10.1103/PhysRevD.84.045004, [arXiv:1103.3383 [hep-ph]].


\bibitem{Horvat:2011wh}
R.~Horvat and J.~Trampetic,
{\it A bound on the scale of spacetime noncommutativity from the reheating phase after inflation,}
Phys. Lett. B \textbf{710} (2012), 219-222, doi:10.1016/j.physletb.2012.02.062, [arXiv:1111.6436 [hep-ph]].

   \bibitem{Horvat:2011qn}
R.~Horvat, A.~Ilakovac, P.~Schupp, J.~Trampeti\'{c}, and J.~You,
{\it Yukawa couplings and seesaw neutrino masses in noncommutative gauge theory},
 {\em Phys. Lett.} {\bf B715}, 340-347 (2012).

\bibitem{Horvat:2011qg}
R.~Horvat, A.~Ilakovac, P.~Schupp, J.~Trampetic and J.~You,
{\it Neutrino propagation in noncommutative spacetimes,}
JHEP \textbf{04} (2012), 108, doi:10.1007/JHEP04(2012)108, [arXiv:1111.4951 [hep-th]].

\bibitem{Horvat:2012vn}
  R.~Horvat, A.~Ilakovac, D.~Kekez, J.~Trampetic and J.~You,
{\it Forbidden and Invisible Z Boson Decays in Covariant theta-exact Noncommutative Standard Model},
J.\ Phys.\ G: Nucl. Part. Phys. {\bf 41} (2014) 055007, arXiv:1204.6201 [hep-ph].

\bibitem{Trampetic:2015zma}
J.~Trampetic and J.~You,
{\it $\theta$-exact Seiberg-Witten maps at order $e^3$,}
Phys. Rev. D \textbf{91} (2015) no.12, 125027
doi:10.1103/PhysRevD.91.125027
[arXiv:1501.00276 [hep-th]].

\bibitem{Horvat:2017gfm}
  R.~Horvat, J.~Trampetic and J.~You,
 {\it Spacetime Deformation Effect on the Early Universe and the PTOLEMY Experiment,}
  Phys.\ Lett.\ B {\bf 772} (2017) 130,  doi:10.1016.j.physletb.2017.06.028, [arXiv:1703.04800 [hep-ph]].

\bibitem{Horvat:2017aqf}
  R.~Horvat, J.~Trampetic and J.~You,
{\it Inferring type and scale of noncommutativity from the PTOLEMY experiment,}
Eur.\ Phys.\ J.\ C {\bf 78} (2018) no.7,  572, doi:10.1140/epjc/s10052-018-6052-1, [arXiv:1711.09643 [hep-ph]].

\bibitem{Selvaganapathy:2019jkm}
J.~Selvaganapathy, P.~Konar and P.~K.~Das,
{\it Inferring the covariant $\Theta$-exact noncommutative coupling in the top quark pair production at linear colliders,}
JHEP \textbf{06} (2019), 108, doi:10.1007/JHEP06(2019)108, [arXiv:1903.03478 [hep-ph]].

\bibitem{Horvat:2020ycy}
  R.~Horvat, D.~Latas, J.~Trampetic and J.~You,
 {\it Light-by-Light Scattering and Spacetime Noncommutativity,}
   Phys.\ Rev.\ D {\bf 101} (2020)  095035,  doi: 10.1103/PhysRevD.101.095035,
  arXiv:2002.01829 [hep-ph].

\bibitem{Trampetic:2023qfv}
J.~Trampeti\'c and J.~You,
{\it Remarks on Compton forward scattering singularity in the NCQED,}
Eur. Phys. J. C \textbf{85} (2025) no.3, 268
doi:10.1140/epjc/s10052-025-13951-2, [arXiv:2301.06343 [hep-ph]].

\bibitem{Lust:2017wrl}
  D.~Lust and E.~Palti,
 {\it Scalar Fields, Hierarchical UV/IR Mixing and The Weak Gravity Conjecture,}
  JHEP {\bf 1802} (2018) 040, doi:10.1007/JHEP02(2018)040,  [arXiv:1709.01790 [hep-th]].

\bibitem{Craig:2019zbn}
N.~Craig and S.~Koren,
{\it IR Dynamics from UV Divergences: UV/IR Mixing, NCFT, and the Hierarchy Problem,}
JHEP \textbf{03} (2020), 037, doi:10.1007/JHEP03(2020)037, [arXiv:1909.01365 [hep-ph]].

\bibitem{Cribiori:2025oek}
N.~Cribiori and F.~Tonioni,
{\it Cosmological constraints from UV/IR mixing,}
[arXiv:2507.02738 [hep-th]].

\bibitem{Martin:2025hzm}
C.~P.~Martin,
{\it Entanglement through high-energy scattering in noncommutative quantum electrodynamics,}
[arXiv:2506.15350 [hep-th]].



\bibitem{Fradkin:2002qw}
  E.~Fradkin, V.~Jejjala and R.~G.~Leigh,
{\it Noncommutative Chern-Simons for the quantum Hall system and duality,}
  Nucl.\ Phys.\ B {\bf 642} (2002) 483,  doi:10.1016/S0550-3213(02)00616-8
  [cond-mat/0205653 [cond-mat.mes-hall]].

\bibitem{Velo}
G.~Velo and A.~S.~Wightman,
{\it Renormalization Theory. Proceedings, NATO Advanced Study Institute: Erice, 17-31 August, 1975,}
Springer, Dordrecht (1976),
doi:10.1007/978-94-010-1490-8.

\bibitem{Rudin}
W. Rudin, "Real and Complex Analysis" 3rd Edition, McGraw-Hill 1986.

\bibitem{Leoni:2010az}
M.~Leoni and A.~Mauri,
{\it On the infrared behaviour of 3d Chern-Simons theories in N=2 superspace,}
JHEP \textbf{11} (2010), 128, doi:10.1007/JHEP11(2010)128, arXiv:1006.2341 [hep-th]].

\bibitem{Ossola:2006us}
G.~Ossola, C.~G.~Papadopoulos and R.~Pittau,
{\it Reducing full one-loop amplitudes to scalar integrals at the integrand level,}
Nucl. Phys. B \textbf{763} (2007), 147-169, doi:10.1016/j.nuclphysb.2006.11.012
[arXiv:hep-ph/0609007 [hep-ph]].




\end{thebibliography}
\end{document}